\documentclass[aps,prb,twocolumn]{revtex4-1}
\usepackage{amsmath,amssymb,bm,epsfig,graphicx,longtable}
\begin{document}

\title{Assessment of the GLLB-SC potential for solid-state properties and attempts for improvement}
\author{Fabien Tran}
\author{Sohaib Ehsan}
\author{Peter Blaha}
\affiliation{Institute of Materials Chemistry, Vienna University of Technology,
Getreidemarkt 9/165-TC, A-1060 Vienna, Austria}

\begin{abstract}

Based on the work of Gritsenko \textit{et al}. (GLLB) [Phys. Rev. A \textbf{51},
1944 (1995)], the method of Kuisma \textit{et al}. [Phys. Rev. B \textbf{82},
115106 (2010)] to calculate the band gap in solids was shown to be much more
accurate than the common local density approximation (LDA) and
generalized gradient approximation (GGA). The main feature of the GLLB-SC
potential (SC stands for solid and correlation) is to lead to a nonzero
derivative discontinuity that can be conveniently
calculated and then added to the Kohn-Sham band gap for a comparison with the
experimental band gap.
In this work, a thorough comparison of GLLB-SC with other methods,
e.g., the modified Becke-Johnson (mBJ) potential [F. Tran and P. Blaha,
Phys. Rev. Lett. \textbf{102}, 226401 (2009)], for electronic, magnetic, and
density-related properties is presented. It is shown that for the band gap,
GLLB-SC does not perform as well as mBJ for systems with a small band gap
and strongly correlated systems, but is on average of similar accuracy
as hybrid functionals. The results on itinerant metals indicate that
GLLB-SC overestimates significantly the magnetic moment (much more than mBJ does),
but leads to excellent results for the electric field gradient, for which
mBJ is in general not recommended. In the aim of improving the results,
variants of the GLLB-SC potential are also tested.

\end{abstract}

\maketitle

\section{\label{introduction}Introduction}

The great success of the Kohn-Sham (KS) density functional theory (DFT) method
\cite{HohenbergPR64,KohnPR65} for the calculation of properties of electronic
systems is due to the fact that in many circumstances, results of sufficient
accuracy can be obtained at much lower cost compared to the supposedly more
reliable post-Hartree-Fock\cite{Jensen} or Green's function based methods.\cite{Bechstedt}
However, in KS-DFT the exchange (x) and correlation (c) effects are approximated and
among the hundreds of approximations available,\cite{MarquesCPC12} one has to
choose an appropriate one for the system at hand.
This choice is crucial when the trends in the results may depend significantly
on the approximation, but the main problem is that it is by far not always obvious
which approximation to choose. Therefore, the search for approximations
that are more broadly accurate is a very active research topic,
\cite{KuemmelRMP08,CohenCR12,BurkeJCP12,BeckeJCP14} in particular since KS-DFT
is used in many areas of science.

The focus of the present work is on the properties which depend directly on the
xc potential $v_{\text{xc},\sigma}$ ($\sigma$ is the spin index) in the KS
equations, namely, the electronic structure, magnetic moment, and electron density.
Given an xc energy functional $E_{\text{xc}}$, the variational principle
requires $v_{\text{xc},\sigma}$ to be the functional derivative of $E_{\text{xc}}$.
Depending on the type of approximation chosen for $E_{\text{xc}}$ and
the way the functional derivative is taken (with respect to the electron density
$\rho_{\sigma}$ or the orbitals $\psi_{i\sigma}$),
the potential $v_{\text{xc},\sigma}$ can be of
different nature: multiplicative or nonmultiplicative.\cite{KuemmelRMP08,YangJCP12}
Strictly speaking, the potential $v_{\text{xc},\sigma}$ in the KS method is
multiplicative, while the generalized KS framework\cite{SeidlPRB96} (gKS) includes also
nonmultiplicative potentials.

It is well-known that with the \textit{exact multiplicative potential},
the KS band gap $E_{\text{g}}^{\text{KS}}$, defined as the conduction band minimum
(CBM) minus the valence band maximum (VBM), is not equal to the true
experimental (i.e., quasi-particle) band gap $E_{\text{g}}=I-A$
(ionization potential $I$ minus electron affinity $A$) since they differ by
the so-called xc derivative discontinuity
$\Delta_{\text{xc}}$,\cite{PerdewPRL82,ShamPRL83}
\begin{equation}
E_{\text{g}} = E_{\text{g}}^{\text{KS}} + \Delta_{\text{xc}},
\label{Eg}
\end{equation}
which can be of the same order of magnitude as the
gap.\cite{GodbyPRL86,GruningJCP06,GruningPRB06,KlimesJCP14}
Since $\Delta_{\text{xc}}$ is positive, the exact KS gap
$E_{\text{g}}^{\text{KS}}$ is (much) smaller than $E_{\text{g}}$.
Therefore, within the KS framework a comparison
with the experimental gap should formally be done only when
$\Delta_{\text{xc}}$ is added to the KS band gap.
With the functionals of the local density approximation (LDA) and
generalized gradient approximation (GGA) that are commonly used
for structure optimization or binding energy calculation
(e.g., PBE\cite{PerdewPRL96}),
$E_{\text{g}}^{\text{KS}}$ is usually much smaller than
$E_{\text{g}}$ (see, e.g., Ref.~\onlinecite{HeydJCP05}),
while adding $\Delta_{\text{xc}}$ (calculated in some way,
which is possible for finite systems\cite{AndradePRL11,ChaiPRL13,KraislerJCP14})
improves the agreement with experiment.
Note that interestingly, the LDA and standard GGA methods
lead to KS band gaps that do not differ
that much from accurate KS band
gaps.\cite{GodbyPRL86,GruningJCP06,GruningPRB06,KlimesJCP14}

Semilocal multiplicative xc potentials that are more useful for band gap
calculation have been proposed,
\cite{EngelPRB93,ZhaoPRB99,BeckeJCP06,TranJPCM07,FerreiraPRB08,TranPRL09,ArmientoPRL13,SinghPRB16,FinzelIJQC17,VermaJPCL17,MoralesGarciaJPCC17}
however since usually this is still $E_{\text{g}}^{\text{KS}}$ that is compared to
the experimental value of $E_{\text{g}}$ (no $\Delta_{\text{xc}}$ added
to $E_{\text{g}}^{\text{KS}}$), the better agreement
is achieved at the cost of having a potential $v_{\text{xc}}$ that may show
features that are most likely unphysical and not present in the exact
KS potential (see, e.g., Fig.~2 of Ref.~\onlinecite{TranJPCA17}).
Then, this may possibly lead to a bad description of
properties other than the band gap. For instance, the modified Becke-Johnson
(mBJ) potential,\cite{TranPRL09} which has been very successful for band gap
prediction,\cite{SinghPRB10,BotanaPRB12,ZhuPRB12,KollerPRB12,CamargoMartinezPRB12,JiangJCP13,DixitJPCM12,WaroquiersPRB13,TranJPCA17}
has also shown to be sometimes rather inaccurate
for other properties, e.g., band widths\cite{WaroquiersPRB13} or the
magnetic moment of itinerant metals.\cite{KollerPRB11}
This is the consequence of constructing a potential that is not well-founded
from the physical point of view.

Thus, when using a multiplicative potential
the proper calculation of $E_{\text{g}}=I-A$ should consist of a nonzero
derivative discontinuity that is added to
$E_{\text{g}}^{\text{KS}}$ [Eq.~(\ref{Eg})].
In Refs.~\onlinecite{KuismaPRB10,AndradePRL11,ChaiPRL13,KraislerJCP14},
methods to calculate the derivative discontinuity were proposed,
however among these works only the one from Kuisma \textit{et al}.\cite{KuismaPRB10}
can be used for solids. They showed how to calculate the exchange
part $\Delta_{\text{x}}$ of the derivative discontinuity from quantities that are
obtained from a standard ground-state KS calculation. $\Delta_{\text{x}}$ is
nonzero since they used a xc potential that is based on the one proposed by
Gritsenko \textit{et al}. (GLLB),\cite{GritsenkoPRA95,GritsenkoIJQC97}
which exhibits a jump (step structure) when the lowest unoccupied orbital
starts to be occupied. The GLLB potential is a simplified
version of the Krieger-Li-Iafrate (KLI) approximation \cite{KriegerPRA92a} to
the optimized effective potential (OEP).\cite{SharpPR53}
The potential of Kuisma \textit{et al}.,\cite{KuismaPRB10} called
GLLB-SC (SC for solid and correlation), has been shown to be much
more accurate than LDA and standard GGA for the calculation of
band gaps in solids
(see Refs.~\onlinecite{CastelliEES12,YanPRB12,HuserPRB13,MiroCSR14,PilaniaSR16,KimJPCC16,PandeyJPCC17}
for extensive tests) and to reach an accuracy similar to the $GW$
methods.\cite{YanPRB12,HuserPRB13}

Very recently,\cite{PandeyJPCC17} GLLB-SC and mBJ band gaps of the
chalcopyrite, kesterite, and wurtzite polymorphs of II-IV-V$_{2}$ and
III-III-V$_{2}$ semiconductors were compared. It was shown that in
most cases the GLLB-SC and mBJ band gaps are rather similar, however,
in a few cases rather large differences were obtained. The experimental values
were not known for a sufficient number of systems to draw a clear conclusion
about the relative accuracy of the GLLB-SC and mBJ methods.
To our knowledge, this is the only work which reports a direct comparison
between the GLLB-SC method and other semilocal potentials that were also shown
to be useful for band gap calculation in solids. The aim of the present work is
to provide such a comparison, and for the band gap the large test set of 76
solids considered in our recent work\cite{TranJPCA17} has been chosen.
Since the band gap is obviously not the only interesting property of a solid to
consider, results for ground-state quantities like electron densities,
electric field gradients, and magnetic moments will also be shown and discussed.

The paper is organized as follows. In Sec. \ref{methodology}, a description of
the methods and the computational details are given. In Sec. \ref{results},
the results are presented and discussed, and in Sec. \ref{summary} the summary
of the work is given.

\section{\label{methodology}Methodology}

We begin by describing briefly the GLLB method and its slightly modified
version GLLB-SC. More details can be found in the original
works\cite{GritsenkoPRA95,GritsenkoIJQC97,KuismaPRB10} or in a
recent review.\cite{BaerendsPCCP17} The xc energy $E_{\text{xc}}$
can be expressed with the pair-correlation function
and this leads naturally to the following partitioning for the
functional derivative $v_{\text{xc},\sigma}=\delta E_{\text{xc}}/\delta\rho_{\sigma}$:
\begin{equation}
v_{\text{xc},\sigma}(\bm{r}) = v_{\text{xc,hole},\sigma}(\bm{r}) +
v_{\text{xc,resp},\sigma}(\bm{r}),
\label{vxc}
\end{equation}
where the first term is twice the xc energy density per particle,
$v_{\text{xc,hole},\sigma}=2\varepsilon_{\text{xc},\sigma}=
2\left(\varepsilon_{\text{x},\sigma}+\varepsilon_{\text{c}}\right)$ with
$\varepsilon_{\text{x},\sigma}$ and $\varepsilon_{\text{c}}$ defined as
\begin{equation}
E_{\text{xc}} =
\sum_{\sigma}\int\varepsilon_{\text{x},\sigma}(\bm{r})\rho_{\sigma}(\bm{r})d^{3}r +
\int\varepsilon_{\text{c}}(\bm{r})\rho(\bm{r})d^{3}r,
\label{Exc}
\end{equation}
and is called the hole
term since it is the Coulomb potential produced by the xc hole.
$v_{\text{xc,resp},\sigma}$ is the response term which accounts for
the response of the pair correlation function to a variation in the electron density.
The exchange part $v_{\text{x,hole},\sigma}$ of the
hole term is the Slater potential,\cite{SlaterPR51} i.e., twice
the Hartree-Fock (HF) exchange energy density, which reduces to
$\left(3/2\right)v_{\text{x},\sigma}^{\text{LDA}}$ for a constant electron density,
where $v_{\text{x},\sigma}^{\text{LDA}}$ is the exchange potential for
constant density [$v_{\text{x,resp},\sigma}$ reduces to
$-\left(1/2\right)v_{\text{x},\sigma}^{\text{LDA}}$ such that overall Eq.~(\ref{vxc})
recovers the LDA limit for constant density].

Neglecting correlation, Gritsenko \textit{et al}.\cite{GritsenkoPRA95} proposed
two exchange potentials based on the partitioning given by Eq.~(\ref{vxc})
which differ by the hole term. Their first potential uses the exact
(i.e., Slater) hole term, while the second one uses the exchange energy density of
the B88 GGA functional\cite{BeckePRA88}
($v_{\text{x,hole},\sigma}^{\text{B88}}=2\varepsilon_{\text{x},\sigma}^{\text{B88}}$).
For both potentials the exchange response term is given by
\begin{equation}
v_{\text{x,resp},\sigma}^{\text{GLLB}}(\bm{r}) =
K_{\text{x}}\sum_{i=1}^{N_{\sigma}}\sqrt{\epsilon_{\text{H}}-\epsilon_{i\sigma}}
\frac{\left\vert\psi_{i\sigma}(\bm{r})\right\vert^{2}}{\rho_{\sigma}(\bm{r})},
\label{vxrespGLLB}
\end{equation}
where the sum runs over the $N_{\sigma}$ occupied orbitals of spin $\sigma$,
$\epsilon_{\text{H}}$ is the highest (H) occupied orbital,
and $K_{\text{x}}$ was either
chosen to be $K_{\text{x}}^{\text{LDA}}=8\sqrt{2}/\left(3\pi^2\right)$
in order to satisfy the correct LDA limit for constant electron density or
determined to satisfy the virial relation for exchange.\cite{LevyPRA85}
Eq.~(\ref{vxrespGLLB})
is a simplified and computationally faster version of the
KLI\cite{KriegerPRA92a} response term.\cite{GritsenkoPRA95,KohutJCP14}
In a subsequent work,\cite{GritsenkoIJQC97} correlation was also
included in the GLLB potential. For the hole term this was done by adding
$v_{\text{c,hole}}^{\text{PW91}}=2\varepsilon_{\text{c}}^{\text{PW91}}$
(PW91 is the GGA from Perdew and Wang\cite{PerdewPRB92b}), while
for the response term $K_{\text{x}}$ was replaced
by $K_{\text{xc}}$ that was determined from different schemes, e.g.,
satisfying the virial relation for exchange and correlation.

The GLLB-SC potential\cite{KuismaPRB10} uses the GGA PBEsol\cite{PerdewPRL08} for
the exchange hole term, but also for the total (hole plus response) correlation:
\begin{eqnarray}
v_{\text{xc},\sigma}^{\text{GLLB-SC}}(\bm{r}) & = &
2\varepsilon_{\text{x},\sigma}^{\text{PBEsol}}(\bm{r}) \nonumber \\
 & & + K_{\text{x}}^{\text{LDA}}\sum_{i=1}^{N_{\sigma}}\sqrt{\epsilon_{\text{H}}-\epsilon_{i\sigma}}
\frac{\left\vert\psi_{i\sigma}(\bm{r})\right\vert^{2}}{\rho_{\sigma}(\bm{r})} \nonumber \\
 & & + v_{\text{c},\sigma}^{\text{PBEsol}}(\bm{r}),
\label{vxcGLLBSC}
\end{eqnarray}
where $v_{\text{c},\sigma}^{\text{PBEsol}} =
\delta E_{\text{c}}^{\text{PBEsol}}/\delta\rho_{\sigma}$ is the total
(hole plus response) correlation potential.

The most important feature of the GLLB(-SC) potentials is to vary abruptly
when the lowest (L) unoccupied orbital ($\psi_{\text{L}}$) starts to be
occupied by an infinitesimal amount $\delta$ and leads to the
replacement of $\epsilon_{\text{H}}$ by $\epsilon_{\text{L}}$ in
Eq.~(\ref{vxrespGLLB}).
This is the so-called step structure that
is also exhibited by the exact xc potential, but not by most LDA and
GGA potentials. The BJ \cite{BeckeJCP06,ArmientoPRB08} and
Armiento-K\"{u}mmel\cite{ArmientoPRL13} potentials are examples of semilocal
potentials that show such step structure.
Kuisma \textit{et al}.\cite{KuismaPRB10} showed that the step structure of the
GLLB(-SC) potential leads to an expression for the exchange component of the
derivative discontinuity that is given by (see Ref.~\onlinecite{BaerendsPCCP17}
for a detailed derivation)
\begin{eqnarray}
\Delta_{\text{x}}^{\text{GLLB(-SC)}} & = &
\int\psi_{\text{L}}^{*}(\bm{r})\left[
\sum_{i=1}^{N_{\sigma_{\text{L}}}}K_{\text{x}}^{(\text{LDA})}\left(
\sqrt{\epsilon_{\text{L}}-\epsilon_{i\sigma_{\text{L}}}} \right.\right. \nonumber\\
& &
-\left.\left.
\sqrt{\epsilon_{\text{H}}-\epsilon_{i\sigma_{\text{L}}}}
\right)
\frac{\left\vert\psi_{i\sigma_{\text{L}}}(\bm{r})
\right\vert^{2}}{\rho_{\sigma_{\text{L}}}(\bm{r})}
\right]\psi_{\text{L}}(\bm{r})d^{3}r, \nonumber\\
\label{deltax}
\end{eqnarray}
where $\sigma_{\text{L}}$ is the spin of $\psi_{\text{L}}$ and
the integration is performed in the unit cell. It should be noted that in
the case of metals, i.e., when $\epsilon_{\text{H}}=\epsilon_{\text{L}}$,
Eq.~(\ref{deltax}) is zero, and therefore if Eq.~(\ref{vxcGLLBSC}) falsely
predicts a system to be metallic (e.g., for InSb or FeO,
see Sec.~\ref{results}), then Eq.~(\ref{deltax}) is of no use.

As underlined in Sec.~\ref{introduction}, the formally correct way to
calculate the true band gap within the KS theory is to add
a discontinuity to the KS band gap, and this is what is done with
the GLLB(-SC) method. This is the very nice feature of GLLB(-SC),
but it is also clear that this method is only
half satisfying since the discontinuity is calculated only for exchange.
In principle correlation effects should be much smaller than
exchange, however it was shown that the xc discontinuity $\Delta_{\text{xc}}$ calculated
in the random phase (RPA) OEP approximation for correlation\cite{GruningJCP06,KlimesJCP14}
is much smaller (by at least 50\%) than $\Delta_{\text{x}}$ calculated
with exact exchange (EXX) OEP.
Therefore, agreement with experiment for the band gap
can still not be fully justified from the formal point of view with GLLB(-SC).

For the present work, the GLLB-SC potential and its associated derivative
discontinuity, Eqs.~(\ref{vxcGLLBSC}) and (\ref{deltax}), have been implemented
in WIEN2k,\cite{WIEN2k} which is an all-electron code based on the linearized
augmented plane-wave method.\cite{AndersenPRB75,Singh}
From the technical point of view, we only mention that the sums in
Eqs.~(\ref{vxcGLLBSC}) and (\ref{deltax}) include both the band and core
electrons. The results of calculations with the GLLB-SC potential
on various properties will be compared to those obtained
with other multiplicative potentials of the LDA, GGA, or meta-GGA (MGGA) type,
which are the following. The LDA\cite{KohnPR65,PerdewPRB92a} is exact for
the homogenous electron gas, while Sloc (abbreviation for local Slater
potential\cite{FinzelIJQC17}) consists of an enhanced exchange LDA
(compare $v_{\text{x},\sigma}^{\text{Sloc}}=-1.67\left(2\rho_{\sigma}\right)^{0.3}$ to
$v_{\text{x},\sigma}^{\text{LDA}}\simeq-0.7386\left(2\rho_{\sigma}\right)^{1/3}$)
with no correlation added.
The GGAs are the xc PBE from Perdew \textit{et al}.,\cite{PerdewPRL96}
the exchange of Engel and Vosko \cite{EngelPRB93} (EV93PW91, combined with
PW91 correlation\cite{PerdewPRB92b} as done previously in
Ref.~\onlinecite{TranJPCM07}), the exchange from
Armiento and K\"{u}mmel\cite{ArmientoPRL13,VlcekPRB15,LindmaaPRB16}
(AK13, no correlation added as done in Refs.~\onlinecite{ArmientoPRL13,VlcekPRB15}),
and HLE16\cite{VermaJPCL17} which is a modification of HCTH/407\cite{BoeseJCP01}
(the exchange and correlation components are multiplied by
1.25 and 0.5, respectively). Note that all GGA potentials depend on $\rho_{\sigma}$ and
its first two derivatives, while the xc potential LB94 of van Leeuwen and
Baerends,\cite{vanLeeuwenPRA94} also considered in the present work,
depends only on $\rho_{\sigma}$ and its first derivative. Therefore, LB94 is
neither a LDA nor a GGA, but lies in between (note that correlation in LB94 is
LDA\cite{PerdewPRB92a}).
The tested MGGA are the aforementioned BJ\cite{BeckeJCP06} and mBJ\cite{TranPRL09}
potentials that are both combined with LDA for correlation\cite{PerdewPRB92a}
(BJLDA and mBJLDA). Note that the LDA and GGA potentials are obtained as
functional derivative $v_{\text{xc},\sigma}=\delta E_{\text{xc}}/\delta\rho_{\sigma}$
of energy functionals, while this is not the case for the GLLB-SC, LB94,
and (m)BJLDA potentials.\cite{KarolewskiJCTC09,GaidukJCP09,KarolewskiPRA13}
We also mention that among these potentials, (m)BJLDA,
AK13, and LB94 were recently shown to lead to severe numerical problems in
finite systems.\cite{AschebrockPRB17a,AschebrockPRB17b}

For completeness, calculations with a hybrid functional, YS-PBE0,\cite{TranPRB11}
were also done. In YS-PBE0 (YS stands for Yukawa screened), the
Coulomb operator in the HF exchange is exponentially screened
(i.e., Yukawa potential) and it was shown\cite{TranPRB11}
(see also Ref.~\onlinecite{ShimazakiCPL08})
that YS-PBE0 leads to the same band gaps as the popular HSE06 from Heyd,
Scuseria, and Ernzerhof\cite{HeydJCP03,KrukauJCP06} which uses an error-function
for the screening of the HF exchange. In the following, the acronym HSE06
will be used for all results that were obtained with YS-PBE0.
Since hybrid functionals contain a
fraction of HF exchange,\cite{BeckeJCP93} (25\% in YS-PBE0/HSE06)
that is usually implemented in the gKS framework
(as done in WIEN2k\cite{TranPRB11}), the potential is nonmultiplicative.
With nonmultiplicative potentials (a part of) the discontinuity
$\Delta_{\text{xc}}$ is included in the orbital energies,\cite{YangJCP12,PerdewPNAS17}
which means that the gap CBM minus VBM should in principle be in better
agreement with the experimental gap $E_{\text{g}}$.
However, note that hybrid functionals are much more expensive than semilocal
functionals such that they can not be applied routinely to very large systems,
in particular with codes based on plane-waves basis functions.

All calculations presented in this work were done with WIEN2k and the
convergence parameters of the calculations, like the size of the basis set or
the number of $\bm{k}$-points for the integrations in the Brillouin zone,
were chosen such that the results are well converged (e.g., within
$\sim0.03$~eV for the band gap). The solids of the test sets are listed in
Table~S1 of the Supplemental Material,\cite{SM_GLLB} along with their space
group and experimental geometry that was used for the
calculations. For most calculations, the deep-lying core states (those which
are below the Fermi energy by at least $\sim6$~Ry) were treated fully
relativistically, i.e., by including spin-orbit coupling (SOC),
while the band states (semicore, valence, and unoccupied)
were treated at the scalar-relativistic level. The only exceptions are the
results for the effective masses of III-V semiconductors, which were obtained
with SOC included also for the band electrons in a second-variational
step.\cite{KoellingJPC77,MacDonaldJPC80}

\section{\label{results}Results and discussion}

\subsection{\label{electronicstructure}Electronic structure}

\begin{table*} 
\caption{\label{table_band_gap}Summary statistics for the error in the
calculated band gaps in Table~S2 of the Supplemental Material\cite{SM_GLLB}
for the set of 76 solids. M(R)E, MA(R)E, and STD(R)E denote the mean
(relative) error, the mean absolute (relative) error, and the standard deviation
of the (relative) error, respectively. The calculations were done at the
experimental geometry specified in Table~S1 of the Supplemental
Material.\cite{SM_GLLB} All results except those obtained with the
BJLDA, LB94, and GLLB-SC methods are taken from Ref.~\onlinecite{TranJPCA17}.}
\begin{ruledtabular}
\begin{tabular}{lccccccccccc} 
           & LDA   & PBE & EV93PW91 & AK13 & Sloc  & HLE16 & BJLDA & mBJLDA & LB94 & GLLB-SC & HSE06 \\ 
\hline
ME (eV)    & -2.17 & -1.99 & -1.55 & -0.28 & -0.76 & -0.82 & -1.53 & -0.30 & -1.87 &  0.20 & -0.68  \\ 
MAE (eV)   &  2.17 &  1.99 &  1.55 &  0.75 &  0.90 &  0.90 &  1.53 &  0.47 &  1.88 &  0.64 &  0.82  \\ 
STDE (eV)  &  1.63 &  1.56 &  1.55 &  0.89 &  0.93 &  1.07 &  1.24 &  0.57 &  1.23 &  0.81 &  1.21  \\ 
MRE (\%)   &   -58 &   -53 &   -35 &    -6 &   -21 &   -20 &   -41 &    -5 &   -54 &    -4 &    -7  \\ 
MARE (\%)  &    58 &    53 &    36 &    24 &    30 &    25 &    41 &    15 &    55 &    24 &    17  \\ 
STDRE (\%) &    23 &    23 &    23 &    31 &    37 &    28 &    22 &    22 &    29 &    34 &    22  \\ 
\end{tabular}
\end{ruledtabular}
\end{table*}

\begin{figure}
\includegraphics[scale=0.6]{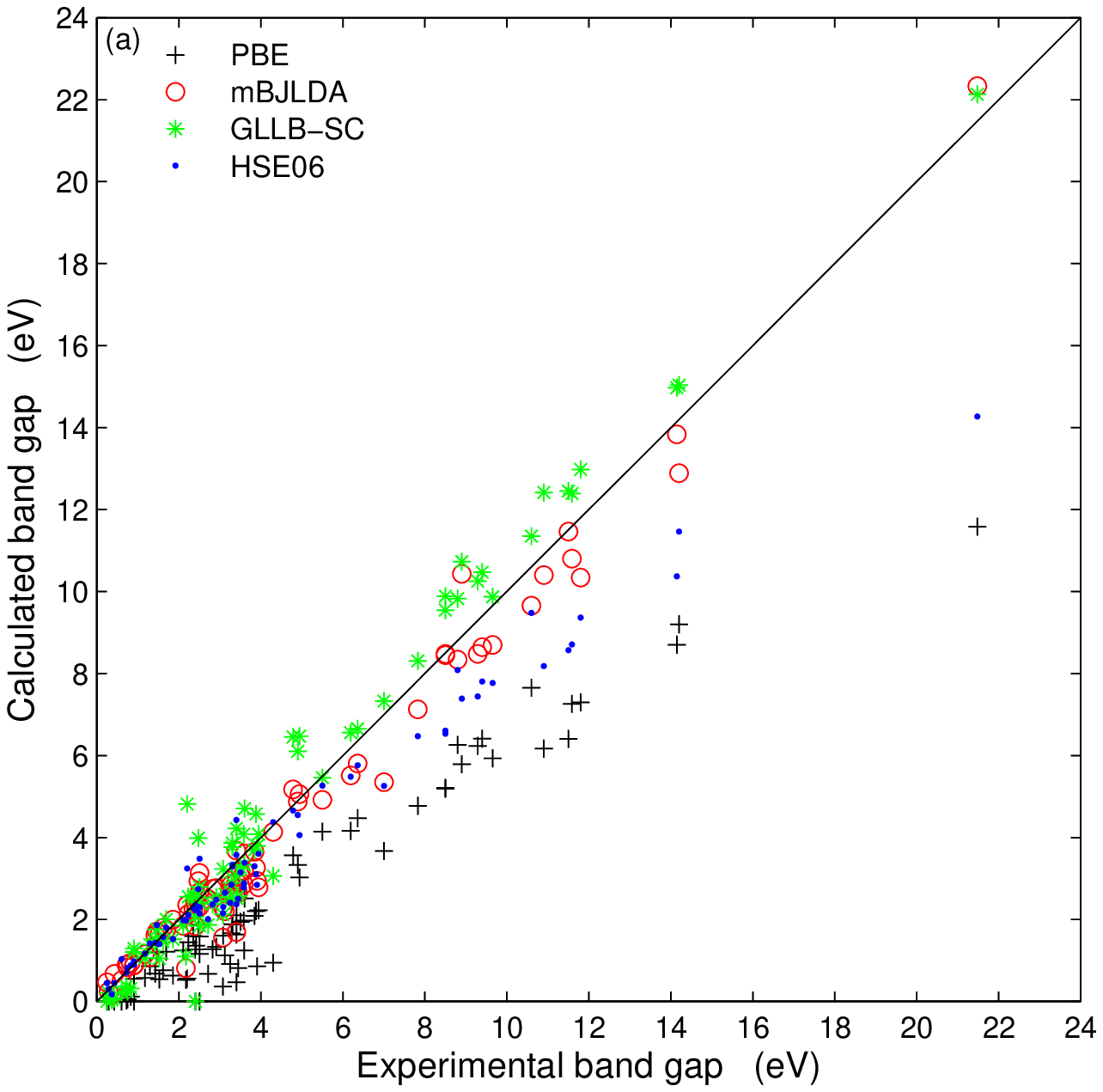}
\includegraphics[scale=0.6]{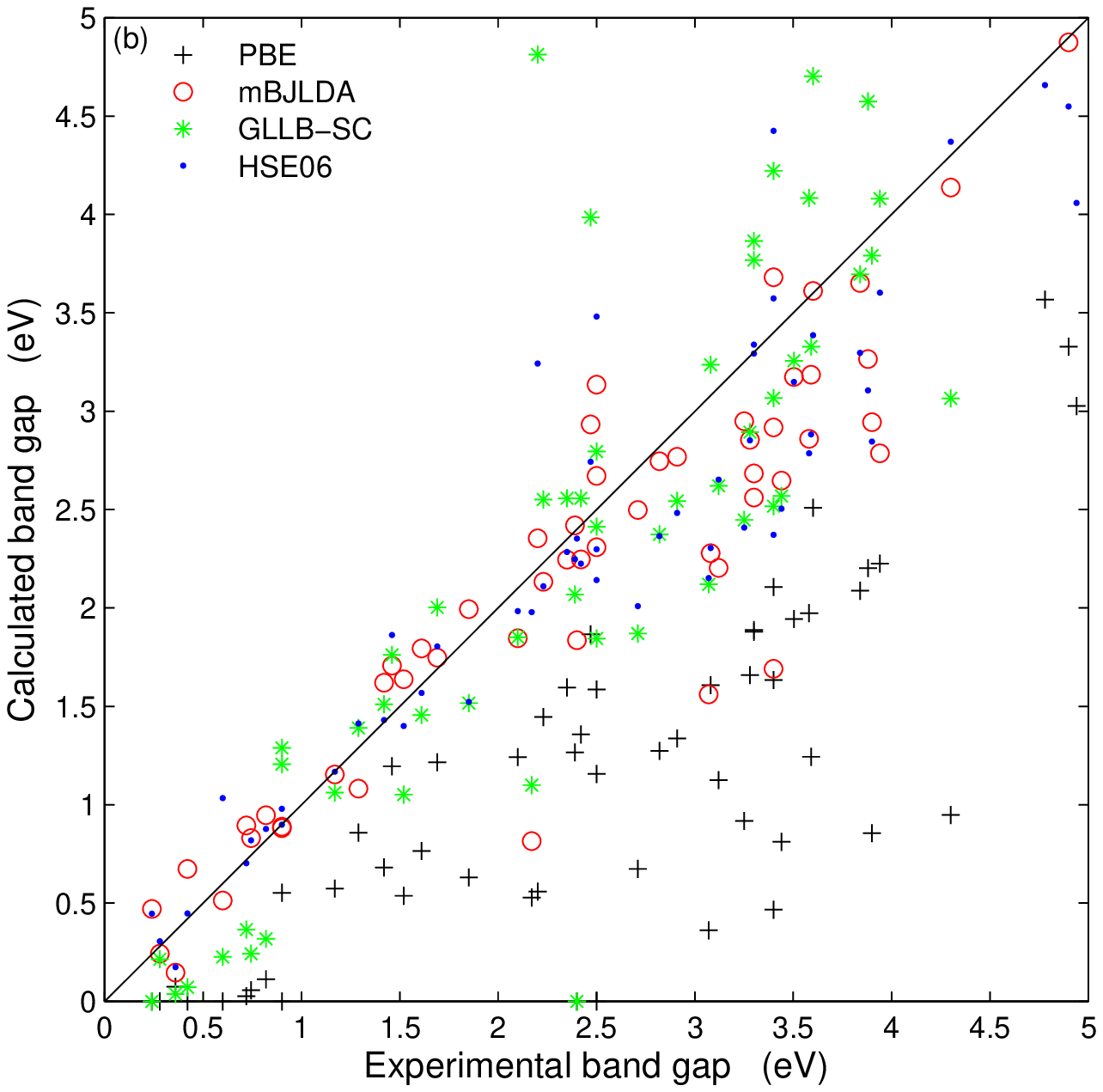}
\caption{\label{fig_band_gap}Calculated versus experimental fundamental band
gaps for the set of 76 solids. The values are given in Table~S2 of the
Supplemental Material.\cite{SM_GLLB} The lower panel is a zoom of the upper
panel focusing on band gaps smaller than 5~eV.}
\end{figure}

We start the discussion of the results with the fundamental band gap,
whose values are shown in Table~S2 of the Supplemental Material\cite{SM_GLLB}
for the 76 solids of our test set, which are of
various types: $sp$-semiconductors, ionic insulators, rare gases,
and strongly correlated solids. The contribution of the exchange
discontinuity $\Delta_{\text{x}}$ to the GLLB-SC band gap is indicated
in parenthesis. The summary statistics for the error is given in
Table~\ref{table_band_gap}. Note that the results obtained with all methods
except BJLDA, LB94, and GLLB-SC are from Ref.~\onlinecite{TranJPCA17}.
The worst agreements with experiment are obtained with the
standard LDA and PBE,
as well as with LB94 (LB94$\alpha$\cite{SchipperJCP00} leads to quite similar results
on average), which
strongly underestimate the band gap and lead to MAE and MARE around
2~eV and $55\%$, respectively, while EV93PW91 and BJLDA are
slightly more accurate. Much better results are obtained with
the other methods, since the MAE (MARE) of AK13,
Sloc, HLE16, GLLB-SC, and the hybrid HSE06 are in the range
0.64-0.90~eV (17-30\%).
However, the best agreement with experiment is obtained with the mBJLDA potential, which
leads to the smallest MAE (0.47~eV) and MARE (15\%).
As discussed in Refs.~\onlinecite{TranPRL09,KollerPRB11,TranJPCA17},
the very good performance of the mBJLDA potential can be attributed
to its dependency on two ingredients: the kinetic-energy density
$t_{\sigma}=\left(1/2\right)
\sum_{i=1}^{N_{\sigma}}\nabla\psi_{i\sigma}^{*}\cdot\nabla\psi_{i\sigma}$, which
seems particularly important for solids with strongly correlated
$3d$-electrons, and the average of $\nabla\rho/\rho$ in the unit cell that
is able to somehow account for screening effects
(see the discussion on metals in Sec.~\ref{results}).

The band gaps calculated with the GLLB-SC method,
which consist of the sum of the KS band gap (CBM minus VBM) and the exchange
discontinuity [Eq.~(\ref{deltax})] are pretty accurate in most cases.
Indeed, the MAE of 0.64~eV is smaller than the value for the hybrid
functional HSE06 (and also B3PW91 which leads to 0.73~eV\cite{TranJPCA17}),
and only mBJLDA has a smaller MAE.
The MARE is 24\%, which is a rather fair value in comparison to the other
methods, since it is similar to the values for the GGAs AK13 and
HLE16, but larger than what mBJLDA and HSE06 give. The ME and MRE, which are
the smallest among all tested methods, indicate that GLLB-SC shows the least
pronounced tendency to underestimate or overestimate the band gaps on average.
However, by looking at Fig.~\ref{fig_band_gap}, which shows graphically the
band gaps for a few selected methods, we can see that there is a noticeable
tendency to underestimate many of the band gaps smaller than 3~eV, while an
overestimation is observed for band gaps larger than 4~eV.
This is more or less the opposite of what is observed for mBJLDA, as seen in
Fig.~\ref{fig_band_gap}.

The main conclusion from the statistics in Table~\ref{table_band_gap} for the
band gap is that mBJLDA is more accurate than GLLB-SC, since the most important
quantities, the MAE and MARE, are the smallest, which is also the case for the
STDE and STDRE. The GLLB-SC results should also be considered as very good
since the overall performance is very similar to hybrid functionals.
However, note that there are some cases where GLLB-SC
gives a band gap that is clearly too small, and from Fig.~\ref{fig_band_gap} and
Table~S2 we can see that this concerns mainly band gaps that are
(experimentally) below 1~eV and FeO. The worst cases are
InAs, InSb, SnTe, and FeO that are described as (nearly) metallic by GLLB-SC, which
is in contradiction with experiment, while mBJLDA leads to reasonably good results.

Concerning the strongly correlated solids, which are known to be very difficult
cases for the standard functionals,\cite{TerakuraPRB84} the GLLB-SC results
seem to be particularly disparate. For Cr$_{2}$O$_{3}$, MnO, and CoO the
agreement with experiment is very good. However, as mentioned above,
GLLB-SC leads to no gap in FeO (from experiment it should be around 2.4~eV)
and in NiO there is an underestimation of more than 1~eV. On the other hand,
the gap is 4.81~eV in Fe$_{2}$O$_{3}$, which is too large by 2.6~eV and should
be the consequence of an exchange splitting that is too large (as shown in
Sec.~\ref{magnetism}, the magnetic moments of metals are by far overestimated).
The mBJLDA potential leads to much more consistent results, since the
largest discrepancy is an underestimation of about 1~eV for MnO.
As mentioned in Ref.~\onlinecite{TranJPCA17}, all LDA and GGA methods lead
to severely underestimated band gaps for the strongly correlated solids,
the only exception being Sloc for MnO. LB94 is even worse since no band gap
is obtained for most strongly correlated solids.

On the other hand, in Ref.~\onlinecite{TranJPCA17} we underlined that mBJLDA
underestimates by a rather large amount (1-1.7~eV, see Table~S2) the band gap
of the Cu$^{1+}$ compounds, with CuCl being one of the worst case.
For these systems, the GLLB-SC band gaps are, with respect to mBJLDA, larger
by 0.3-1.3~eV such that the agreement with experiment is improved.
However, with the exception of CuSCN, a sizeable underestimation is still obtained.

\begin{figure}
\includegraphics[scale=0.7]{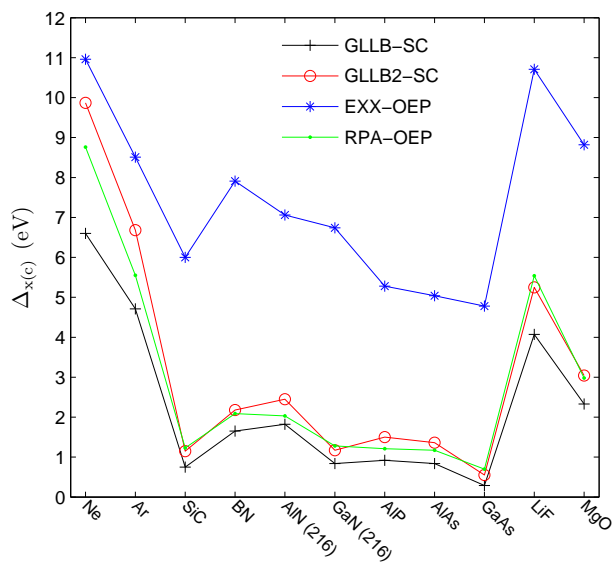}
\caption{\label{fig_delta}GLLB-SC and GLLB2-SC derivative discontinuities compared
to the values calculated in the framework of the EXX-OEP and RPA-OEP
methods.\cite{KlimesJCP14}}
\end{figure}

Discussing now the derivative discontinuity $\Delta_{\text{x}}^{\text{GLLB-SC}}$,
Table~S2 shows that its contribution to the total GLLB-SC band gap is in the
range 25-35\% for most solids, which is rather substantial.
Without $\Delta_{\text{x}}^{\text{GLLB-SC}}$, the GLLB-SC band gaps would be
still larger than the PBE band gaps, but clearly smaller than experiment.
In Ref.~\onlinecite{KlimesJCP14}, discontinuities were calculated in the framework
of the EXX-OEP and RPA-OEP methods, and it was shown that the sum of the
RPA-OEP KS band gap and $\Delta_{\text{xc}}^{\text{RPA-OEP}}$ is in relatively
fair agreement with the experimental band gap for many of the solids that were
considered. Thus, the order of magnitude of $\Delta_{\text{xc}}^{\text{RPA-OEP}}$
should be similar to the exact one in those cases where agreement with experiment
is good. Figure~\ref{fig_delta} compares the discontinuities
calculated with the GLLB-SC and OEP methods, and one can clearly see that the
GLLB-SC values are much closer to RPA-OEP than to EXX-OEP, despite the fact that
$\Delta_{\text{x}}^{\text{GLLB-SC}}$ is supposed to be only for exchange.
We can also see that $\Delta_{\text{x}}^{\text{GLLB-SC}}$ is 
smaller than $\Delta_{\text{xc}}^{\text{RPA-OEP}}$ by 0.2-2~eV.
For the sake of consistency, it would be more preferable to have agreement
with $\Delta_{\text{xc}}^{\text{RPA-OEP}}$ with a GLLB-type discontinuity which also
includes correlation. As mentioned in the introduction, a (conveniently easy)
way to do it was proposed in Ref.~\onlinecite{GritsenkoIJQC97},
and we followed a similar strategy for the construction of a
potential, called GLLB2-SC, that also includes correlation in the discontinuity
(see Sec.~\ref{variants} for details). The values of
$\Delta_{\text{xc}}^{\text{GLLB2-SC}}$, also shown in Fig.~\ref{fig_delta},
show a surprisingly nice agreement with the RPA-OEP values for most solids.
The largest difference between $\Delta_{\text{xc}}^{\text{GLLB2-SC}}$ and
$\Delta_{\text{xc}}^{\text{RPA-OEP}}$ are found for Ne and Ar, and are
about 1~eV. Nevertheless, we do not expect such a good agreement for
strongly correlated systems, since in the case of FeO, for instance, the
KS band gap, and therefore the discontinuity, is still zero with GLLB2-SC.
Furthermore, as discussed later in Sec.~\ref{variants}, the band gaps
with GLLB2-SC are much less accurate than with the original GLLB-SC potential,
such that GLLB2-SC is not really interesting for band gap calculation.

\begin{table*}
\footnotesize
\caption{\label{table_effective_masses}Effective hole and electron masses at
the $\Gamma$ point in units of the electron rest mass ${m_e}$ calculated along
the $\Gamma$X [100] direction. The calculations were done at the experimental
geometry (see Table~S1 of the Supplemental Material\cite{SM_GLLB}) and include
SOC. The experimental values were calculated from the
Luttinger parameters tabulated in Ref.~\onlinecite{VurgaftmanJAP01}.
The values which agree the best (worst) with experiment are in bold (underlined).
In some cases, because the shape of the band at $\Gamma$ is of
nonparabolic type, the effective mass can not be calculated.}
\begin{ruledtabular}
\begin{tabular}{cccccc}
\multicolumn{1}{l}{Solid} & \multicolumn{1}{c}{Method} &
\multicolumn{1}{c}{$\left\vert m^{*}_{\text{split-off}}/m_{e}\right\vert$} &
\multicolumn{1}{c}{$\left\vert m^{*}_{\text{light-hole}}/m_{e}\right\vert$} &
\multicolumn{1}{c}{$\left\vert m^{*}_{\text{heavy-hole}}/m_{e}\right\vert$} &
\multicolumn{1}{c}{$\left\vert m^{*}_{\text{electron}}/m_{e}\right\vert$} \\
\hline 
\multicolumn{1}{l}{InP} & \multicolumn{1}{c}{LDA} & \multicolumn{1}{c}{\underline{0.095}} & \multicolumn{1}{c}{\underline{0.051}}  & \multicolumn{1}{c}{\underline{0.404}}  & \multicolumn{1}{c}{\underline{0.036}} \\
\multicolumn{1}{l}{} & \multicolumn{1}{c}{PBE} & \multicolumn{1}{c}{\underline{0.127}}  & \multicolumn{1}{c}{\underline{0.073}}  & \multicolumn{1}{c}{\underline{0.418}}  & \multicolumn{1}{c}{\underline{0.052}} \\
\multicolumn{1}{l}{} & \multicolumn{1}{c}{EV93PW91} & \multicolumn{1}{c}{\bf 0.200}  & \multicolumn{1}{c}{\bf 0.128}  & \multicolumn{1}{c}{0.453}  & \multicolumn{1}{c}{\bf 0.096} \\
\multicolumn{1}{l}{} & \multicolumn{1}{c}{AK13} & \multicolumn{1}{c}{0.250}  & \multicolumn{1}{c}{0.170}  & \multicolumn{1}{c}{0.489}  & \multicolumn{1}{c}{0.138} \\
\multicolumn{1}{l}{} & \multicolumn{1}{c}{Sloc} & \multicolumn{1}{c}{\underline{0.133}}  & \multicolumn{1}{c}{\underline{0.070}}  & \multicolumn{1}{c}{0.600}  & \multicolumn{1}{c}{\underline{0.050}} \\
\multicolumn{1}{l}{} & \multicolumn{1}{c}{HLE16} & \multicolumn{1}{c}{\bf 0.212}  & \multicolumn{1}{c}{\bf 0.130}  & \multicolumn{1}{c}{\bf 0.538}  & \multicolumn{1}{c}{\bf 0.097} \\
\multicolumn{1}{l}{} & \multicolumn{1}{c}{BJLDA} & \multicolumn{1}{c}{0.150}  & \multicolumn{1}{c}{0.086}  & \multicolumn{1}{c}{\underline{0.427}}  & \multicolumn{1}{c}{\bf 0.065} \\
\multicolumn{1}{l}{} & \multicolumn{1}{c}{mBJLDA} & \multicolumn{1}{c}{0.231}  & \multicolumn{1}{c}{0.153}  & \multicolumn{1}{c}{0.476}  & \multicolumn{1}{c}{0.115} \\
\multicolumn{1}{l}{} & \multicolumn{1}{c}{LB94} & \multicolumn{1}{c}{\underline{0.030}}  & \multicolumn{1}{c}{}  & \multicolumn{1}{c}{\underline{0.408}}  & \multicolumn{1}{c}{\underline{0.049}} \\
\multicolumn{1}{l}{} & \multicolumn{1}{c}{GLLB-SC} & \multicolumn{1}{c}{0.174}  & \multicolumn{1}{c}{\bf 0.107}  & \multicolumn{1}{c}{0.442}  & \multicolumn{1}{c}{\bf 0.079} \\
\multicolumn{1}{l}{} & \multicolumn{1}{c}{Expt.} & \multicolumn{1}{c}{0.210}  & \multicolumn{1}{c}{0.121}  & \multicolumn{1}{c}{0.531}  & \multicolumn{1}{c}{0.080} \\
\multicolumn{1}{l}{InAs} & \multicolumn{1}{c}{LDA} & \multicolumn{1}{c}{\underline{0.028}}  & \multicolumn{1}{c}{}  & \multicolumn{1}{c}{\bf 0.317}  & \multicolumn{1}{c}{\underline{0.062}} \\
\multicolumn{1}{l}{} & \multicolumn{1}{c}{PBE} & \multicolumn{1}{c}{}  & \multicolumn{1}{c}{}  & \multicolumn{1}{c}{\bf 0.324}  & \multicolumn{1}{c}{\bf 0.036} \\
\multicolumn{1}{l}{} & \multicolumn{1}{c}{EV93PW91} & \multicolumn{1}{c}{0.094}  & \multicolumn{1}{c}{\bf 0.027}  & \multicolumn{1}{c}{\bf 0.344}  & \multicolumn{1}{c}{\bf 0.022} \\
\multicolumn{1}{l}{} & \multicolumn{1}{c}{AK13} & \multicolumn{1}{c}{\bf 0.148}  & \multicolumn{1}{c}{\underline{0.062}}  & \multicolumn{1}{c}{0.387}  & \multicolumn{1}{c}{0.049} \\
\multicolumn{1}{l}{} & \multicolumn{1}{c}{Sloc} & \multicolumn{1}{c}{\underline{0.026}}  & \multicolumn{1}{c}{}  & \multicolumn{1}{c}{\underline{0.472}}  & \multicolumn{1}{c}{\underline{0.062}} \\
\multicolumn{1}{l}{} & \multicolumn{1}{c}{HLE16} & \multicolumn{1}{c}{0.081}  & \multicolumn{1}{c}{0.012}  & \multicolumn{1}{c}{0.417}  & \multicolumn{1}{c}{0.011} \\
\multicolumn{1}{l}{} & \multicolumn{1}{c}{BJLDA} & \multicolumn{1}{c}{\underline{0.023}}  & \multicolumn{1}{c}{}  & \multicolumn{1}{c}{\bf 0.337}  & \multicolumn{1}{c}{\bf 0.023} \\
\multicolumn{1}{l}{} & \multicolumn{1}{c}{mBJLDA} & \multicolumn{1}{c}{\bf 0.132}  & \multicolumn{1}{c}{0.053}  & \multicolumn{1}{c}{0.368}  & \multicolumn{1}{c}{0.041} \\
\multicolumn{1}{l}{} & \multicolumn{1}{c}{LB94} & \multicolumn{1}{c}{0.104}  & \multicolumn{1}{c}{}  & \multicolumn{1}{c}{\bf 0.338}  & \multicolumn{1}{c}{0.200} \\
\multicolumn{1}{l}{} & \multicolumn{1}{c}{GLLB-SC} & \multicolumn{1}{c}{\underline{0.049}}  & \multicolumn{1}{c}{}  & \multicolumn{1}{c}{0.354}  & \multicolumn{1}{c}{} \\
\multicolumn{1}{l}{} & \multicolumn{1}{c}{Expt.} & \multicolumn{1}{c}{0.140}  & \multicolumn{1}{c}{0.027}  & \multicolumn{1}{c}{0.333}  & \multicolumn{1}{c}{0.026} \\
\multicolumn{1}{l}{InSb} & \multicolumn{1}{c}{LDA} & \multicolumn{1}{c}{\underline{0.014}}  & \multicolumn{1}{c}{}  & \multicolumn{1}{c}{\underline{0.225}}  & \multicolumn{1}{c}{0.056} \\
\multicolumn{1}{l}{} & \multicolumn{1}{c}{PBE} & \multicolumn{1}{c}{0.044}  & \multicolumn{1}{c}{\underline{0.111}}  & \multicolumn{1}{c}{\underline{0.200}}  & \multicolumn{1}{c}{0.036} \\
\multicolumn{1}{l}{} & \multicolumn{1}{c}{EV93PW91} & \multicolumn{1}{c}{0.097}  & \multicolumn{1}{c}{}  & \multicolumn{1}{c}{0.213}  & \multicolumn{1}{c}{} \\
\multicolumn{1}{l}{} & \multicolumn{1}{c}{AK13} & \multicolumn{1}{c}{\bf 0.120}  & \multicolumn{1}{c}{\bf 0.024}  & \multicolumn{1}{c}{0.230}  & \multicolumn{1}{c}{0.022} \\
\multicolumn{1}{l}{} & \multicolumn{1}{c}{Sloc} & \multicolumn{1}{c}{\underline{0.030}}  & \multicolumn{1}{c}{}  & \multicolumn{1}{c}{\underline{0.305}}  & \multicolumn{1}{c}{\underline{0.094}} \\
\multicolumn{1}{l}{} & \multicolumn{1}{c}{HLE16} & \multicolumn{1}{c}{0.069}  & \multicolumn{1}{c}{0.031}  & \multicolumn{1}{c}{\bf 0.267}  & \multicolumn{1}{c}{0.023} \\
\multicolumn{1}{l}{} & \multicolumn{1}{c}{BJLDA} & \multicolumn{1}{c}{\underline{0.048}}  & \multicolumn{1}{c}{\underline{0.080}}  & \multicolumn{1}{c}{0.211}  & \multicolumn{1}{c}{0.034} \\
\multicolumn{1}{l}{} & \multicolumn{1}{c}{mBJLDA} & \multicolumn{1}{c}{\bf 0.114}  & \multicolumn{1}{c}{\bf 0.020}  & \multicolumn{1}{c}{0.229}  & \multicolumn{1}{c}{\bf 0.018} \\
\multicolumn{1}{l}{} & \multicolumn{1}{c}{LB94} & \multicolumn{1}{c}{0.086}  & \multicolumn{1}{c}{}  & \multicolumn{1}{c}{0.205}  & \multicolumn{1}{c}{\underline{0.213}} \\
\multicolumn{1}{l}{} & \multicolumn{1}{c}{GLLB-SC} & \multicolumn{1}{c}{0.060}  & \multicolumn{1}{c}{0.033}  & \multicolumn{1}{c}{0.210}  & \multicolumn{1}{c}{0.023} \\
\multicolumn{1}{l}{} & \multicolumn{1}{c}{Expt.} & \multicolumn{1}{c}{0.110}  & \multicolumn{1}{c}{0.015}  & \multicolumn{1}{c}{0.263}  & \multicolumn{1}{c}{0.014} \\
\multicolumn{1}{l}{GaAs} & \multicolumn{1}{c}{LDA} & \multicolumn{1}{c}{\underline{0.083}}  & \multicolumn{1}{c}{\underline{0.018}}  & \multicolumn{1}{c}{0.331}  & \multicolumn{1}{c}{\underline{0.015}} \\
\multicolumn{1}{l}{} & \multicolumn{1}{c}{PBE} & \multicolumn{1}{c}{0.111}  & \multicolumn{1}{c}{\underline{0.039}}  & \multicolumn{1}{c}{0.335}  & \multicolumn{1}{c}{\underline{0.031}}\\
\multicolumn{1}{l}{} & \multicolumn{1}{c}{EV93PW91} & \multicolumn{1}{c}{\bf 0.169}  & \multicolumn{1}{c}{\bf 0.083}  & \multicolumn{1}{c}{\bf 0.348}  & \multicolumn{1}{c}{\bf 0.067} \\
\multicolumn{1}{l}{} & \multicolumn{1}{c}{AK13} & \multicolumn{1}{c}{0.200}  & \multicolumn{1}{c}{0.107}  & \multicolumn{1}{c}{0.371}  & \multicolumn{1}{c}{0.088} \\
\multicolumn{1}{l}{} & \multicolumn{1}{c}{Sloc} & \multicolumn{1}{c}{\underline{0.066}}  & \multicolumn{1}{c}{}  & \multicolumn{1}{c}{\underline{0.481}}  & \multicolumn{1}{c}{} \\
\multicolumn{1}{l}{} & \multicolumn{1}{c}{HLE16} & \multicolumn{1}{c}{\bf 0.165}  & \multicolumn{1}{c}{0.072}  & \multicolumn{1}{c}{0.421}  & \multicolumn{1}{c}{\bf 0.057} \\
\multicolumn{1}{l}{} & \multicolumn{1}{c}{BJLDA} & \multicolumn{1}{c}{0.135}  & \multicolumn{1}{c}{0.056}  & \multicolumn{1}{c}{\bf 0.351}  & \multicolumn{1}{c}{0.044} \\
\multicolumn{1}{l}{} & \multicolumn{1}{c}{mBJLDA} & \multicolumn{1}{c}{0.212}  & \multicolumn{1}{c}{0.118}  & \multicolumn{1}{c}{0.379}  & \multicolumn{1}{c}{0.094} \\
\multicolumn{1}{l}{} & \multicolumn{1}{c}{LB94} & \multicolumn{1}{c}{\underline{0.055}}  & \multicolumn{1}{c}{}  & \multicolumn{1}{c}{\bf 0.355}  & \multicolumn{1}{c}{\underline{0.107}} \\
\multicolumn{1}{l}{} & \multicolumn{1}{c}{GLLB-SC} & \multicolumn{1}{c}{0.136}  & \multicolumn{1}{c}{0.057}  & \multicolumn{1}{c}{\bf 0.358}  & \multicolumn{1}{c}{0.045} \\
\multicolumn{1}{l}{} & \multicolumn{1}{c}{Expt.} & \multicolumn{1}{c}{0.172}  & \multicolumn{1}{c}{0.090}  & \multicolumn{1}{c}{0.350}  & \multicolumn{1}{c}{0.067} \\
\multicolumn{1}{l}{GaSb} & \multicolumn{1}{c}{LDA} & \multicolumn{1}{c}{\underline{0.057}}  & \multicolumn{1}{c}{\bf 0.045}  & \multicolumn{1}{c}{\underline{0.206}}  & \multicolumn{1}{c}{0.028} \\
\multicolumn{1}{l}{} & \multicolumn{1}{c}{PBE} & \multicolumn{1}{c}{0.079}  & \multicolumn{1}{c}{0.100}  & \multicolumn{1}{c}{\underline{0.207}}  & \multicolumn{1}{c}{\underline{0.009}} \\
\multicolumn{1}{l}{} & \multicolumn{1}{c}{EV93PW91} & \multicolumn{1}{c}{\bf 0.120}  & \multicolumn{1}{c}{0.029}  & \multicolumn{1}{c}{0.214}  & \multicolumn{1}{c}{0.026} \\
\multicolumn{1}{l}{} & \multicolumn{1}{c}{AK13} & \multicolumn{1}{c}{\bf 0.136}  & \multicolumn{1}{c}{\bf 0.040}  & \multicolumn{1}{c}{0.228}  & \multicolumn{1}{c}{\bf 0.036} \\
\multicolumn{1}{l}{} & \multicolumn{1}{c}{Sloc} & \multicolumn{1}{c}{\underline{0.011}}  & \multicolumn{1}{c}{\underline{0.014}}  & \multicolumn{1}{c}{\underline{0.313}}  & \multicolumn{1}{c}{\underline{0.074}} \\
\multicolumn{1}{l}{} & \multicolumn{1}{c}{HLE16} & \multicolumn{1}{c}{\bf 0.101}  & \multicolumn{1}{c}{}  & \multicolumn{1}{c}{\bf 0.267}  & \multicolumn{1}{c}{} \\
\multicolumn{1}{l}{} & \multicolumn{1}{c}{BJLDA} & \multicolumn{1}{c}{0.093}  & \multicolumn{1}{c}{}  & \multicolumn{1}{c}{0.219}  & \multicolumn{1}{c}{} \\
\multicolumn{1}{l}{} & \multicolumn{1}{c}{mBJLDA} & \multicolumn{1}{c}{0.148}  & \multicolumn{1}{c}{\bf 0.051}  & \multicolumn{1}{c}{\bf 0.235}  & \multicolumn{1}{c}{\bf 0.045} \\
\multicolumn{1}{l}{} & \multicolumn{1}{c}{LB94} & \multicolumn{1}{c}{0.074}  & \multicolumn{1}{c}{}  & \multicolumn{1}{c}{0.219}  & \multicolumn{1}{c}{\underline{0.188}} \\
\multicolumn{1}{l}{} & \multicolumn{1}{c}{GLLB-SC} & \multicolumn{1}{c}{0.089}  & \multicolumn{1}{c}{}  & \multicolumn{1}{c}{0.213}  & \multicolumn{1}{c}{} \\
\multicolumn{1}{l}{} & \multicolumn{1}{c}{Expt.} & \multicolumn{1}{c}{0.120}  & \multicolumn{1}{c}{0.044}  & \multicolumn{1}{c}{0.250}  & \multicolumn{1}{c}{0.039} \\
\end{tabular}
\end{ruledtabular}
\end{table*}

Table~\ref{table_effective_masses} shows the effective hole and electron masses
of zinc-blende III-V semiconductors calculated at $\Gamma$ along the $\Gamma$X [100]
direction in the Brillouin zone. These semiconductors are those that we considered in
Ref.~\onlinecite{KimPRB10} to compare the accuracy of various
methods for effective masses. The first observation that can be made about the
present results is that there is no potential which is systematically among the
best ones for all systems. Nevertheless, it is still possible to make a
distinction between the most and least accurate methods.
By considering the number of values
which show the best and worst agreement with experiment (the values in bold and
underlined, respectively), as well as the cases where the effective mass
can not be calculated (when the band at $\Gamma$ is of nonparabolic type), the
most reliable methods are EV93PW91 (11 accurate and 2 non-calculable),
mBJLDA (7 accurate), HLE16 (9 accurate and 2 non-calculable),
and AK13 (5 accurate and 1 inaccurate).
EV93PW91 is particularly good for InP, InAs, and GaAs, HLE16 for InP, while mBJLDA
and AK13 are very accurate for InSb and GaSb.

The other potentials are less accurate since
as indicated in Table~\ref{table_effective_masses}, there is a (much) larger
number of cases where either the value is very inaccurate or can not be calculated.
For instance, in the case of the GLLB-SC potential there are 3 accurate values,
1 inaccurate, and 4 that can not be calculated.
The large errors usually correspond to underestimations at
the split-off-hole and light-hole VBM, while no particular
trend is observed for the heavy-hole VBM and electron CBM.

\begin{figure}
\includegraphics[scale=0.88]{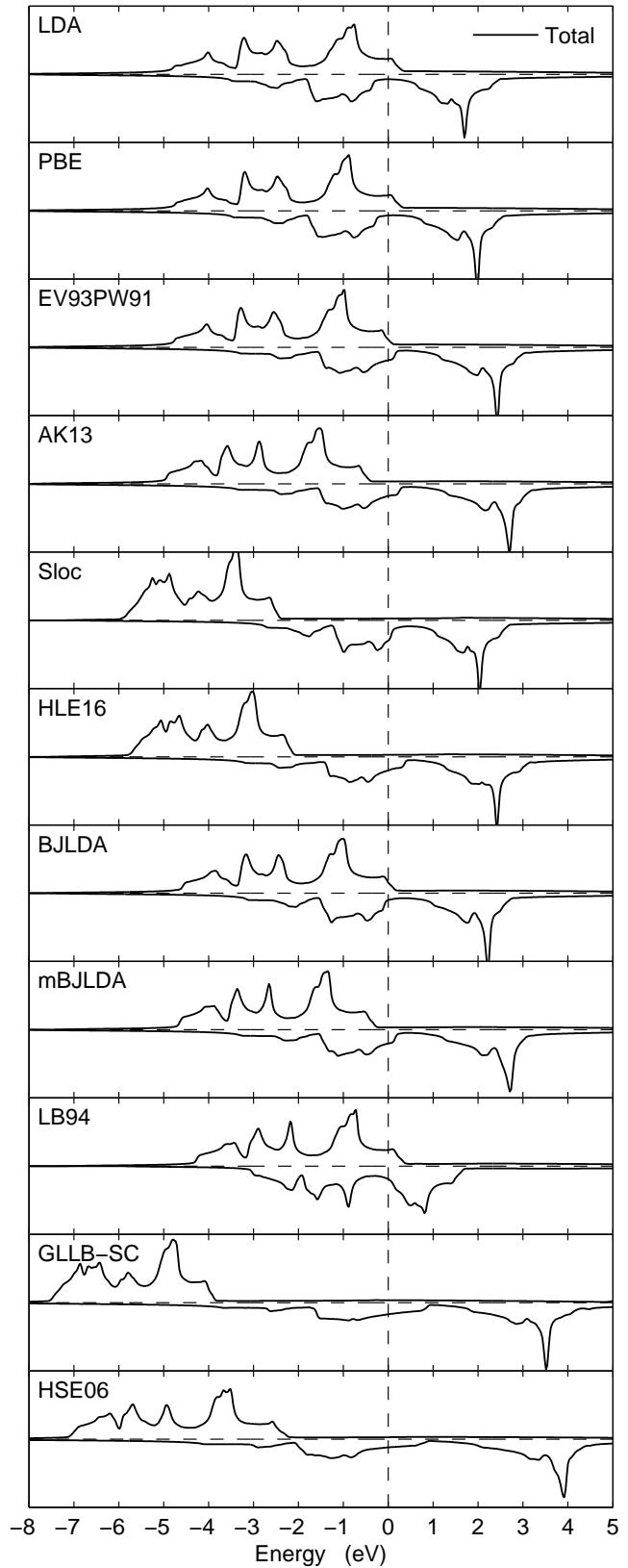}
\caption{\label{fig_Fe_DOS}Calculated spin-up (upper curve) and
spin-down (lower curve) total DOS of ferromagnetic Fe.
The Fermi energy is set at zero.}
\end{figure}

To finish the discussion about the electronic structure, we show in
Fig.~\ref{fig_Fe_DOS} the density of states (DOS) of the ferromagnetic metal Fe.
Compared to the case of $3d$ transition-metal oxides discussed above,
the LDA and standard GGAs should be more reliable for itinerant transition
metals where the $3d$ electrons are weakly correlated. In the case of Fe,
the DOS obtained with LDA was shown to be in rather good agreement with
experiment (spin-resolved X-ray photoelectron spectroscopy\cite{SeeSS95})
for the valence band, and the same can be said for the PBE DOS which differs
very little from the LDA DOS. However, the DOS obtained with some of the other
methods differ substantially from the LDA/PBE DOS. The largest
differences are obtained with GLLB-SC and HSE06, which lead to severely
overestimated exchange splittings as shown in Fig.~\ref{fig_Fe_DOS} and,
consequently, to magnetic moments that are by far too large
(see Sec.~\ref{magnetism}). The overestimation of the exchange splitting is
also very large with the Sloc and HLE16 potentials.
The results for Co and Ni are quite similar with an exchange splitting
that is the largest with GLLB-SC and HSE06.
The failure of hybrid functionals for Fe, Co, and Ni has already been
pointed out in recent studies.\cite{PaierJCP06,JangJPSJ12,JanthonJCTC14,GaoSSC16}

\subsection{\label{magnetism}Magnetism}

\begin{table*}
\caption{\label{table_MMI}Calculated atomic spin magnetic moment $\mu_{S}$
(in $\mu_{\text{B}}$) of antiferromagnetic $3d$-transition metal oxides
compared to experimental values of the total atomic magnetic moment $\mu_{S}+\mu_{L}$.
The orbital moment $\mu_{L}$ is estimated to be in the range
0.6-1~$\mu_{\text{B}}$ for FeO,
\cite{SvanePRL90,TranPRB06,RadwanskiPB08,SchronJPCM13}
1-1.6~$\mu_{\text{B}}$ for CoO,
\cite{SvanePRL90,SolovyevPRL98,ShishidouJPSJ98,NeubeckJPCS01,JauchPRB02,GhiringhelliPRB02,RadwanskiPB04,TranPRB06,RadwanskiPB08,BoussendelPRB10,SchronJPCM13}
0.3-0.45~$\mu_{\text{B}}$ for NiO,\cite{SvanePRL90,FernandezPRB98,NeubeckJPCS01,RadwanskiPB04,RadwanskiPB08}
and much smaller in other systems.
The values of $\mu_{S}$ are those inside the atomic
sphere of radius (in bohr) 2.02 (MnO), 2.00 (FeO), 2.00 (CoO),
1.92 (NiO), 1.97 (CuO), 1.94 (Cr$_{2}$O$_{3}$), and 1.96 (Fe$_{2}$O$_{3}$)
of the transition-metal atom.
The calculations were done at the experimental geometry
specified in Table~S1 of the Supplemental Material.\cite{SM_GLLB}}
\begin{ruledtabular}
\begin{tabular}{lccccccc}
Method   & MnO  & FeO  & CoO  & NiO  & CuO  & Cr$_{2}$O$_{3}$ & Fe$_{2}$O$_{3}$ \\
\hline
LDA      & 4.11 & 3.33 & 2.36 & 1.21 & 0.12 & 2.36            & 3.34            \\
PBE      & 4.17 & 3.39 & 2.43 & 1.38 & 0.38 & 2.44            & 3.53            \\
EV93PW91 & 4.24 & 3.44 & 2.49 & 1.47 & 0.46 & 2.53            & 3.71            \\
AK13     & 4.39 & 3.51 & 2.59 & 1.57 & 0.54 & 2.68            & 3.93            \\
Sloc     & 4.55 & 3.59 & 2.53 & 1.40 & 0.32 & 3.11            & 3.97            \\
HLE16    & 4.51 & 3.62 & 2.59 & 1.48 & 0.40 & 2.96            & 4.02            \\
BJLDA    & 4.19 & 3.40 & 2.48 & 1.48 & 0.50 & 2.43            & 3.59            \\
mBJLDA   & 4.41 & 3.58 & 2.71 & 1.75 & 0.74 & 2.60            & 4.09            \\
LB94     & 3.93 & 3.02 & 1.75 & 0.67 & 0.00 & 2.27            & 1.50            \\
GLLB-SC  & 4.56 & 3.74 & 2.73 & 1.65 & 0.55 & 2.99            & 4.43            \\
HSE06    & 4.36 & 3.55 & 2.65 & 1.68 & 0.67 & 2.61            & 4.08            \\
Expt.    &
4.58\footnotemark[1] &
3.32,\footnotemark[2]4.2,\footnotemark[3]4.6\footnotemark[4] &
3.35,\footnotemark[5]3.8,\footnotemark[2]\footnotemark[6]3.98\footnotemark[7] &
1.9,\footnotemark[1]\footnotemark[2]2.2\footnotemark[8]\footnotemark[9] &
0.65\footnotemark[10] &
2.44,\footnotemark[11]2.48,\footnotemark[12]2.76\footnotemark[13] &
4.17,\footnotemark[14]4.22\footnotemark[15] \\
\end{tabular}
\end{ruledtabular}
\footnotetext[1]{Ref.~\onlinecite{CheethamPRB83}.} 
\footnotetext[2]{Ref.~\onlinecite{RothPR58}.} 
\footnotetext[3]{Ref.~\onlinecite{BattleJPC79}.} 
\footnotetext[4]{Ref.~\onlinecite{FjellvagJSSC96}.} 
\footnotetext[5]{Ref.~\onlinecite{KhanPRB70}.} 
\footnotetext[6]{Ref.~\onlinecite{HerrmannRonzaudJPC78}.} 
\footnotetext[7]{Ref.~\onlinecite{JauchPRB01}.} 
\footnotetext[8]{Ref.~\onlinecite{FernandezPRB98}.} 
\footnotetext[9]{Ref.~\onlinecite{NeubeckJAP99}.} 
\footnotetext[10]{Ref.~\onlinecite{ForsythJPC88}.} 
\footnotetext[11]{Ref.~\onlinecite{GolosovaJAC17}.} 
\footnotetext[12]{Ref.~\onlinecite{BrownJPCM02}.} 
\footnotetext[13]{Ref.~\onlinecite{CorlissJAP65}.} 
\footnotetext[14]{Ref.~\onlinecite{BaronSSS05}.} 
\footnotetext[15]{Ref.~\onlinecite{HillCM08}.} 
\end{table*}

Turning now to the magnetic properties of systems with $3d$ electrons,
Table~\ref{table_MMI} shows the atomic spin magnetic moment $\mu_{S}$ in
antiferromagnetic transition-metal oxides. The comparison
with experiment should be done by keeping in mind that there is a non-negligible
orbital contribution $\mu_{L}$ for FeO, CoO, and NiO (see caption of Table~\ref{table_MMI}).

In addition of being particularly inaccurate to describe the electronic
structure of strongly correlated solids, the LDA and commonly used GGAs like
PBE also lead to magnetic moments that are too small for this class of solids.
\cite{TerakuraPRB84,DufekPRB94a} DFT+$U$\cite{AnisimovPRB91} and the hybrid
functionals\cite{BredowPRB00,MoreiraPRB02,FranchiniPRB05,TranPRB06,MarsmanJPCM08}
lead to much improved results and are therefore commonly used nowadays for such systems.
However, those multiplicative potentials which are more accurate than LDA/PBE
for the band gap also improve the results for the magnetic moment in most cases.
From Table~\ref{table_MMI} we can see that in most cases, all tested potentials
except LB94 increase the value of $\mu_{S}$ compared to LDA/PBE.
EV93PW91 and BJLDA lead to moments which are only moderately larger, and
AK13, Sloc, and HLE16 further improve the results, but the agreement with experiment
is still not always satisfying.
For instance, AK13 leads to a moment that is too small by 0.2~$\mu_{\text{B}}$
in MnO, while too large values are obtained with Sloc and HLE16 in the case of
Cr$_{2}$O$_{3}$. GLLB-SC is pretty accurate for all monoxides, but less
for Cr$_{2}$O$_{3}$ and Fe$_{2}$O$_{3}$ since the moments are overestimated by
at least 0.2~$\mu_{\text{B}}$.
Overall, the most reliable multiplicative potential seems to be mBJLDA, since it is the
only one which leads to an error in the magnetic moment that should be below
$\sim0.2$~$\mu_{\text{B}}$ when a quantitative comparison with experiment is
possible. As already pointed out in Ref.~\onlinecite{KollerPRB11},
the moment of CuO obtained with mBJLDA is too large by at least 0.1~$\mu_{\text{B}}$.
Furthermore, we note that for most systems the mBJLDA results are very
close to the results obtained with HSE06.
The worst results are obtained with LB94 which leads to the smallest magnetic
moments in all cases.

\begin{table*}
\caption{\label{table_MMTOT}Calculated unit cell spin magnetic moment
$\mu_{S}$ (in $\mu_{\text{B}}$/atom) of $3d$-transition metals.
The experimental values are also spin magnetic moments.
The calculations were done at the experimental geometry
specified in Table~S1 of the Supplemental Material.\cite{SM_GLLB}}
\begin{ruledtabular}
\begin{tabular}{lccc}
Method   & Fe   & Co   & Ni   \\
\hline
LDA      & 2.21 & 1.59 & 0.61 \\
PBE      & 2.22 & 1.62 & 0.64 \\
EV93PW91 & 2.48 & 1.68 & 0.68 \\
AK13     & 2.58 & 1.70 & 0.69 \\
Sloc     & 2.69 & 1.63 & 0.50 \\
HLE16    & 2.72 & 1.72 & 0.63 \\
BJLDA    & 2.39 & 1.63 & 0.62 \\
mBJLDA   & 2.51 & 1.69 & 0.73 \\
LB94     & 2.02 & 1.39 & 0.41 \\
GLLB-SC  & 3.08 & 1.98 & 0.81 \\
HSE06    & 2.79 & 1.90 & 0.88 \\
Expt.    &
1.98,\footnotemark[1]2.05,\footnotemark[2]2.08\footnotemark[3] &
1.52,\footnotemark[3]1.58,\footnotemark[2]\footnotemark[4]1.55-1.62\footnotemark[1] &
0.52,\footnotemark[3]0.55\footnotemark[2]\footnotemark[5] \\
\end{tabular}
\end{ruledtabular}
\footnotetext[1]{Ref.~\onlinecite{ChenPRL95}.}
\footnotetext[2]{Ref.~\onlinecite{Scherz03}.}
\footnotetext[3]{Ref.~\onlinecite{ReckPR69}.}
\footnotetext[4]{Ref.~\onlinecite{MoonPR64}.}
\footnotetext[5]{Ref.~\onlinecite{MookJAP66}.}
\end{table*}

We also calculated the unit cell spin magnetic moment of the ferromagnetic metals
Fe, Co, and Ni, and Table~\ref{table_MMTOT} shows the results that are
compared with experimental values that do not include the orbital
component $\mu_{L}$. It is well known that the
simple LDA is relatively accurate for the magnetic moment of itinerant metals,
while the trend of standard GGAs is to slightly overestimate the values
(see Refs.~\onlinecite{BarbielliniJPCM90,SinghPRB91b,LeungPRB91,AmadorPRB92}
for early results on Fe, Co, and Ni).

Our results in Table~\ref{table_MMTOT} follow the same trends observed above
for the transition-metal oxides. The LDA and PBE magnetic moments are
(aside from the results with LB94 and Sloc) the smallest, however, the major
difference is that for
the metals, the agreement with experiment deteriorates if another potential is
used, since LDA and PBE already overestimate the values (albeit slightly).
For the three metals, GLLB-SC leads to magnetic moments which are by far the
largest among the multiplicative potentials
and too large with experiment by about 50\% for Fe and Ni and 25\% for Co,
which is clearly worse than the overestimations obtained with the
mBJLDA potential (see also Ref.~\onlinecite{KollerPRB11})
that are about 25\% (Fe), 10\% (Co), and 35\% (Ni).

In Ref.~\onlinecite{TranPRB12} we showed that the screened hybrid functional YS-PBE0
($\sim$ HSE06) leads to a ground-state solution in fcc Rh, Pd, and Pt that is
ferromagnetic instead of being nonmagnetic as determined experimentally,
and the same was obtained with PBE for Pd. In general, such problems are more likely to
occur with strong potentials like mBJLDA, AK13, or GLLB-SC, but probably
not with LDA which is the weakest potential. For Pd for instance, the
unit cell spin magnetic moment at the experimental geometry ($a=3.887$~\AA)
is 0.25~$\mu_{\text{B}}$ for PBE, 0.36~$\mu_{\text{B}}$ for HLE16,
0.39-0.40~$\mu_{\text{B}}$ for EV93PW91, AK13, and mBJLDA, and
0.44~$\mu_{\text{B}}$ for GLLB-SC, which is similar to
0.43~$\mu_{\text{B}}$ obtained with YS-PBE0/HSE06.\cite{TranPRB12}
This ferromagnetic state is more stable than the nonmagnetic one
for all these methods.
No energy functional exists for mBJLDA and GLLB-SC,
but independently of the one that is used to evaluate the total energy
(except maybe LDA), the results show that the ferromagnetic state has a more
negative total energy than the nonmagnetic one. Thus, such potentials should be
used with care also in nonmagnetic metals.

In general, the use of the HF or EXX-OEP methods is not recommended for
itinerant metals,
\cite{KotaniJMMM98,SchnellPRB03,PaierJCP06,TranPRB12,JangJPSJ12,GaoSSC16}
since for instance, even the use of only 25\% of screened HF, as in HSE06,
leads to very large overestimations
\cite{PaierJCP06,JangJPSJ12,JanthonJCTC14,GaoSSC16}
in the magnetic moment (see our HSE06 results in Table~\ref{table_MMTOT}).
Compared to GLLB-SC, the HSE06 magnetic moment is much smaller for Fe,
but more similar for Co and Ni. It is only when EXX-OEP is used in combination
with the RPA for correlation that reasonable values can be obtained for the
magnetic moments of Fe, Co, and Ni.\cite{KotaniJMMM98,FukazawaJPCM15}
Concerning $GW$, a recent study reported large overestimations
with self-consistent $GW$,\cite{KutepovJPCM17} while a good
agreement with experiment was obtained with quasi-self-consistent $GW$.
\cite{vanSchilfgaardePRL06}

\subsection{\label{EFG}Electric field gradient}

\begin{table*}
\caption{\label{table_EFG}EFG (in $10^{21}$ V/m$^{2}$) in elemental metals and
at the Cu site in CuO, Cu$_{2}$O, and Cu$_{2}$Mg. The error bars of the
experimental values are calculated from the uncertainty in the quadrupole
moment and quadrupole coupling constants when available. The values which
are far outside the range of experimental estimates are underlined.
The calculations were done at the experimental geometry
specified in Table~S1 of the Supplemental Material.\cite{SM_GLLB}}
\begin{ruledtabular}
\begin{tabular}{lccccccccc}
Method                & Ti               & Zn               & Zr               & Tc                & Ru                & Cd               & CuO                & Cu$_{2}$O         & Cu$_{2}$Mg        \\
\hline
LDA                   & \underline{1.80} & 3.50             & 4.21             & -1.65             & \underline{-1.56} & 7.47             & \underline{-1.86}  & \underline{-5.27} & -5.70             \\
PBE                   & 1.73             & 3.49             & 4.19             & -1.61             & \underline{-1.46} & 7.54             & \underline{-2.83}  & \underline{-5.54} & -5.70             \\
EV93PW91              & 1.61             & 3.43             & 4.13             & -1.57             & \underline{-1.33} & 7.63             & \underline{-3.17}  & \underline{-6.53} & -5.82             \\
AK13                  & 1.65             & 3.86             & 4.17             & \underline{-1.28} & -1.13             & \underline{8.53} & \underline{-3.56}  & \underline{-7.92} & -5.44             \\
Sloc                  & 1.44             & \underline{3.93} & \underline{2.75} & \underline{-0.52} & \underline{-0.35} & 8.01             & \underline{-3.97}  & -11.97            & \underline{-4.10} \\
HLE16                 & 1.70             & 3.29             & \underline{3.78} & \underline{-0.95} & -0.73             & 7.66             & \underline{-4.18}  & -10.10            & \underline{-4.59} \\
BJLDA                 & \underline{1.97} & 3.51             & 4.25             & \underline{-1.27} & -1.16             & 7.61             & \underline{-5.42}  & \underline{-7.74} & -5.20             \\
mBJLDA                & \underline{1.99} & 3.35             & 4.33             & \underline{-1.20} & -0.90             & 7.56             & \underline{-13.93} & \underline{-7.40} & \underline{-4.89} \\
LB94                  & \underline{0.94} & 3.78             & \underline{1.83} & \underline{-0.72} & -1.05             & 7.47             & \underline{-1.23}  & -11.16            & \underline{-4.97} \\
GLLB-SC               & 1.62             & 3.72             & 4.42             & -1.66             & -1.26             & 8.05             & \underline{-4.65}  & -9.99             & -5.58             \\
HSE06                 & 1.5              & \underline{4.4}  & 4.5              & -2.0              & \underline{-1.3}  & \underline{9.4}  & -8.9               & -8.3              & \underline{-6.3}  \\
Expt.\footnotemark[1] & 1.57(12)         & 3.40(35)         & 4.39(15)         & 1.83(9)           & 0.97(11)          & 7.60(75)         & 7.55(52)           & 10.08(69)         & 5.76(39)          \\
\end{tabular}
\end{ruledtabular}
\footnotetext[1]{Calculated using the nuclear quadrupole
moments\cite{StoneADNDT16} (in barn) of
0.302(10) ($^{47}$Ti, 5/2-),
0.220(15) ($^{63}$Cu, 3/2-),
0.150(15) ($^{67}$Zn, 5/2-),
0.176(3) ($^{91}$Zr, 5/2+),
0.129(6) ($^{99}$Tc, 9/2+),
0.231(13) ($^{99}$Ru, 3/2+), and
0.74(7) ($^{111}$Cd, 5/2+),
and the nuclear quadrupole coupling constants (in MHz) of
11.5(5) (Ti),\cite{EbertJPF86}
12.34(3) (Zn),\cite{ViandenHI87}
18.7(3) (Zr),\cite{ViandenHI87}
5.716 (Tc),\cite{ViandenHI87}
5.4(3) (Ru),\cite{ViandenHI87}
136.02(41) (Cd),\cite{ChristiansenZPB76}
40.14 (CuO),\cite{GrahamJPCM91}
53.60 (Cu$_{2}$O),\cite{GrahamJPCM91} and
30.66 (Cu$_{2}$Mg).\cite{AzevedoJPF81}}
\end{table*}

Now we consider the EFG, which is a measure of the accuracy of the
electron density, and Table~\ref{table_EFG} shows the values for
elemental metals and at the Cu site in CuO, Cu$_{2}$O, and Cu$_{2}$Mg.
In our recent works,\cite{TranPRB11,KollerPRB11,TranPRB15,TranPRB16} we showed
that in the case of Cu$_{2}$O, the standard semilocal functionals, DFT+$U$, and
on-site hybrids (similar to DFT+$U$) lead to magnitudes of the EFG that
are by far too small compared to experiment, while
the mBJLDA value is much too large.
Better results could be obtained with hybrid functionals\cite{TranPRB11}
or with nonstandard semilocal methods like AK13 or other variants of the
BJ potential.\cite{TranPRB15,TranPRB16} In the case of CuO,
\cite{KollerPRB11} PBE and mBJLDA underestimates and
overestimates significantly the EFG, respectively, while the on-site hybrid
used in Ref.~\onlinecite{RocquefelteJPCM10} was pretty accurate.
A study by Haas and Correia\cite{HaasHI07} on many other Cu$^{2+}$ compounds
showed that it is necessary to use DFT+$U$ in order to get a reasonable
agreement with experiment.

The results of the present work indicate that the GLLB-SC potential is overall
the most accurate for the EFG. Indeed, it is only in the case of CuO that
GLLB-SC leads to an EFG that differs noticeably from the experimental value,
while all other potentials are clearly inaccurate in more than one case.
Furthermore, despite that the error for CuO with GLLB-SC is rather large, it is still
one of the smallest. The most inaccurate methods are LDA, Sloc,
HLE16, mBJLDA, and LB94, which lead to large errors in four or five cases and are
therefore not recommended for EFG calculations no matter what the system is
(a semiconductor or a metal). In particular, Sloc leads to extremely large
underestimation of the magnitude of the EFG in Zr, Tc, and Ru.
Regarding the hybrid functional HSE06, the results for the metals
seem to be reasonable for Ti, Zr, and Tc, while large errors are
obtained for the others (see also Haas \textit{et al}.\cite{HaasHI16} for
previous results). Thus, as mentioned above for the magnetic moment
and in previous works,\cite{PaierJCP06,TranPRB12,JangJPSJ12,GaoSSC16}
the hybrid functionals are not especially recommended for itinerant metals.

In conclusion, the GLLB-SC potential seems to be the most reliable method for
the calculation of the EFG in solids. Noteworthy, in contrast to the strong
overestimation of the magnetic moment of Fe, Co, and Ni with GLLB-SC,
the accuracy for the EFG in metals is very good and apparently superior
to LDA and PBE that were supposed to lead to qualitatively correct results
in metals.\cite{BlahaPRB88a}

\subsection{\label{density}Electron density of Si}

\begin{table}
\caption{\label{table_rho}
Difference between the experimental and calculated electron densities of Si as
measured by the $R$-factor (in $\%$) and GoF defined as
$R=100\sum_{i=1}^{N}
\left\vert f_{i}^{\text{calc}}-f_{i}^{\text{exp}}\right\vert/
\sum_{i=1}^{N}\left\vert f_{i}^{\text{exp}}\right\vert$ and
$\text{GoF}=\left(1/N\right)\sum_{i=1}^{N}
\left(f_{i}^{\text{calc}}-f_{i}^{\text{exp}}\right)^{2}/\sigma_{i}^{2}$,
where the sums are over the $N=31$ form factors $f_{i}$
of Table~S3 and the GoF is calculated with the average variance $\sigma=0.0022$.
The calculated form factors are multiplied by a temperature factor
(see Ref.~\onlinecite{ZuoJPCM97} for details).}
\begin{ruledtabular}
\begin{tabular}{lcc}
Method & $R$ & GoF \\ 
\hline 
LDA                   &   0.25 &    34.6 \\
PBE                   &   0.13 &    10.1 \\
EV93PW91              &   0.14 &    12.4 \\
AK13                  &   0.31 &    67.3 \\
Sloc                  &   2.22 &  2028.9 \\
HLE16                 &   1.64 &  1087.8 \\
BJLDA                 &   0.16 &    14.1 \\
mBJLDA                &   0.20 &    32.4 \\
LB94                  &   0.55 &   200.8 \\
GLLB-SC               &   0.75 &   240.1 \\
HSE06                 &   0.10 &     5.5 \\
\end{tabular}
\end{ruledtabular}
\end{table}
\begin{figure}
\includegraphics[scale=0.76]{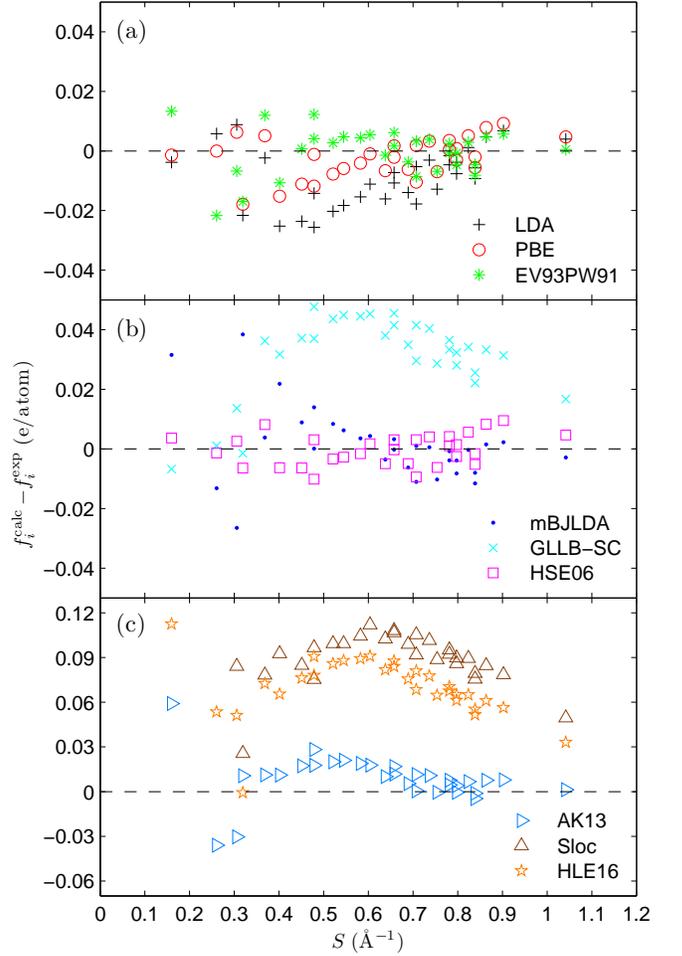}
\caption{\label{fig_form_factors_Si}Difference between the experimental and
calculated form factors of Si in Table~S3 plotted as a function of
$S=\sin\left(\theta\right)/\lambda$. The errors with Sloc for the 1st (111)
and 2nd (220) form factors are 0.22 and 0.13~e/atom, respectively, and
outside the range of the $y$-axis range.
Note that the scale of the $y$-axis in panel (c) is different from
the one in panels (a) and (b).
The results for BJLDA and LB94 are omitted.}
\end{figure}

\begin{figure}
\includegraphics[scale=0.62]{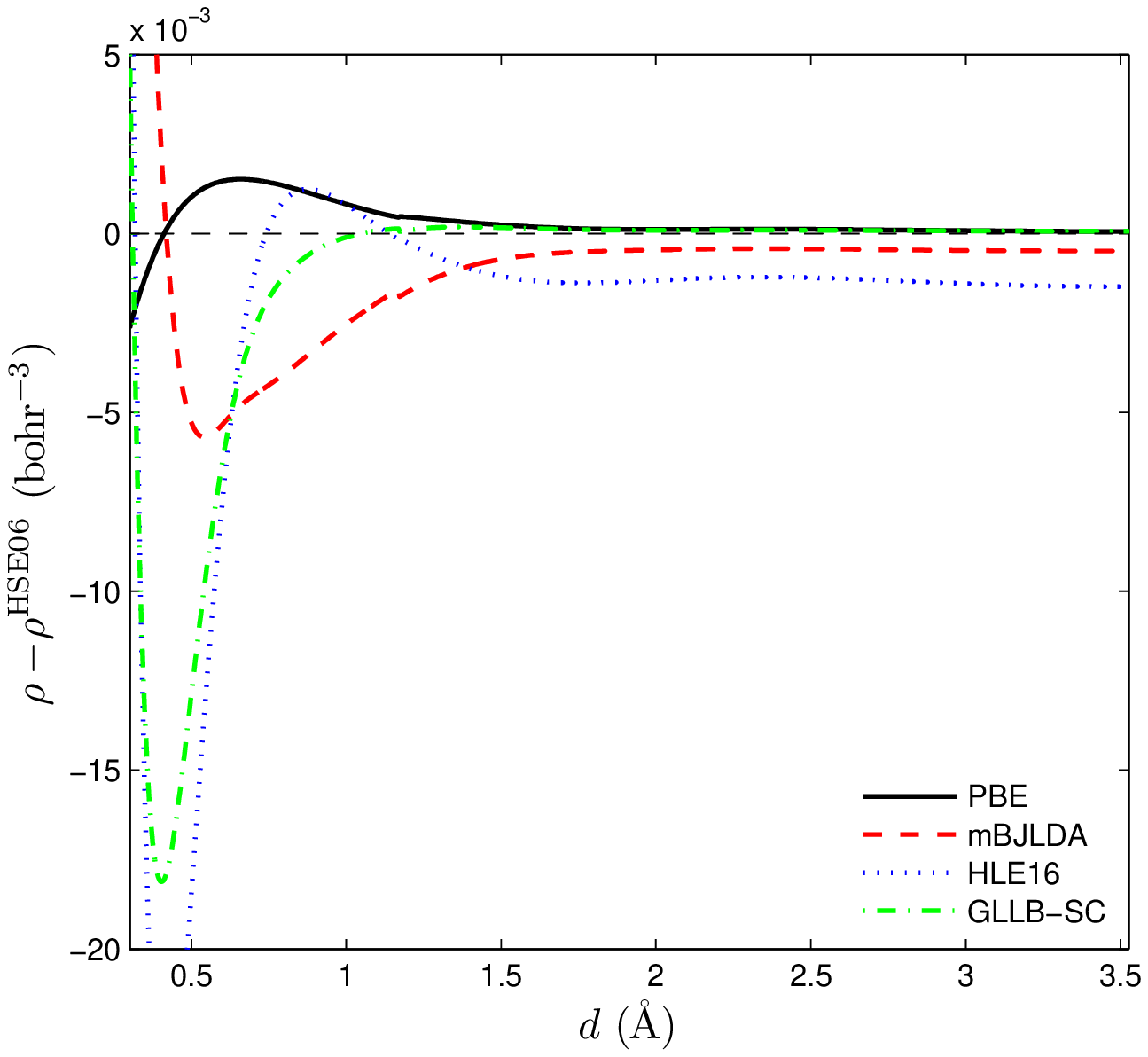}
\caption{\label{fig_rho_Si}
Difference between the electron density $\rho$ in Si obtained with various
potentials and the one from HSE06 which is considered as the reference.
The plot is along a path starting at a distance of
$d=0.3$~\AA~from the atom at $\left(\frac{1}{8},\frac{1}{8},\frac{1}{8}\right)$
in the direction of the center of the unit cell which is at $d=3.527$~\AA.}
\end{figure}

The last property that we want to consider in order to judge the quality of the xc
potentials is the electron density of Si, for which X-ray structure factors have been
experimentally measured for the reflections from (111) to
(880) and the Si form factors derived from them.\cite{SakaAC86,CummingAJP88}
The calculated values are given in Table~S3
of the Supplemental Material\cite{SM_GLLB} and the deviations with experiment are
shown graphically in Fig.~\ref{fig_form_factors_Si}. As done in previous works (see, e.g.,
Refs.~\onlinecite{LuPRB93,ZuoJPCM97}) the agreement with experiment is
quantified in terms of $R$-factor and goodness-of-fit (GoF) (defined in the
caption of Table~\ref{table_rho}). The results in Table~\ref{table_rho} show
that the lowest errors, $R=0.10$\% and $\text{GoF}=5.5$, are obtained with the
hybrid functional HSE06. The best nonhybrid methods are PBE, EV93PW91, and BJLDA which
lead to values for $R$ and GoF that are slightly larger than with HSE06.
Next come LDA and mBJLDA which lead to very similar values for $R$ and GoF,
that are roughly two (for $R$) or three (for GoF) times larger than with PBE
and EV93PW91. The most inaccurate electron densities are obtained with Sloc and
HLE16, since the values for $R$ and GoF are one and two orders
of magnitude larger, respectively. The errors obtained with GLLB-SC are also
significant since $R=0.75$\% and $\text{GoF}=240$.

It is also instructive to look at the form factors individually in order to
have a clue about which part of the electron density (shown in
Fig.~\ref{fig_rho_Si}) is described (in)accurately by a given potential.
The bonding/valence region is revealed by the low-order form factors (very
roughly, corresponding to $S\lesssim0.3$~\AA$^{-1}$ in Fig.~\ref{fig_form_factors_Si}),
while the high-order ones correspond to the high density of the semicore and
core electrons that are localized around the nucleus.
As discussed in Ref.~\onlinecite{ZuoJPCM97}, the use of EV93 for exchange
(that was combined with LDA for correlation in that work)
improves the description of the core density (in particular of the
$2s$- and $2p$-electron subshells), but deteriorates the accuracy of the
bonding region compared to LDA and the GGA PW91\cite{PerdewPRB92b} (very similar
to PBE). This is confirmed in Fig.~\ref{fig_form_factors_Si}(a), where we can see that the
deviations from experiment with EV93PW91 are larger for the first five
form factors, but smaller on average for the higher ones compared to LDA and PBE.
An explanation for this may be that the exchange EV93 functional was fitted
to the EXX-OEP in atoms, which is possibly more accurate than
standard LDA/PBE for core electrons.

The mBJLDA potential [see Figs.~\ref{fig_form_factors_Si}(b) and \ref{fig_rho_Si}] is
quite inaccurate for the bonding region, since for the first few low-order form factors
the errors are quite large, and slightly larger than with EV93PW91. However,
for the (semi)core density, mBJLDA is of similar accuracy as PBE and EV93PW91,
and therefore quite accurate. The GLLB-SC potential shows the opposite trend compared
to mBJLDA: accurate bonding/valence electron density (small errors for the first four
form factors) and very inaccurate core density (large errors for all other
form factors).
The HSE06 functionals leads to errors that are small for all form factors.
Figure~\ref{fig_form_factors_Si}(c) shows that the Sloc and HLE16 potentials lead
to extremely inaccurate electron density in general. The errors
$f_{i}^{\text{calc}}-f_{i}^{\text{exp}}$ are in the range 0.05-0.12~e/atom for
most form factors, while the errors with LDA, PBE, and EV93PW91 are all below
0.02~e/atom. Figure~\ref{fig_rho_Si} shows indeed that the error
in the electron density (HSE06 is chosen as the reference) 
with HLE16 is much larger than with PBE in the semicore and valence regions.
Concerning the AK13 potential, the errors are very
large for the first three form factors (i.e., the valence $3s$ and $3p$ electrons),
but more or less
of the same magnitude as LDA (but with opposite sign) for the others, which represent
the core density.

The main conclusion of this section is that among the semilocal methods,
PBE is the most accurate for the electron density of Si. The other methods
are less accurate for the core and/or valence parts of the electron density.
The GLLB-SC potential seems to be as accurate as PBE for the valence/bonding
density, which is consistent with the observations made in Sec.~\ref{EFG} for
the EFG that is determined mainly by the valence electron density.
The mBJLDA potential describes quite well the core density, but not the
valence density. Overall, the best performance is obtained with the hybrid HSE06.

\subsection{\label{further}Further discussion}

\subsubsection{\label{plot}Visualization of the xc potentials}

The results presented above should provide some guidance when choosing
(within the KS method) an exchange-correlation potential that is adequate
for the problem at hand. However, it has also
been clearly shown that none of the tested potentials leads to sufficiently
accurate results in all circumstances, which is hardly surprising
with semilocal and hybrid methods. The search for a fast semilocal multiplicative xc
potential which is more universally accurate than those presented above is
certainly not an easy task. However, in this respect it may be helpful
to try to understand what is going on in terms of the shape of the potentials
considered in this work.

\begin{figure}
\includegraphics[scale=0.72]{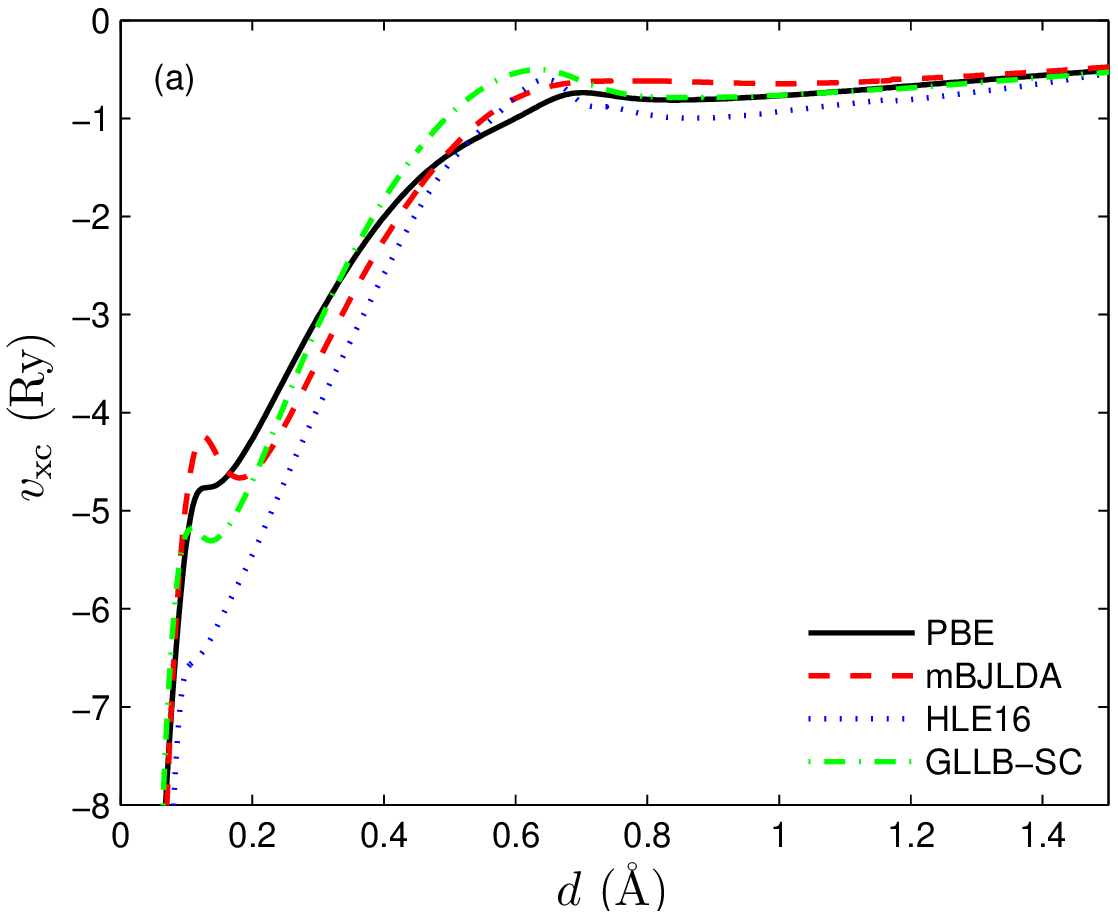}
\includegraphics[scale=0.72]{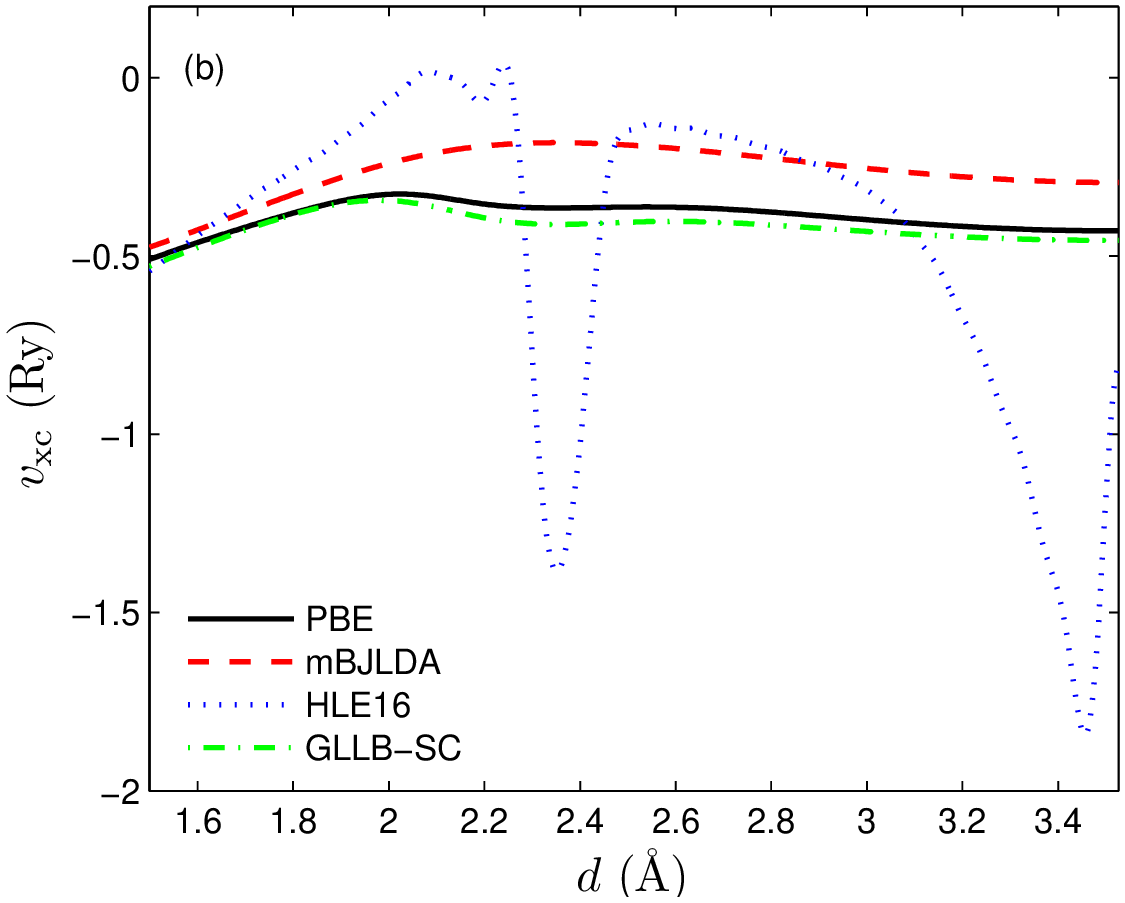}
\caption{\label{fig_vxc_Si}
Potentials $v_{\text{xc}}$ in Si plotted from the atom
at $(\frac{1}{8},\frac{1}{8},\frac{1}{8})$ ($d=0$) to the center of the
unit cell which is at $d=3.527$~\AA.}
\end{figure}

In previous works,
\cite{StaedelePRL97,StaedelePRB99,AulburPRB00,QteishPRB06,TranJPCM07,BetzingerPRB11,KollerPRB11,TranPRB15,TranJCTC15,TranPRB16}
trends in the results could be understood by comparing the shape of the
potentials. Basically, the magnitude of the band gap and magnetic moment are
directly related to the inhomogeneities in the potential, and it was observed
and rationalized that more pronounced inhomogeneities favor larger values of
the band gap and magnetic moment. In order to understand some of the results
discussed in the previous sections, two cases are now studied in
more detail.

In Sec.~\ref{density}, we showed that some of the potentials lead to very
inaccurate electron density in Si. For instance, the densities obtained
with Sloc, HLE16, and GLLB-SC differ quite significantly from the
reference HSE06 density in the region close to the nucleus
($d\lesssim0.7$~\AA~in Fig.~\ref{fig_rho_Si}). Taking a look at the
potentials should help us to understand the reason for this, and actually
Fig.~\ref{fig_vxc_Si}(a) shows that from $d=0.1$ to 0.7~\AA~the magnitudes of the
HLE16 and GLLB-SC potentials (Sloc is not shown but similar) vary faster than for
PBE and mBJLDA, which are much more accurate for the (semi)core density.
It is this (too) fast variation in $v_{\text{xc}}$ which leads to
inaccurate density in the (semi)core region.

In our previous works,\cite{TranPRB15,TranJPCA17} we showed that the
nonstandard GGA potentials EV93PW91, AK13, and HLE16 show large oscillations
in the middle of the interstitial region [visible for $d$ larger than
1.5~\AA~in Si, see Fig.~\ref{fig_vxc_Si}(b)], which should mainly be a
consequence of their strong dependence on the second derivative of $\rho$.
This is not the case for the BJ-type potentials and EXX-OEP
which were shown to be rather flat (as LDA and PBE) in the interstitial.
The RPA-OEP also seems to be smooth according to Ref.~\onlinecite{KlimesJCP14}.
Since the GLLB-SC potential depends on the second derivative of $\rho$
only via the correlation potential $v_{\text{c}}^{\text{PBEsol}}$, i.e. weakly,
it is also smooth in the interstitial and very close to PBE (and LDA)
as shown in Fig.~\ref{fig_vxc_Si}(b).

\begin{figure}
\includegraphics[scale=0.79]{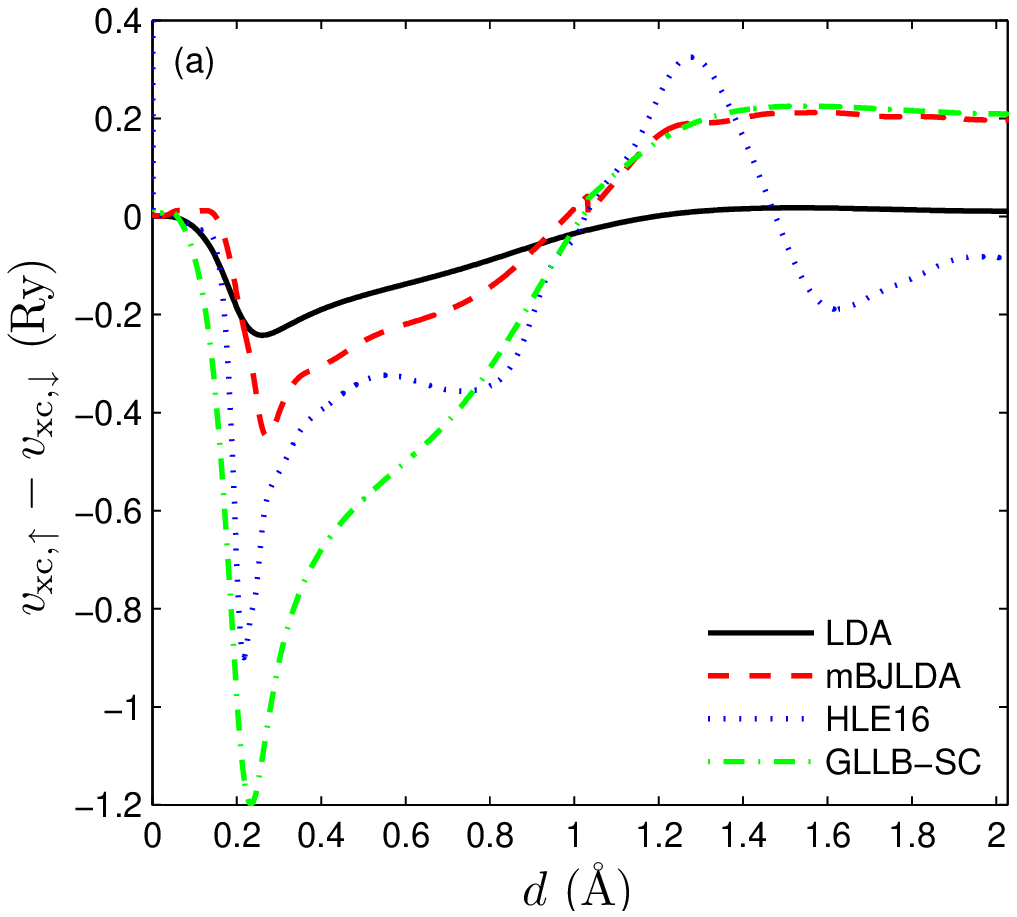}
\includegraphics[scale=0.79]{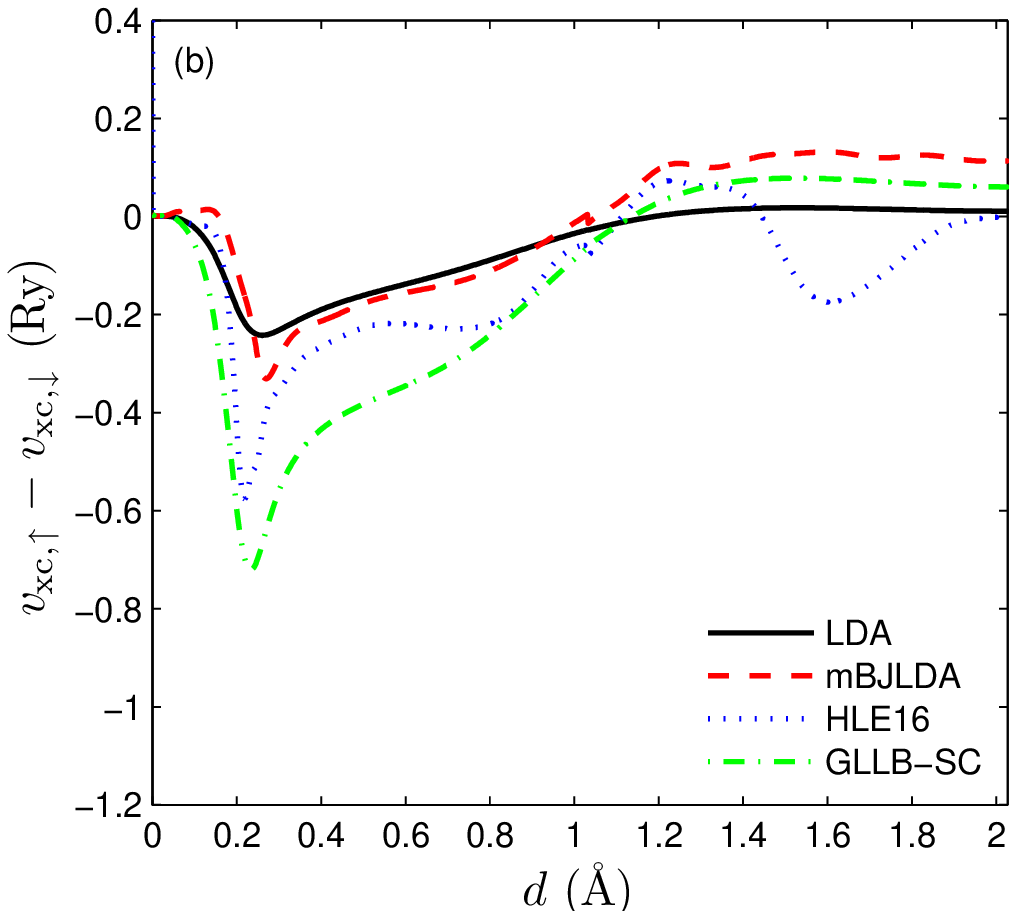}
\caption{\label{fig_vxcup_vxcdn_Fe}
$v_{\text{xc},\uparrow}-v_{\text{xc},\downarrow}$ in Fe plotted from the atom
at (0,0,0) of the body-centered cubic cell to
$\left(\frac{1}{2},\frac{1}{2},0\right)$ (middle of a face of the cell).
(a) shows the potentials of the self-consistent calculations,
while (b) shows the potentials calculated non-self-consistently
using $\rho_{\sigma}$, $t_{\sigma}$, and $\psi_{i\sigma}$ from the LDA calculation.}
\end{figure}

One of the problems of GLLB-SC is to overestimate the exchange splitting in
metals and, therefore, the magnetic moment in Fe, Co, and Ni.
Figure~\ref{fig_vxcup_vxcdn_Fe}(a) shows the difference
$v_{\text{xc},\uparrow}-v_{\text{xc},\downarrow}$ between the 
self-consistent spin-up and spin-down xc potentials in Fe, where we can see that
it is the largest for GLLB-SC. The large exchange splitting observed in
Fig.~\ref{fig_Fe_DOS} for HLE16 can also be understood from the large
magnitude of $v_{\text{xc},\uparrow}-v_{\text{xc},\downarrow}$.
Note that for $d\gtrsim1$~\AA, the mBJLDA and GLLB-SC potentials coincide
very closely. Besides this, we can also see that
$v_{\text{xc},\uparrow}-v_{\text{xc},\downarrow}$ is negative until
$d\sim1$~\AA~and then positive in some cases. The negative region
is where the $3d$ electrons, which are the main contributors to the
magnetic moment, are located, while the positive region is the interstitial
where the $s$ and $p$ electrons, which also contribute
to the magnetic moment but with opposite sign
and a much smaller magnitude,\cite{MoonPR64,MookJAP66} are found.
Just to give some examples, the $d$ ($d$ inside the atomic sphere) and $sp$
($sp$ inside the atomic sphere and total from interstitial) contributions to the
spin magnetic moment of Fe are 2.21 and 0.00~$\mu_{\text{B}}$ with LDA,
2.76 and -0.25~$\mu_{\text{B}}$ with mBJLDA, and 3.41 and -0.34~$\mu_{\text{B}}$
with GLLB-SC. Figure~\ref{fig_vxcup_vxcdn_Fe}(b) also shows
$v_{\text{xc},\uparrow}-v_{\text{xc},\downarrow}$, but this time
evaluated non-self-consistently with the LDA density, orbitals, etc.,
where we can see that the magnitude is much smaller, indicating that
the exchange splitting is strongly enhanced by the self-consistent field
procedure.

\subsubsection{\label{variants}Variants of GLLB-SC: Attempts of improvement}

The results of the present and previous works
\cite{KuismaPRB10,CastelliEES12,YanPRB12,HuserPRB13,MiroCSR14,PilaniaSR16,KimJPCC16,PandeyJPCC17}
have shown that the GLLB-SC potential is much more reliable for band gap
calculation than all LDA and GGA methods that have been considered so far for
comparison, and of quite similar accuracy as mBJLDA, the hybrids, and $GW$.
Nevertheless, among the few problems of GLLB-SC that were pointed out, the most
important are (1) a clear underestimation of most
band gaps smaller than $\sim1$~eV, (2) some unpredictable behavior for strongly
correlated systems (for which mBJLDA is much more reliable), and (3)
a very large overestimation of the magnetic moment of metals.

\begin{figure}
\includegraphics[scale=0.6]{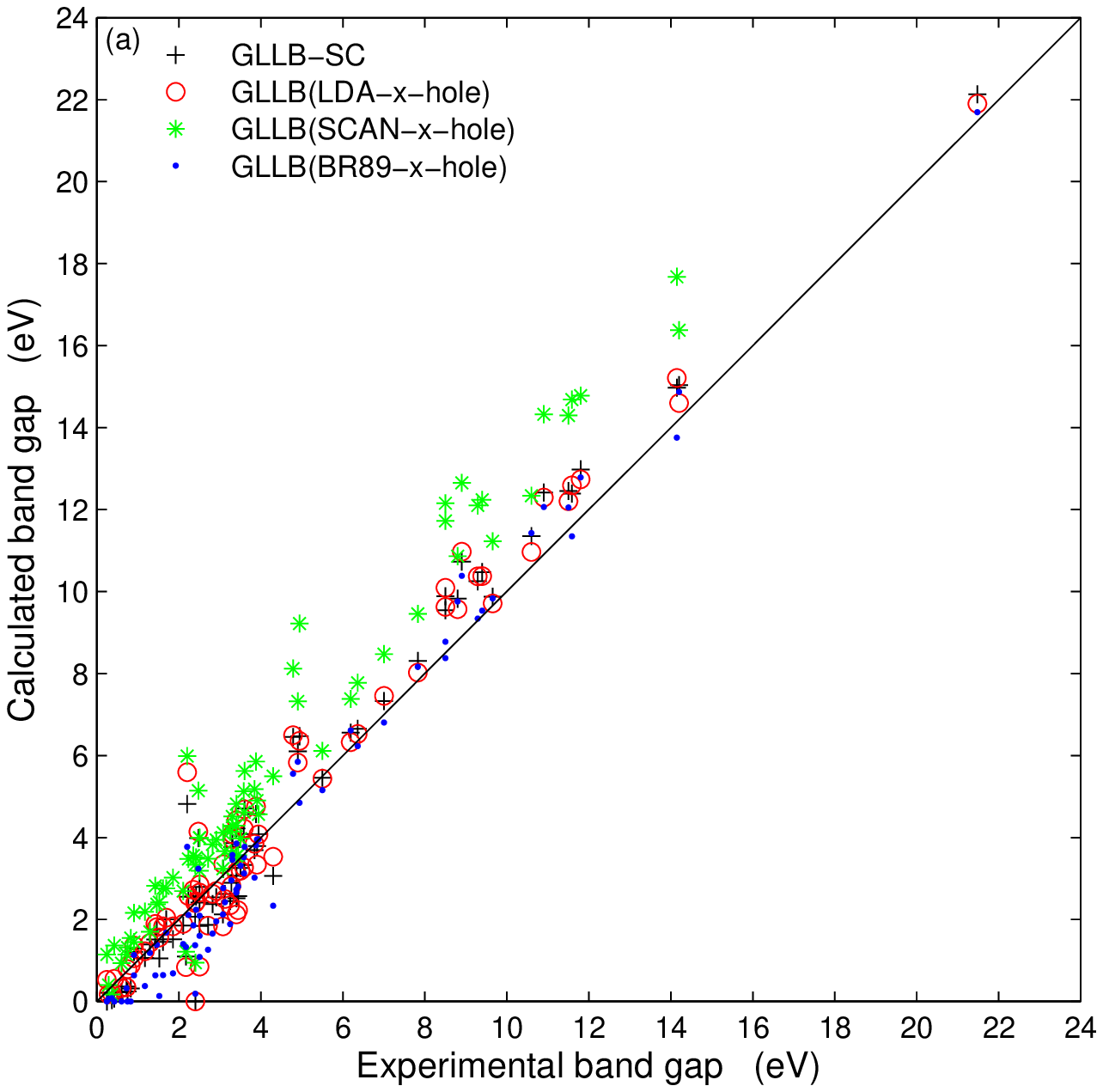}
\includegraphics[scale=0.6]{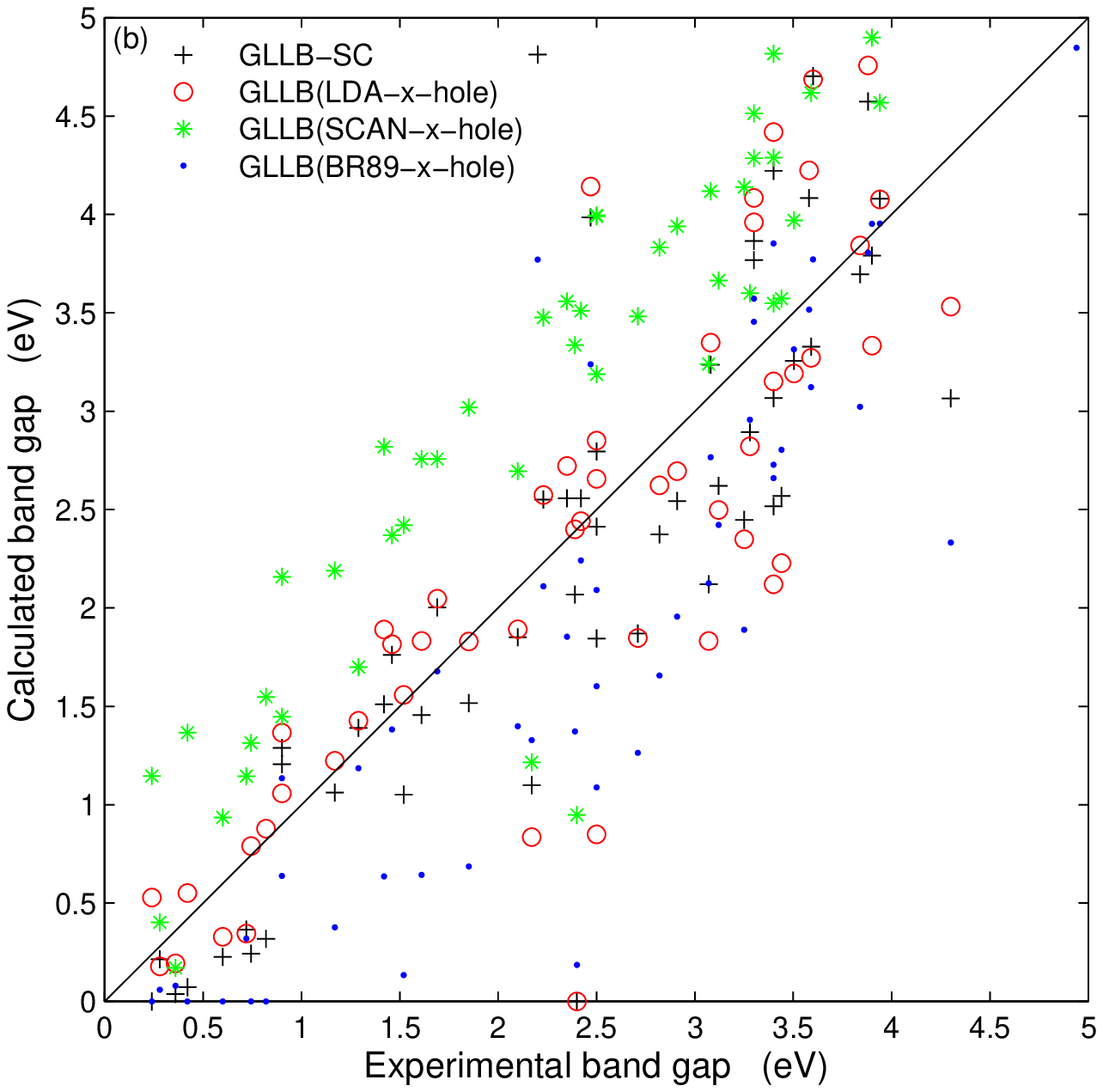}
\caption{\label{fig_band_gap2}Calculated versus experimental fundamental band
gaps for the set of 76 solids. The calculated values were obtained with GLLB-SC
and various variants which differ in the exchange hole term. The lower panel is
a zoom of the upper panel focusing on band gaps smaller than 5~eV.}
\end{figure}

\begin{figure}
\includegraphics[scale=0.6]{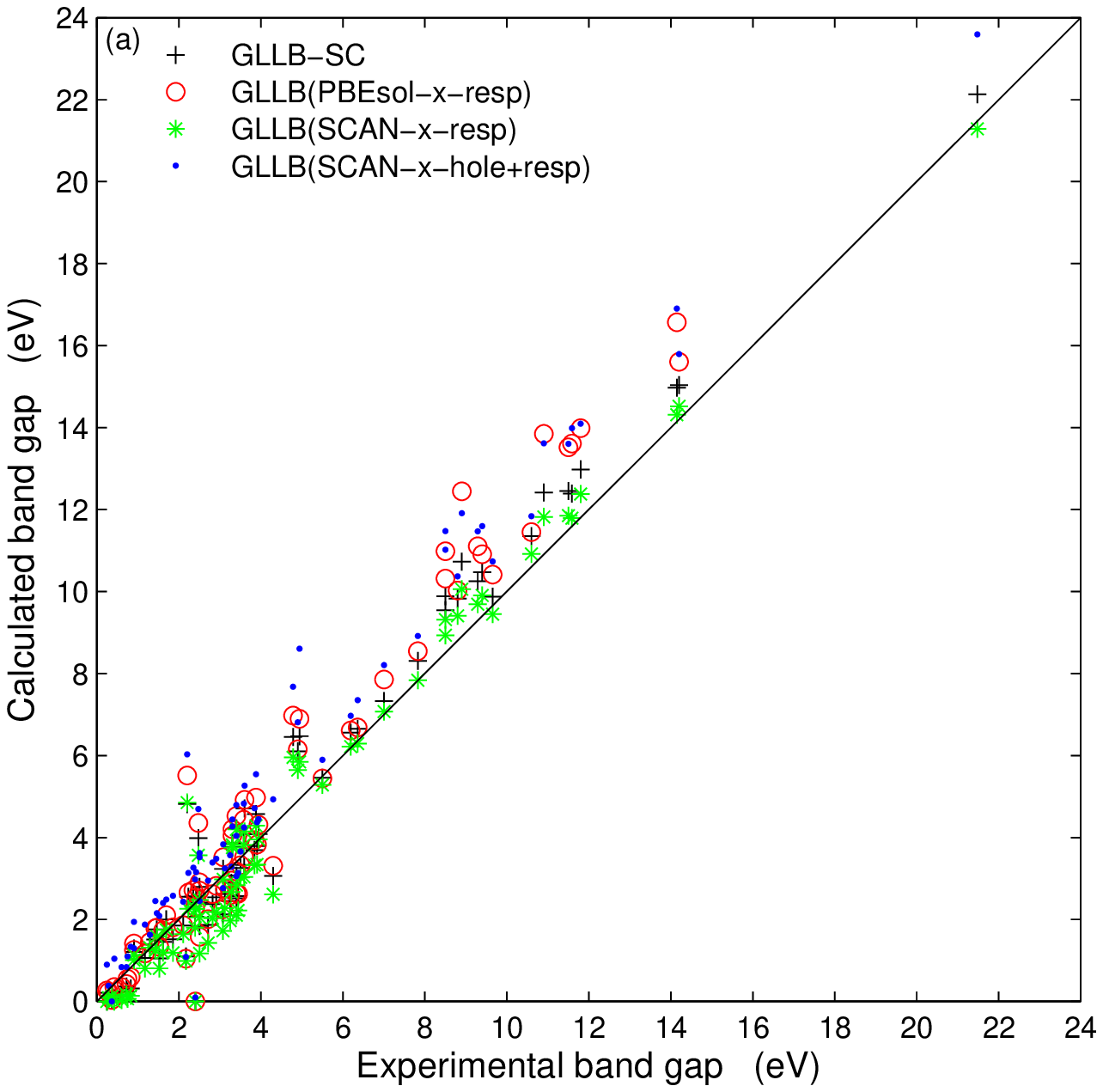}
\includegraphics[scale=0.6]{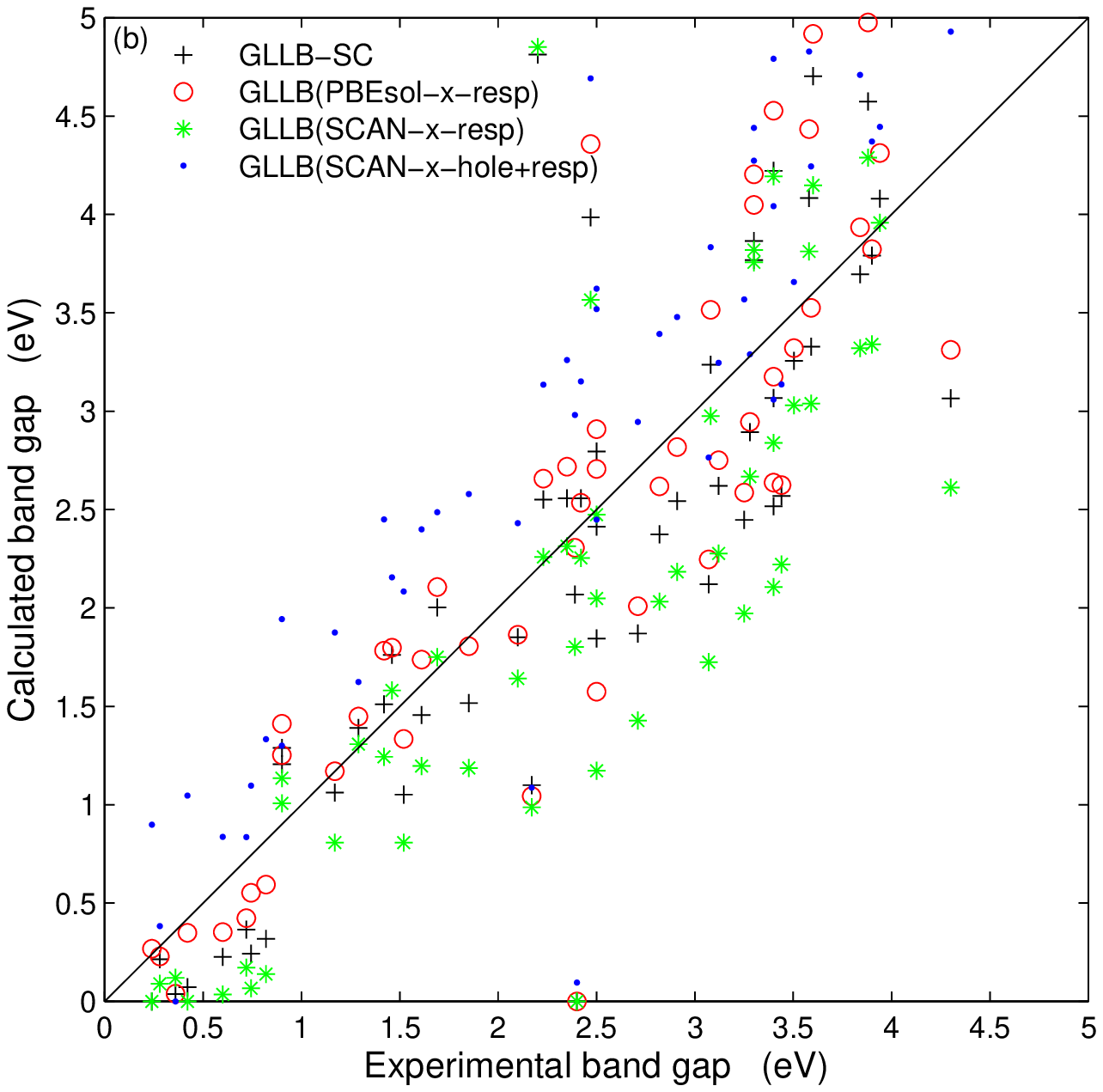}
\caption{\label{fig_band_gap3}Calculated versus experimental fundamental band
gaps for the set of 76 solids. The calculated values were obtained with GLLB-SC
and various variants which differ in the exchange response term or in both
the exchange hole and response terms. The lower panel
is a zoom of the upper panel focusing on band gaps smaller than 5~eV.}
\end{figure}
\begin{table*} 
\caption{\label{table_band_gap2}Summary statistics for the error in the
calculated band gaps in Table~S4 of the Supplemental Material\cite{SM_GLLB}
for the set of 76 solids obtained with GLLB-SC and various
variants which differ either in the exchange hole term, in the exchange
response term, or both. The calculations were done at the experimental
geometry specified in Table~S1 of the Supplemental Material.\cite{SM_GLLB}}
\begin{ruledtabular}
\begin{tabular}{lccccccccccc} 
\multicolumn{1}{l}{} &
\multicolumn{7}{c}{GLLB} \\
\cline{2-8}
\multicolumn{1}{l}{} &
\multicolumn{1}{c}{SC} &
\multicolumn{1}{c}{LDA-x-hole} &
\multicolumn{1}{c}{SCAN-x-hole} &
\multicolumn{1}{c}{BR89-x-hole} &
\multicolumn{1}{c}{PBEsol-x-resp} &
\multicolumn{1}{c}{SCAN-x-resp} &
\multicolumn{1}{c}{SCAN-x-hole+resp} \\
\hline
ME (eV)    & 0.20 & 0.23 &  1.41 & -0.24 & 0.55 & -0.13 & 1.01 \\ 
MAE (eV)   & 0.64 & 0.65 &  1.48 &  0.60 & 0.89 &  0.58 & 1.14 \\ 
STDE (eV)  & 0.81 & 0.88 &  1.13 &  0.73 & 1.15 &  0.76 & 1.05 \\ 
MRE (\%)   &   -4 &    4 &    46 &   -21 & 6    &  -15  &  31  \\ 
MARE (\%)  &   24 &   22 &    50 &    28 & 24   &  26   &  38  \\ 
STDRE (\%) &   34 &   35 &    56 &    36 & 33   &  36   &  47  \\ 
\end{tabular}
\end{ruledtabular}
\end{table*}

As said in the introduction, the idea behind the construction of the GLLB(-SC)
potential is very interesting, but also very promising since it allows for a proper
calculation of the band gap at a computational cost that is similar to semilocal methods.
Therefore, it is certainly worth to explore further the GLLB idea, and
the important question is to which extent is it possible to improve upon
GLLB-SC without deteriorating the results which are already good.
To try to answer this question we have considered several variants of
Eq.~(\ref{vxcGLLBSC}), and one of the most obvious modifications
consists of choosing an alternative to $2\varepsilon_{\text{x},\sigma}^{\text{PBEsol}}$ for
the exchange hole term $v_{\text{x,hole},\sigma}$, while keeping the second and third
terms the same as in GLLB-SC. Among the numerous choices that we have tried for
$v_{\text{x,hole},\sigma}$, three of them will be discussed, namely,
$2\varepsilon_{\text{x},\sigma}^{\text{LDA}}$, $2\varepsilon_{\text{x},\sigma}^{\text{SCAN}}$,
and $v_{\text{x},\sigma}^{\text{BR89}}$. The latter two are MGGA since
$\varepsilon_{\text{x},\sigma}^{\text{SCAN}}$ is $t_{\sigma}$-dependent,\cite{SunPRL15}
while $v_{\text{x},\sigma}^{\text{BR89}}$ is $t_{\sigma}$- and
$\nabla^{2}\rho_{\sigma}$-dependent.\cite{BeckePRA89}
BR89 was constructed to be similar to the Slater
potential\cite{BeckePRA89,TranJCTC15} and is the hole term in the BJ
potential.\cite{BeckeJCP06} The results for the band gap
can be found in Table~S4 of the Supplemental
Material\cite{SM_GLLB} and are shown in Fig.~\ref{fig_band_gap2}, while
the average errors are in Table~\ref{table_band_gap2}. Compared to the
original version of the potential, the results obtained
with GLLB(LDA-x-hole) and GLLB(BR89-x-hole) are rather similar in terms of
MA(R)E and STD(R)E. However, GLLB(BR89-x-hole) leads to negative
M(R)E, which is due to a clear underestimation of band gaps smaller than 4~eV
[see Fig.~\ref{fig_band_gap2}(b)], and actually the band gap is zero
for five solids (Ge, GaSb, InAs, InSb, and VO$_{2}$), while this was the case for only two
solids with GLLB-SC. The opposite trend is observed with GLLB(LDA-x-hole), which is
more accurate than GLLB-SC for the band gaps smaller than $\sim1$~eV, such that
only one system (FeO) is described as metallic. On the other hand, the
band gaps in the range 2-5~eV are more underestimated with GLLB(LDA-x-hole)
than with GLLB-SC. The band gaps obtained with GLLB(SCAN-x-hole) are too
large with respect to experiment and overall the results are very inaccurate
since the MAE and MARE are 1.48~eV and 50\%, respectively.

Thus, replacing PBEsol by something else for the exchange hole term does not
really help in improving the results overall, and the same conclusion can be
drawn with the other choices for $v_{\text{x,hole},\sigma}$ that we have tested
(results not shown), namely, $2\varepsilon_{\text{x},\sigma}$ of the exchange functionals
EV93,\cite{EngelPRB93} revTPSS,\cite{PerdewPRL09} MVS,\cite{SunPNAS15} and
TM,\cite{TaoPRL16} with the latter three being MGGA. The general observation is
that a clear improvement for a group of band gaps that are, e.g., underestimated with
GLLB-SC is necessarily accompanied by a clear deterioration for another group.
Also, the case of the iron oxides FeO and Fe$_{2}$O$_{3}$ is particularly problematic.
While GLLB-SC leads to no band gap in FeO (experiment is 2.4~eV)
and strongly overestimates the value for Fe$_{2}$O$_{3}$ (4.81 eV versus 2.2~eV for
experiment), GLLB-SC(SCAN-x-hole) improves the result for FeO (0.95~eV),
but overestimates even more than GLLB-SC for Fe$_{2}$O$_{3}$ (5.99~eV).

The other type of variants of Eq.~(\ref{vxcGLLBSC}) that we have considered
consists of an exchange response term that is multiplied by a function $F_{\sigma}$:
\begin{equation}
F_{\sigma}(\bm{r})K_{\text{x}}^{\text{LDA}}
\sum_{i=1}^{N_{\sigma}}\sqrt{\epsilon_{\text{H}}-\epsilon_{i\sigma}}
\frac{\left\vert\psi_{i\sigma}(\bm{r})\right\vert^{2}}{\rho_{\sigma}(\bm{r})}
\label{2ndterm}
\end{equation}
and similarly in Eq.~(\ref{deltax}) for the associated derivative discontinuity.
In order to be reasonable from the formal point of view, $F_{\sigma}$ should satisfy two
constraints. The first one is $F_{\sigma}=1$ for a constant $\rho_{\sigma}$ such that the
potential recovers LDA as GLLB-SC does. The second constraint requires $F_{\sigma}$ to be
constructed such that the scaling property of the exchange
potential\cite{Ou-YangPRL90}
[$v_{\text{x}}([\rho_{\lambda}];\bm{r})=\lambda v_{\text{x}}([\rho];\lambda\bm{r})$,
where $\rho_{\lambda}(\bm{r})=\lambda^{3}\rho(\lambda\bm{r})$] is satisfied,
which is the case if $F_{\sigma}$ depends only on the reduced density gradient
$s_{\sigma}=\left\vert\nabla\rho_{\sigma}\right\vert/
\left(2\left(6\pi^{2}\right)^{1/3}\rho_{\sigma}^{4/3}\right)$
for a GGA-type $F_{\sigma}$, and on quantities like $t_{\sigma}^{\text{W}}/t_{\sigma}$ or
$t_{\sigma}/t_{\sigma}^{\text{TF}}$ for a $t_{\sigma}$-dependent MGGA-type $F_{\sigma}$
[$t_{\sigma}^{\text{W}}=\left\vert\nabla\rho_{\sigma}\right\vert^{2}/\left(8\rho_{\sigma}\right)$ and
$t_{\sigma}^{\text{TF}}=\left(3/10\right)\left(6\pi^{2}\right)^{2/3}\rho_{\sigma}^{5/3}$ are
the von Weizs\"{a}cker\cite{vonWeizsackerZP35} and
Thomas-Fermi kinetic-energy density\cite{ThomasPCPS27,FermiRANL27}].
Thus, a possible choice for $F_{\sigma}$ in Eq.~(\ref{2ndterm}) would be to use (blindly) any
of the GGA or MGGA exchange enhancement factors
$F_{\text{x},\sigma}^{\text{(M)GGA}}=
\varepsilon_{\text{x},\sigma}^{\text{(M)GGA}}/\varepsilon_{\text{x},\sigma}^{\text{LDA}}$
that are available in the literature. Such a choice could seem quite empirical and not
justified, but it should just be considered as the first attempt to
improve the results by making Eq.~(\ref{2ndterm}) $\nabla\rho_{\sigma}$- or $t_{\sigma}$-dependent. 

The results for the band gap obtained with two different exchange enhancement factors
for $F_{\sigma}$ in Eq.~(\ref{2ndterm}), namely $F_{\text{x},\sigma}^{\text{PBEsol}}$ and
$F_{\text{x},\sigma}^{\text{SCAN}}$ (as in GLLB-SC, we keep PBEsol for
$v_{\text{x,hole},\sigma}$ and $v_{\text{c},\sigma}$),
are shown in Fig.~\ref{fig_band_gap3} and Tables~S4 and \ref{table_band_gap2}. 
The results are rather disappointing and the trends are similar to those
discussed above when different exchange hole terms were considered.
Compared to GLLB-SC, GLLB(PBEsol-x-resp) and GLLB(SCAN-x-resp) lead to band
gaps which are larger and smaller, respectively. Since the direction of change
in the band gap is the same in basically all cases, an improvement for the
underestimated band gaps (e.g., the small ones) is associated with an overestimation
for most other solids, or vice versa.

Figure~\ref{fig_band_gap3} and Tables~S4 and \ref{table_band_gap2} also show the results
obtained with GLLB(SCAN-x-hole+resp) which consists of using SCAN exchange for
the hole and response terms simultaneously.
The accuracy of GLLB(SCAN-x-hole+resp) is in
between those of GLLB(SCAN-x-hole) and GLLB(SCAN-x-resp), and
overall rather low since the MAE and MARE are quite large (1.14~eV and 38\%).
Not shown, the results obtained with the MGGA exchange MVS\cite{SunPNAS15} or
TM\cite{TaoPRL16} instead of SCAN for the hole and response terms indicate
that TM leads to reduced errors (but still larger than GLLB-SC),
while with MVS the errors are similar to SCAN. 

The other possibilities for $F_{\sigma}$ in Eq.~(\ref{2ndterm}) that we have tried are
simple MGGA functions $F_{\sigma}=\left(t_{\sigma}/t_{\sigma}^{\text{TF}}\right)^{p}$ or
$F_{\sigma}=\left(\left(t_{\sigma}-t_{\sigma}^{\text{W}}\right)/t_{\sigma}^{\text{TF}}\right)^{p}$,
which are somehow similar to the response terms
$v_{\text{x,resp},\sigma}^{\text{BJ}}\propto\sqrt{t_{\sigma}/\rho_{\sigma}}$ and
$v_{\text{x,resp},\sigma}^{\text{BJUC}}\propto
\sqrt{\left(t_{\sigma}-t_{\sigma}^{\text{W}}\right)/\rho_{\sigma}}$
of the BJ potential\cite{BeckeJCP06} and its gauge-invariant version (universal
correction),\cite{RasanenJCP10} respectively. Using such functions should be
promising since this is a way to apply the GLLB idea to a potential whose
response term is somehow similar to the BJ potential. However, the results that
we have obtained so far with such functions $F$ are not very encouraging and
are not worth to be discussed in detail. Nevertheless, we believe that the
search for such simple functions depending on the kinetic-energy density
that could lead to interesting results is worth to be pursued.

The last variant of GLLB-type potential that we have considered is given by
\begin{eqnarray}
v_{\text{xc},\sigma}^{\text{GLLB2-SC}}(\bm{r}) & = &
2\varepsilon_{\text{xc},\sigma}^{\text{PBEsol}}(\bm{r}) +
F_{\text{xc},\sigma}^{\text{PBEsol}}(\bm{r}) \nonumber \\
& & \times
K_{\text{x}}^{\text{LDA}}\sum_{i=1}^{N_{\sigma}}\sqrt{\epsilon_{\text{H}}-\epsilon_{i\sigma}}
\frac{\left\vert\psi_{i\sigma}(\bm{r})\right\vert^{2}}{\rho_{\sigma}(\bm{r})},
\label{vxcGLLB2}
\end{eqnarray}
where $F_{\text{xc},\sigma}^{\text{PBEsol}}=
\varepsilon_{\text{xc},\sigma}^{\text{PBEsol}}/\varepsilon_{\text{x},\sigma}^{\text{LDA}}$
is the total PBEsol enhancement factor.\cite{PerdewPRL08} Compared to GLLB-SC
[Eq.~(\ref{vxcGLLBSC})], correlation is now treated the same way as exchange
and, therefore, also contributes to the derivative discontinuity.
Equation~(\ref{vxcGLLB2}) should be considered as a simple, but still
rather meaningful way to extend GLLB-SC to correlation.
As discussed in Sec.~\ref{electronicstructure}, GLLB2-SC leads to a
xc discontinuity that agrees very well with the RPA-OEP value for simple solids.
However, the calculated band gaps with GLLB2-SC are quite inaccurate such that
the ME and MAE are 1.35 and 1.47~eV, respectively. A large overestimation is
obtained for most solids, except for those with a band gap smaller than 1~eV
and a few others like FeO for which the band gap is still zero.
Thus, for band gap calculation GLLB2-SC is of no interest and does not solve the
problems found with GLLB-SC. Replacing PBEsol by SCAN in Eq.~(\ref{vxcGLLB2})
also does not lead to any interesting improvement in the band gaps.

In this section, numerous attempts to improve upon the original GLLB-SC method
have been presented. However, the results are rather disappointing since none
of the variants of GLLB-SC that we have tested could really solve the problems
of GLLB-SC for the band gap. We also mention that it has
not been possible to really reduce the large overestimation of the magnetic
moment in ferromagnetic metals compared to GLLB-SC. The fact that absolutely
all variants of the GLLB-SC potential mentioned above lead to such large
overestimation of $\mu_{S}$ strongly suggests that this problem is due to the
dependency on the orbital energies $\epsilon_{i}$ of the second term of
Eq.~(\ref{vxcGLLBSC}) which makes $v_{\text{xc},\uparrow}^{\text{GLLB-SC}}-
v_{\text{xc},\downarrow}^{\text{GLLB-SC}}$ too large when 
$\epsilon_{i\uparrow}\neq\epsilon_{i\downarrow}$.
Without entering into details, reducing the magnitude of the
second term in Eq.~(\ref{vxcGLLBSC}) (instead of increasing it as done above)
with various schemes did not lead to satisfying results. Work is under way
in order to find such a scheme that reduces the exchange splitting of metals
without deteriorating too much the results for other systems.

As shown above, the mBJLDA potential leads to much more balanced results for
the magnetic moment (very accurate for strongly correlated solids and
moderately overestimated for metals). This is mainly due to the use of the
average of $\nabla\rho/\rho$ in the unit cell,
which is larger in the transition-metal oxides (1.9-1.95 bohr$^{-1/2}$) than
in the ferromagnetic metals (1.4-1.5 bohr$^{-1/2}$) and therefore provides
a way to make the difference between the two classes of solids
(see Ref.~\onlinecite{KollerPRB11} for related discussions).
Thus, the use of the average of $\nabla\rho/\rho$ or another similar quantity
in a GLLB-type potential should also be considered in future works.

\section{\label{summary}Summary and conclusion}

In this work, the GLLB-SC potential has been tested and compared to other
methods for the description of the electronic and magnetic properties of
solids, as well as properties directly related to the electron density.
Concerning the band gap, GLLB-SC is, as expected, much more accurate than
the LDA and GGA methods and of similar accuracy as hybrid functionals.
However, GLLB-SC is on average not as accurate as mBJLDA, and the two main
problems are (1) a clear underestimation of band gaps smaller than 1~eV and
(2) very large variations in the error for strongly correlated solids.
mBJLDA is overall less prone to large errors than GLLB-SC, in
particular for the strongly correlated solids. However, mBJLDA clearly
underestimates the band gap of Cu$^{1+}$ compounds like Cu$_{2}$O.

The magnetic moment in antiferromagnetic insulators is accurately described
by GLLB-SC, mBJLDA, and the hybrid functional HSE06, while for the
ferromagnetic metals GLLB-SC and HSE06 lead to very large overestimations
of the magnetic moment. The mBJLDA potential also overestimates the magnetic
moment in the metals, but to a much lesser extent.
Concerning the EFG, it has been shown that GLLB-SC is the method
leading to the best agreement with experiment, meaning that
the valence electron density should be described accurately by GLLB-SC.
This is, however, not the case with mBJLDA which is not recommended for EFG
calculations.

Focusing on the band gap, the goal was then to improve the results with
respect to GLLB-SC by modifying either the hole term or the response term
(or both) in the potential. However, our numerous attempts have
remained fruitless, and actually it was not possible to improve significantly the
results for a group of solids (e.g., those with a small band gap) without
significantly worsening the results for other compounds. This is rather disappointing,
in particular since more was expected by bringing in a dependency on the kinetic-energy
density into a GLLB-type potential.

Nevertheless, a multiplicative xc potential that has the same features as GLLB-SC,
namely, to be computationally fast and leads to a nonzero derivative discontinuity,
is ideal from the formal and practical points of view, similarly as the
nonmultiplicative potentials that are functional derivatives of MGGA functionals.
\cite{PeveratiJCP12,YangPRB16} Therefore, it is certainly worth to pursue the
development of such potentials, possibly by trying to incorporate more features
of other successful semilocal potentials like mBJLDA or AK13, or by learning more
from very accurate ab initio potentials.
\cite{KlimesJCP14,GrabowskiJCP14,OspadovJCP17}
In this respect, we
should remind that the mBJLDA potential uses an ingredient, the average of
$\nabla\rho/\rho$ in the unit cell that has not been used in
other potentials except hybrid functionals.
\cite{BylanderPRB90,SeidlPRB96,MarquesPRB11}
Alternatively, the dielectric function could be used,
\cite{MarquesPRB11,KollerJPCM13,SkonePRB14,ShimazakiJCP15} however
this requires the use of unoccupied orbitals.
Also not yet explored, is the
possibility to use the step structure of the BJ potential,\cite{ArmientoPRB08}
which in principle should lead to a derivative discontinuity (which is not
the case with the AK13 potential as shown in Ref.~\onlinecite{AschebrockPRB17b}).

\begin{acknowledgments}

This work was supported by the project F41 (SFB ViCoM) of the Austrian Science
Fund (FWF). S.E. and P.B. acknowledge financial support from Higher Education
Commission (HEC), Pakistan.

\end{acknowledgments}

\bibliography{references}

\begin{thebibliography}{182}%
\makeatletter
\providecommand \@ifxundefined [1]{%
 \@ifx{#1\undefined}
}%
\providecommand \@ifnum [1]{%
 \ifnum #1\expandafter \@firstoftwo
 \else \expandafter \@secondoftwo
 \fi
}%
\providecommand \@ifx [1]{%
 \ifx #1\expandafter \@firstoftwo
 \else \expandafter \@secondoftwo
 \fi
}%
\providecommand \natexlab [1]{#1}%
\providecommand \enquote  [1]{``#1''}%
\providecommand \bibnamefont  [1]{#1}%
\providecommand \bibfnamefont [1]{#1}%
\providecommand \citenamefont [1]{#1}%
\providecommand \href@noop [0]{\@secondoftwo}%
\providecommand \href [0]{\begingroup \@sanitize@url \@href}%
\providecommand \@href[1]{\@@startlink{#1}\@@href}%
\providecommand \@@href[1]{\endgroup#1\@@endlink}%
\providecommand \@sanitize@url [0]{\catcode `\\12\catcode `\$12\catcode
  `\&12\catcode `\#12\catcode `\^12\catcode `\_12\catcode `\%12\relax}%
\providecommand \@@startlink[1]{}%
\providecommand \@@endlink[0]{}%
\providecommand \url  [0]{\begingroup\@sanitize@url \@url }%
\providecommand \@url [1]{\endgroup\@href {#1}{\urlprefix }}%
\providecommand \urlprefix  [0]{URL }%
\providecommand \Eprint [0]{\href }%
\providecommand \doibase [0]{http://dx.doi.org/}%
\providecommand \selectlanguage [0]{\@gobble}%
\providecommand \bibinfo  [0]{\@secondoftwo}%
\providecommand \bibfield  [0]{\@secondoftwo}%
\providecommand \translation [1]{[#1]}%
\providecommand \BibitemOpen [0]{}%
\providecommand \bibitemStop [0]{}%
\providecommand \bibitemNoStop [0]{.\EOS\space}%
\providecommand \EOS [0]{\spacefactor3000\relax}%
\providecommand \BibitemShut  [1]{\csname bibitem#1\endcsname}%
\let\auto@bib@innerbib\@empty
\bibitem [{\citenamefont {Hohenberg}\ and\ \citenamefont
  {Kohn}(1964)}]{HohenbergPR64}%
  \BibitemOpen
  \bibfield  {author} {\bibinfo {author} {\bibfnamefont {P.}~\bibnamefont
  {Hohenberg}}\ and\ \bibinfo {author} {\bibfnamefont {W.}~\bibnamefont
  {Kohn}},\ }\href@noop {} {\bibfield  {journal} {\bibinfo  {journal} {Phys.
  Rev.}\ }\textbf {\bibinfo {volume} {136}},\ \bibinfo {pages} {B864} (\bibinfo
  {year} {1964})}\BibitemShut {NoStop}%
\bibitem [{\citenamefont {Kohn}\ and\ \citenamefont {Sham}(1965)}]{KohnPR65}%
  \BibitemOpen
  \bibfield  {author} {\bibinfo {author} {\bibfnamefont {W.}~\bibnamefont
  {Kohn}}\ and\ \bibinfo {author} {\bibfnamefont {L.~J.}\ \bibnamefont
  {Sham}},\ }\href@noop {} {\bibfield  {journal} {\bibinfo  {journal} {Phys.
  Rev.}\ }\textbf {\bibinfo {volume} {140}},\ \bibinfo {pages} {A1133}
  (\bibinfo {year} {1965})}\BibitemShut {NoStop}%
\bibitem [{\citenamefont {Jensen}(2007)}]{Jensen}%
  \BibitemOpen
  \bibfield  {author} {\bibinfo {author} {\bibfnamefont {F.}~\bibnamefont
  {Jensen}},\ }\href@noop {} {\emph {\bibinfo {title} {Introduction to
  Computational Chemistry, 2nd ed.}}}\ (\bibinfo  {publisher} {John Wiley \&
  Sons Ltd},\ \bibinfo {address} {Chichester},\ \bibinfo {year}
  {2007})\BibitemShut {NoStop}%
\bibitem [{\citenamefont {Bechstedt}(2015)}]{Bechstedt}%
  \BibitemOpen
  \bibfield  {author} {\bibinfo {author} {\bibfnamefont {F.}~\bibnamefont
  {Bechstedt}},\ }\href@noop {} {\emph {\bibinfo {title} {Many-Body Approach to
  Electronic Excitations: Concepts and Applications}}}\ (\bibinfo  {publisher}
  {Springer},\ \bibinfo {address} {Berlin Heidelberg},\ \bibinfo {year}
  {2015})\BibitemShut {NoStop}%
\bibitem [{\citenamefont {Marques}\ \emph {et~al.}(2012)\citenamefont
  {Marques}, \citenamefont {Oliveira},\ and\ \citenamefont
  {Burnus}}]{MarquesCPC12}%
  \BibitemOpen
  \bibfield  {author} {\bibinfo {author} {\bibfnamefont {M.~A.~L.}\
  \bibnamefont {Marques}}, \bibinfo {author} {\bibfnamefont {M.~J.~T.}\
  \bibnamefont {Oliveira}}, \ and\ \bibinfo {author} {\bibfnamefont
  {T.}~\bibnamefont {Burnus}},\ }\href@noop {} {\bibfield  {journal} {\bibinfo
  {journal} {Comput. Phys. Commun.}\ }\textbf {\bibinfo {volume} {183}},\
  \bibinfo {pages} {2272} (\bibinfo {year} {2012})}\BibitemShut {NoStop}%
\bibitem [{\citenamefont {K{\"u}mmel}\ and\ \citenamefont
  {Kronik}(2008)}]{KuemmelRMP08}%
  \BibitemOpen
  \bibfield  {author} {\bibinfo {author} {\bibfnamefont {S.}~\bibnamefont
  {K{\"u}mmel}}\ and\ \bibinfo {author} {\bibfnamefont {L.}~\bibnamefont
  {Kronik}},\ }\href@noop {} {\bibfield  {journal} {\bibinfo  {journal} {Rev.
  Mod. Phys.}\ }\textbf {\bibinfo {volume} {80}},\ \bibinfo {pages} {3}
  (\bibinfo {year} {2008})}\BibitemShut {NoStop}%
\bibitem [{\citenamefont {Cohen}\ \emph {et~al.}(2012)\citenamefont {Cohen},
  \citenamefont {Mori-S{\'{a}}nchez},\ and\ \citenamefont {Yang}}]{CohenCR12}%
  \BibitemOpen
  \bibfield  {author} {\bibinfo {author} {\bibfnamefont {A.~J.}\ \bibnamefont
  {Cohen}}, \bibinfo {author} {\bibfnamefont {P.}~\bibnamefont
  {Mori-S{\'{a}}nchez}}, \ and\ \bibinfo {author} {\bibfnamefont
  {W.}~\bibnamefont {Yang}},\ }\href@noop {} {\bibfield  {journal} {\bibinfo
  {journal} {Chem. Rev.}\ }\textbf {\bibinfo {volume} {112}},\ \bibinfo {pages}
  {289} (\bibinfo {year} {2012})}\BibitemShut {NoStop}%
\bibitem [{\citenamefont {Burke}(2012)}]{BurkeJCP12}%
  \BibitemOpen
  \bibfield  {author} {\bibinfo {author} {\bibfnamefont {K.}~\bibnamefont
  {Burke}},\ }\href@noop {} {\bibfield  {journal} {\bibinfo  {journal} {J.
  Chem. Phys.}\ }\textbf {\bibinfo {volume} {136}},\ \bibinfo {pages} {150901}
  (\bibinfo {year} {2012})}\BibitemShut {NoStop}%
\bibitem [{\citenamefont {Becke}(2014)}]{BeckeJCP14}%
  \BibitemOpen
  \bibfield  {author} {\bibinfo {author} {\bibfnamefont {A.~D.}\ \bibnamefont
  {Becke}},\ }\href@noop {} {\bibfield  {journal} {\bibinfo  {journal} {J.
  Chem. Phys.}\ }\textbf {\bibinfo {volume} {140}},\ \bibinfo {pages} {18A301}
  (\bibinfo {year} {2014})}\BibitemShut {NoStop}%
\bibitem [{\citenamefont {Yang}\ \emph {et~al.}(2012)\citenamefont {Yang},
  \citenamefont {Cohen},\ and\ \citenamefont {Mori-S\'{a}nchez}}]{YangJCP12}%
  \BibitemOpen
  \bibfield  {author} {\bibinfo {author} {\bibfnamefont {W.}~\bibnamefont
  {Yang}}, \bibinfo {author} {\bibfnamefont {A.~J.}\ \bibnamefont {Cohen}}, \
  and\ \bibinfo {author} {\bibfnamefont {P.}~\bibnamefont {Mori-S\'{a}nchez}},\
  }\href@noop {} {\bibfield  {journal} {\bibinfo  {journal} {J. Chem. Phys.}\
  }\textbf {\bibinfo {volume} {136}},\ \bibinfo {pages} {204111} (\bibinfo
  {year} {2012})}\BibitemShut {NoStop}%
\bibitem [{\citenamefont {Seidl}\ \emph {et~al.}(1996)\citenamefont {Seidl},
  \citenamefont {G\"orling}, \citenamefont {Vogl}, \citenamefont {Majewski},\
  and\ \citenamefont {Levy}}]{SeidlPRB96}%
  \BibitemOpen
  \bibfield  {author} {\bibinfo {author} {\bibfnamefont {A.}~\bibnamefont
  {Seidl}}, \bibinfo {author} {\bibfnamefont {A.}~\bibnamefont {G\"orling}},
  \bibinfo {author} {\bibfnamefont {P.}~\bibnamefont {Vogl}}, \bibinfo {author}
  {\bibfnamefont {J.~A.}\ \bibnamefont {Majewski}}, \ and\ \bibinfo {author}
  {\bibfnamefont {M.}~\bibnamefont {Levy}},\ }\href@noop {} {\bibfield
  {journal} {\bibinfo  {journal} {Phys. Rev. B}\ }\textbf {\bibinfo {volume}
  {53}},\ \bibinfo {pages} {3764} (\bibinfo {year} {1996})}\BibitemShut
  {NoStop}%
\bibitem [{\citenamefont {Perdew}\ \emph {et~al.}(1982)\citenamefont {Perdew},
  \citenamefont {Parr}, \citenamefont {Levy},\ and\ \citenamefont
  {Balduz}}]{PerdewPRL82}%
  \BibitemOpen
  \bibfield  {author} {\bibinfo {author} {\bibfnamefont {J.~P.}\ \bibnamefont
  {Perdew}}, \bibinfo {author} {\bibfnamefont {R.~G.}\ \bibnamefont {Parr}},
  \bibinfo {author} {\bibfnamefont {M.}~\bibnamefont {Levy}}, \ and\ \bibinfo
  {author} {\bibfnamefont {J.~L.}\ \bibnamefont {Balduz}, \bibfnamefont
  {Jr.}},\ }\href@noop {} {\bibfield  {journal} {\bibinfo  {journal} {Phys.
  Rev. Lett.}\ }\textbf {\bibinfo {volume} {49}},\ \bibinfo {pages} {1691}
  (\bibinfo {year} {1982})}\BibitemShut {NoStop}%
\bibitem [{\citenamefont {Sham}\ and\ \citenamefont
  {Schl\"uter}(1983)}]{ShamPRL83}%
  \BibitemOpen
  \bibfield  {author} {\bibinfo {author} {\bibfnamefont {L.~J.}\ \bibnamefont
  {Sham}}\ and\ \bibinfo {author} {\bibfnamefont {M.}~\bibnamefont
  {Schl\"uter}},\ }\href@noop {} {\bibfield  {journal} {\bibinfo  {journal}
  {Phys. Rev. Lett.}\ }\textbf {\bibinfo {volume} {51}},\ \bibinfo {pages}
  {1888} (\bibinfo {year} {1983})}\BibitemShut {NoStop}%
\bibitem [{\citenamefont {Godby}\ \emph {et~al.}(1986)\citenamefont {Godby},
  \citenamefont {Schl\"uter},\ and\ \citenamefont {Sham}}]{GodbyPRL86}%
  \BibitemOpen
  \bibfield  {author} {\bibinfo {author} {\bibfnamefont {R.~W.}\ \bibnamefont
  {Godby}}, \bibinfo {author} {\bibfnamefont {M.}~\bibnamefont {Schl\"uter}}, \
  and\ \bibinfo {author} {\bibfnamefont {L.~J.}\ \bibnamefont {Sham}},\
  }\href@noop {} {\bibfield  {journal} {\bibinfo  {journal} {Phys. Rev. Lett.}\
  }\textbf {\bibinfo {volume} {56}},\ \bibinfo {pages} {2415} (\bibinfo {year}
  {1986})}\BibitemShut {NoStop}%
\bibitem [{\citenamefont {Gr\"{u}ning}\ \emph {et~al.}(2006)\citenamefont
  {Gr\"{u}ning}, \citenamefont {Marini},\ and\ \citenamefont
  {Rubio}}]{GruningJCP06}%
  \BibitemOpen
  \bibfield  {author} {\bibinfo {author} {\bibfnamefont {M.}~\bibnamefont
  {Gr\"{u}ning}}, \bibinfo {author} {\bibfnamefont {A.}~\bibnamefont {Marini}},
  \ and\ \bibinfo {author} {\bibfnamefont {A.}~\bibnamefont {Rubio}},\
  }\href@noop {} {\bibfield  {journal} {\bibinfo  {journal} {J. Chem. Phys.}\
  }\textbf {\bibinfo {volume} {124}},\ \bibinfo {pages} {154108} (\bibinfo
  {year} {2006})}\BibitemShut {NoStop}%
\bibitem [{\citenamefont {Gr\"uning}\ \emph {et~al.}(2006)\citenamefont
  {Gr\"uning}, \citenamefont {Marini},\ and\ \citenamefont
  {Rubio}}]{GruningPRB06}%
  \BibitemOpen
  \bibfield  {author} {\bibinfo {author} {\bibfnamefont {M.}~\bibnamefont
  {Gr\"uning}}, \bibinfo {author} {\bibfnamefont {A.}~\bibnamefont {Marini}}, \
  and\ \bibinfo {author} {\bibfnamefont {A.}~\bibnamefont {Rubio}},\
  }\href@noop {} {\bibfield  {journal} {\bibinfo  {journal} {Phys. Rev. B}\
  }\textbf {\bibinfo {volume} {74}},\ \bibinfo {pages} {161103(R)} (\bibinfo
  {year} {2006})}\BibitemShut {NoStop}%
\bibitem [{\citenamefont {Klime\v{s}}\ and\ \citenamefont
  {Kresse}(2014)}]{KlimesJCP14}%
  \BibitemOpen
  \bibfield  {author} {\bibinfo {author} {\bibfnamefont {J.}~\bibnamefont
  {Klime\v{s}}}\ and\ \bibinfo {author} {\bibfnamefont {G.}~\bibnamefont
  {Kresse}},\ }\href@noop {} {\bibfield  {journal} {\bibinfo  {journal} {J.
  Chem. Phys.}\ }\textbf {\bibinfo {volume} {140}},\ \bibinfo {pages} {054516}
  (\bibinfo {year} {2014})}\BibitemShut {NoStop}%
\bibitem [{\citenamefont {Perdew}\ \emph {et~al.}(1996)\citenamefont {Perdew},
  \citenamefont {Burke},\ and\ \citenamefont {Ernzerhof}}]{PerdewPRL96}%
  \BibitemOpen
  \bibfield  {author} {\bibinfo {author} {\bibfnamefont {J.~P.}\ \bibnamefont
  {Perdew}}, \bibinfo {author} {\bibfnamefont {K.}~\bibnamefont {Burke}}, \
  and\ \bibinfo {author} {\bibfnamefont {M.}~\bibnamefont {Ernzerhof}},\
  }\href@noop {} {\bibfield  {journal} {\bibinfo  {journal} {Phys. Rev. Lett.}\
  }\textbf {\bibinfo {volume} {77}},\ \bibinfo {pages} {3865} (\bibinfo {year}
  {1996})},\ \bibinfo {note} {\textbf{78}, 1396(E) (1997)}\BibitemShut
  {NoStop}%
\bibitem [{\citenamefont {Heyd}\ \emph {et~al.}(2005)\citenamefont {Heyd},
  \citenamefont {Peralta}, \citenamefont {Scuseria},\ and\ \citenamefont
  {Martin}}]{HeydJCP05}%
  \BibitemOpen
  \bibfield  {author} {\bibinfo {author} {\bibfnamefont {J.}~\bibnamefont
  {Heyd}}, \bibinfo {author} {\bibfnamefont {J.~E.}\ \bibnamefont {Peralta}},
  \bibinfo {author} {\bibfnamefont {G.~E.}\ \bibnamefont {Scuseria}}, \ and\
  \bibinfo {author} {\bibfnamefont {R.~L.}\ \bibnamefont {Martin}},\
  }\href@noop {} {\bibfield  {journal} {\bibinfo  {journal} {J. Chem. Phys.}\
  }\textbf {\bibinfo {volume} {123}},\ \bibinfo {pages} {174101} (\bibinfo
  {year} {2005})}\BibitemShut {NoStop}%
\bibitem [{\citenamefont {Andrade}\ and\ \citenamefont
  {Aspuru-Guzik}(2011)}]{AndradePRL11}%
  \BibitemOpen
  \bibfield  {author} {\bibinfo {author} {\bibfnamefont {X.}~\bibnamefont
  {Andrade}}\ and\ \bibinfo {author} {\bibfnamefont {A.}~\bibnamefont
  {Aspuru-Guzik}},\ }\href@noop {} {\bibfield  {journal} {\bibinfo  {journal}
  {Phys. Rev. Lett.}\ }\textbf {\bibinfo {volume} {107}},\ \bibinfo {pages}
  {183002} (\bibinfo {year} {2011})}\BibitemShut {NoStop}%
\bibitem [{\citenamefont {Chai}\ and\ \citenamefont {Chen}(2013)}]{ChaiPRL13}%
  \BibitemOpen
  \bibfield  {author} {\bibinfo {author} {\bibfnamefont {J.-D.}\ \bibnamefont
  {Chai}}\ and\ \bibinfo {author} {\bibfnamefont {P.-T.}\ \bibnamefont
  {Chen}},\ }\href@noop {} {\bibfield  {journal} {\bibinfo  {journal} {Phys.
  Rev. Lett.}\ }\textbf {\bibinfo {volume} {110}},\ \bibinfo {pages} {033002}
  (\bibinfo {year} {2013})}\BibitemShut {NoStop}%
\bibitem [{\citenamefont {Kraisler}\ and\ \citenamefont
  {Kronik}(2014)}]{KraislerJCP14}%
  \BibitemOpen
  \bibfield  {author} {\bibinfo {author} {\bibfnamefont {E.}~\bibnamefont
  {Kraisler}}\ and\ \bibinfo {author} {\bibfnamefont {L.}~\bibnamefont
  {Kronik}},\ }\href@noop {} {\bibfield  {journal} {\bibinfo  {journal} {J.
  Chem. Phys.}\ }\textbf {\bibinfo {volume} {140}},\ \bibinfo {pages} {18A540}
  (\bibinfo {year} {2014})}\BibitemShut {NoStop}%
\bibitem [{\citenamefont {Engel}\ and\ \citenamefont
  {Vosko}(1993)}]{EngelPRB93}%
  \BibitemOpen
  \bibfield  {author} {\bibinfo {author} {\bibfnamefont {E.}~\bibnamefont
  {Engel}}\ and\ \bibinfo {author} {\bibfnamefont {S.~H.}\ \bibnamefont
  {Vosko}},\ }\href@noop {} {\bibfield  {journal} {\bibinfo  {journal} {Phys.
  Rev. B}\ }\textbf {\bibinfo {volume} {47}},\ \bibinfo {pages} {13164}
  (\bibinfo {year} {1993})}\BibitemShut {NoStop}%
\bibitem [{\citenamefont {Zhao}\ \emph {et~al.}(1999)\citenamefont {Zhao},
  \citenamefont {Bagayoko},\ and\ \citenamefont {Williams}}]{ZhaoPRB99}%
  \BibitemOpen
  \bibfield  {author} {\bibinfo {author} {\bibfnamefont {G.~L.}\ \bibnamefont
  {Zhao}}, \bibinfo {author} {\bibfnamefont {D.}~\bibnamefont {Bagayoko}}, \
  and\ \bibinfo {author} {\bibfnamefont {T.~D.}\ \bibnamefont {Williams}},\
  }\href@noop {} {\bibfield  {journal} {\bibinfo  {journal} {Phys. Rev. B}\
  }\textbf {\bibinfo {volume} {60}},\ \bibinfo {pages} {1563} (\bibinfo {year}
  {1999})}\BibitemShut {NoStop}%
\bibitem [{\citenamefont {Becke}\ and\ \citenamefont
  {Johnson}(2006)}]{BeckeJCP06}%
  \BibitemOpen
  \bibfield  {author} {\bibinfo {author} {\bibfnamefont {A.~D.}\ \bibnamefont
  {Becke}}\ and\ \bibinfo {author} {\bibfnamefont {E.~R.}\ \bibnamefont
  {Johnson}},\ }\href@noop {} {\bibfield  {journal} {\bibinfo  {journal} {J.
  Chem. Phys.}\ }\textbf {\bibinfo {volume} {124}},\ \bibinfo {pages} {221101}
  (\bibinfo {year} {2006})}\BibitemShut {NoStop}%
\bibitem [{\citenamefont {Tran}\ \emph {et~al.}(2007)\citenamefont {Tran},
  \citenamefont {Blaha},\ and\ \citenamefont {Schwarz}}]{TranJPCM07}%
  \BibitemOpen
  \bibfield  {author} {\bibinfo {author} {\bibfnamefont {F.}~\bibnamefont
  {Tran}}, \bibinfo {author} {\bibfnamefont {P.}~\bibnamefont {Blaha}}, \ and\
  \bibinfo {author} {\bibfnamefont {K.}~\bibnamefont {Schwarz}},\ }\href@noop
  {} {\bibfield  {journal} {\bibinfo  {journal} {J. Phys.: Condens. Matter}\
  }\textbf {\bibinfo {volume} {19}},\ \bibinfo {pages} {196208} (\bibinfo
  {year} {2007})}\BibitemShut {NoStop}%
\bibitem [{\citenamefont {Ferreira}\ \emph {et~al.}(2008)\citenamefont
  {Ferreira}, \citenamefont {Marques},\ and\ \citenamefont
  {Teles}}]{FerreiraPRB08}%
  \BibitemOpen
  \bibfield  {author} {\bibinfo {author} {\bibfnamefont {L.~G.}\ \bibnamefont
  {Ferreira}}, \bibinfo {author} {\bibfnamefont {M.}~\bibnamefont {Marques}}, \
  and\ \bibinfo {author} {\bibfnamefont {L.~K.}\ \bibnamefont {Teles}},\
  }\href@noop {} {\bibfield  {journal} {\bibinfo  {journal} {Phys. Rev. B}\
  }\textbf {\bibinfo {volume} {78}},\ \bibinfo {pages} {125116} (\bibinfo
  {year} {2008})}\BibitemShut {NoStop}%
\bibitem [{\citenamefont {Tran}\ and\ \citenamefont {Blaha}(2009)}]{TranPRL09}%
  \BibitemOpen
  \bibfield  {author} {\bibinfo {author} {\bibfnamefont {F.}~\bibnamefont
  {Tran}}\ and\ \bibinfo {author} {\bibfnamefont {P.}~\bibnamefont {Blaha}},\
  }\href@noop {} {\bibfield  {journal} {\bibinfo  {journal} {Phys. Rev. Lett.}\
  }\textbf {\bibinfo {volume} {102}},\ \bibinfo {pages} {226401} (\bibinfo
  {year} {2009})}\BibitemShut {NoStop}%
\bibitem [{\citenamefont {Armiento}\ and\ \citenamefont
  {K\"{u}mmel}(2013)}]{ArmientoPRL13}%
  \BibitemOpen
  \bibfield  {author} {\bibinfo {author} {\bibfnamefont {R.}~\bibnamefont
  {Armiento}}\ and\ \bibinfo {author} {\bibfnamefont {S.}~\bibnamefont
  {K\"{u}mmel}},\ }\href@noop {} {\bibfield  {journal} {\bibinfo  {journal}
  {Phys. Rev. Lett.}\ }\textbf {\bibinfo {volume} {111}},\ \bibinfo {pages}
  {036402} (\bibinfo {year} {2013})}\BibitemShut {NoStop}%
\bibitem [{\citenamefont {Singh}\ \emph {et~al.}(2016)\citenamefont {Singh},
  \citenamefont {Harbola}, \citenamefont {Hemanadhan}, \citenamefont
  {Mookerjee},\ and\ \citenamefont {Johnson}}]{SinghPRB16}%
  \BibitemOpen
  \bibfield  {author} {\bibinfo {author} {\bibfnamefont {P.}~\bibnamefont
  {Singh}}, \bibinfo {author} {\bibfnamefont {M.~K.}\ \bibnamefont {Harbola}},
  \bibinfo {author} {\bibfnamefont {M.}~\bibnamefont {Hemanadhan}}, \bibinfo
  {author} {\bibfnamefont {A.}~\bibnamefont {Mookerjee}}, \ and\ \bibinfo
  {author} {\bibfnamefont {D.~D.}\ \bibnamefont {Johnson}},\ }\href@noop {}
  {\bibfield  {journal} {\bibinfo  {journal} {Phys. Rev. B}\ }\textbf {\bibinfo
  {volume} {93}},\ \bibinfo {pages} {085204} (\bibinfo {year}
  {2016})}\BibitemShut {NoStop}%
\bibitem [{\citenamefont {Finzel}\ and\ \citenamefont
  {Baranov}(2017)}]{FinzelIJQC17}%
  \BibitemOpen
  \bibfield  {author} {\bibinfo {author} {\bibfnamefont {K.}~\bibnamefont
  {Finzel}}\ and\ \bibinfo {author} {\bibfnamefont {A.~I.}\ \bibnamefont
  {Baranov}},\ }\href@noop {} {\bibfield  {journal} {\bibinfo  {journal} {Int.
  J. Quantum Chem.}\ }\textbf {\bibinfo {volume} {117}},\ \bibinfo {pages} {40}
  (\bibinfo {year} {2017})}\BibitemShut {NoStop}%
\bibitem [{\citenamefont {Verma}\ and\ \citenamefont
  {Truhlar}(2017)}]{VermaJPCL17}%
  \BibitemOpen
  \bibfield  {author} {\bibinfo {author} {\bibfnamefont {P.}~\bibnamefont
  {Verma}}\ and\ \bibinfo {author} {\bibfnamefont {D.~G.}\ \bibnamefont
  {Truhlar}},\ }\href@noop {} {\bibfield  {journal} {\bibinfo  {journal} {J.
  Phys. Chem. Lett.}\ }\textbf {\bibinfo {volume} {8}},\ \bibinfo {pages} {380}
  (\bibinfo {year} {2017})}\BibitemShut {NoStop}%
\bibitem [{\citenamefont {Morales-Garc\'{i}a}\ \emph
  {et~al.}(2017)\citenamefont {Morales-Garc\'{i}a}, \citenamefont {Valero},\
  and\ \citenamefont {Illas}}]{MoralesGarciaJPCC17}%
  \BibitemOpen
  \bibfield  {author} {\bibinfo {author} {\bibfnamefont {{\'A}.}~\bibnamefont
  {Morales-Garc\'{i}a}}, \bibinfo {author} {\bibfnamefont {R.}~\bibnamefont
  {Valero}}, \ and\ \bibinfo {author} {\bibfnamefont {F.}~\bibnamefont
  {Illas}},\ }\href@noop {} {\bibfield  {journal} {\bibinfo  {journal} {J.
  Phys. Chem. C}\ }\textbf {\bibinfo {volume} {121}},\ \bibinfo {pages} {18862}
  (\bibinfo {year} {2017})}\BibitemShut {NoStop}%
\bibitem [{\citenamefont {Tran}\ and\ \citenamefont
  {Blaha}(2017)}]{TranJPCA17}%
  \BibitemOpen
  \bibfield  {author} {\bibinfo {author} {\bibfnamefont {F.}~\bibnamefont
  {Tran}}\ and\ \bibinfo {author} {\bibfnamefont {P.}~\bibnamefont {Blaha}},\
  }\href@noop {} {\bibfield  {journal} {\bibinfo  {journal} {J. Phys. Chem. A}\
  }\textbf {\bibinfo {volume} {121}},\ \bibinfo {pages} {3318} (\bibinfo {year}
  {2017})}\BibitemShut {NoStop}%
\bibitem [{\citenamefont {Singh}(2010)}]{SinghPRB10}%
  \BibitemOpen
  \bibfield  {author} {\bibinfo {author} {\bibfnamefont {D.~J.}\ \bibnamefont
  {Singh}},\ }\href@noop {} {\bibfield  {journal} {\bibinfo  {journal} {Phys.
  Rev. B}\ }\textbf {\bibinfo {volume} {82}},\ \bibinfo {pages} {205102}
  (\bibinfo {year} {2010})}\BibitemShut {NoStop}%
\bibitem [{\citenamefont {Botana}\ \emph {et~al.}(2012)\citenamefont {Botana},
  \citenamefont {Tran}, \citenamefont {Pardo}, \citenamefont {Baldomir},\ and\
  \citenamefont {Blaha}}]{BotanaPRB12}%
  \BibitemOpen
  \bibfield  {author} {\bibinfo {author} {\bibfnamefont {A.~S.}\ \bibnamefont
  {Botana}}, \bibinfo {author} {\bibfnamefont {F.}~\bibnamefont {Tran}},
  \bibinfo {author} {\bibfnamefont {V.}~\bibnamefont {Pardo}}, \bibinfo
  {author} {\bibfnamefont {D.}~\bibnamefont {Baldomir}}, \ and\ \bibinfo
  {author} {\bibfnamefont {P.}~\bibnamefont {Blaha}},\ }\href@noop {}
  {\bibfield  {journal} {\bibinfo  {journal} {Phys. Rev. B}\ }\textbf {\bibinfo
  {volume} {85}},\ \bibinfo {pages} {235118} (\bibinfo {year}
  {2012})}\BibitemShut {NoStop}%
\bibitem [{\citenamefont {Zhu}\ and\ \citenamefont
  {Schwingenschl\"ogl}(2012)}]{ZhuPRB12}%
  \BibitemOpen
  \bibfield  {author} {\bibinfo {author} {\bibfnamefont {Z.}~\bibnamefont
  {Zhu}}\ and\ \bibinfo {author} {\bibfnamefont {U.}~\bibnamefont
  {Schwingenschl\"ogl}},\ }\href@noop {} {\bibfield  {journal} {\bibinfo
  {journal} {Phys. Rev. B}\ }\textbf {\bibinfo {volume} {86}},\ \bibinfo
  {pages} {075149} (\bibinfo {year} {2012})}\BibitemShut {NoStop}%
\bibitem [{\citenamefont {Koller}\ \emph {et~al.}(2012)\citenamefont {Koller},
  \citenamefont {Tran},\ and\ \citenamefont {Blaha}}]{KollerPRB12}%
  \BibitemOpen
  \bibfield  {author} {\bibinfo {author} {\bibfnamefont {D.}~\bibnamefont
  {Koller}}, \bibinfo {author} {\bibfnamefont {F.}~\bibnamefont {Tran}}, \ and\
  \bibinfo {author} {\bibfnamefont {P.}~\bibnamefont {Blaha}},\ }\href@noop {}
  {\bibfield  {journal} {\bibinfo  {journal} {Phys. Rev. B}\ }\textbf {\bibinfo
  {volume} {85}},\ \bibinfo {pages} {155109} (\bibinfo {year}
  {2012})}\BibitemShut {NoStop}%
\bibitem [{\citenamefont {Camargo-Mart\'{\i}nez}\ and\ \citenamefont
  {Baquero}(2012)}]{CamargoMartinezPRB12}%
  \BibitemOpen
  \bibfield  {author} {\bibinfo {author} {\bibfnamefont {J.~A.}\ \bibnamefont
  {Camargo-Mart\'{\i}nez}}\ and\ \bibinfo {author} {\bibfnamefont
  {R.}~\bibnamefont {Baquero}},\ }\href@noop {} {\bibfield  {journal} {\bibinfo
   {journal} {Phys. Rev. B}\ }\textbf {\bibinfo {volume} {86}},\ \bibinfo
  {pages} {195106} (\bibinfo {year} {2012})}\BibitemShut {NoStop}%
\bibitem [{\citenamefont {Jiang}(2013)}]{JiangJCP13}%
  \BibitemOpen
  \bibfield  {author} {\bibinfo {author} {\bibfnamefont {H.}~\bibnamefont
  {Jiang}},\ }\href@noop {} {\bibfield  {journal} {\bibinfo  {journal} {J.
  Chem. Phys.}\ }\textbf {\bibinfo {volume} {138}},\ \bibinfo {pages} {134115}
  (\bibinfo {year} {2013})}\BibitemShut {NoStop}%
\bibitem [{\citenamefont {Dixit}\ \emph {et~al.}(2012)\citenamefont {Dixit},
  \citenamefont {Saniz}, \citenamefont {Cottenier}, \citenamefont {Lamoen},\
  and\ \citenamefont {Partoens}}]{DixitJPCM12}%
  \BibitemOpen
  \bibfield  {author} {\bibinfo {author} {\bibfnamefont {H.}~\bibnamefont
  {Dixit}}, \bibinfo {author} {\bibfnamefont {R.}~\bibnamefont {Saniz}},
  \bibinfo {author} {\bibfnamefont {S.}~\bibnamefont {Cottenier}}, \bibinfo
  {author} {\bibfnamefont {D.}~\bibnamefont {Lamoen}}, \ and\ \bibinfo {author}
  {\bibfnamefont {B.}~\bibnamefont {Partoens}},\ }\href@noop {} {\bibfield
  {journal} {\bibinfo  {journal} {J. Phys.: Condens. Matter}\ }\textbf
  {\bibinfo {volume} {24}},\ \bibinfo {pages} {205503} (\bibinfo {year}
  {2012})}\BibitemShut {NoStop}%
\bibitem [{\citenamefont {Waroquiers}\ \emph {et~al.}(2013)\citenamefont
  {Waroquiers}, \citenamefont {Lherbier}, \citenamefont {Miglio}, \citenamefont
  {Stankovski}, \citenamefont {Ponc\'e}, \citenamefont {Oliveira},
  \citenamefont {Giantomassi}, \citenamefont {Rignanese},\ and\ \citenamefont
  {Gonze}}]{WaroquiersPRB13}%
  \BibitemOpen
  \bibfield  {author} {\bibinfo {author} {\bibfnamefont {D.}~\bibnamefont
  {Waroquiers}}, \bibinfo {author} {\bibfnamefont {A.}~\bibnamefont
  {Lherbier}}, \bibinfo {author} {\bibfnamefont {A.}~\bibnamefont {Miglio}},
  \bibinfo {author} {\bibfnamefont {M.}~\bibnamefont {Stankovski}}, \bibinfo
  {author} {\bibfnamefont {S.}~\bibnamefont {Ponc\'e}}, \bibinfo {author}
  {\bibfnamefont {M.~J.~T.}\ \bibnamefont {Oliveira}}, \bibinfo {author}
  {\bibfnamefont {M.}~\bibnamefont {Giantomassi}}, \bibinfo {author}
  {\bibfnamefont {G.-M.}\ \bibnamefont {Rignanese}}, \ and\ \bibinfo {author}
  {\bibfnamefont {X.}~\bibnamefont {Gonze}},\ }\href@noop {} {\bibfield
  {journal} {\bibinfo  {journal} {Phys. Rev. B}\ }\textbf {\bibinfo {volume}
  {87}},\ \bibinfo {pages} {075121} (\bibinfo {year} {2013})}\BibitemShut
  {NoStop}%
\bibitem [{\citenamefont {Koller}\ \emph {et~al.}(2011)\citenamefont {Koller},
  \citenamefont {Tran},\ and\ \citenamefont {Blaha}}]{KollerPRB11}%
  \BibitemOpen
  \bibfield  {author} {\bibinfo {author} {\bibfnamefont {D.}~\bibnamefont
  {Koller}}, \bibinfo {author} {\bibfnamefont {F.}~\bibnamefont {Tran}}, \ and\
  \bibinfo {author} {\bibfnamefont {P.}~\bibnamefont {Blaha}},\ }\href@noop {}
  {\bibfield  {journal} {\bibinfo  {journal} {Phys. Rev. B}\ }\textbf {\bibinfo
  {volume} {83}},\ \bibinfo {pages} {195134} (\bibinfo {year}
  {2011})}\BibitemShut {NoStop}%
\bibitem [{\citenamefont {Kuisma}\ \emph {et~al.}(2010)\citenamefont {Kuisma},
  \citenamefont {Ojanen}, \citenamefont {Enkovaara},\ and\ \citenamefont
  {Rantala}}]{KuismaPRB10}%
  \BibitemOpen
  \bibfield  {author} {\bibinfo {author} {\bibfnamefont {M.}~\bibnamefont
  {Kuisma}}, \bibinfo {author} {\bibfnamefont {J.}~\bibnamefont {Ojanen}},
  \bibinfo {author} {\bibfnamefont {J.}~\bibnamefont {Enkovaara}}, \ and\
  \bibinfo {author} {\bibfnamefont {T.~T.}\ \bibnamefont {Rantala}},\
  }\href@noop {} {\bibfield  {journal} {\bibinfo  {journal} {Phys. Rev. B}\
  }\textbf {\bibinfo {volume} {82}},\ \bibinfo {pages} {115106} (\bibinfo
  {year} {2010})}\BibitemShut {NoStop}%
\bibitem [{\citenamefont {Gritsenko}\ \emph {et~al.}(1995)\citenamefont
  {Gritsenko}, \citenamefont {van Leeuwen}, \citenamefont {van Lenthe},\ and\
  \citenamefont {Baerends}}]{GritsenkoPRA95}%
  \BibitemOpen
  \bibfield  {author} {\bibinfo {author} {\bibfnamefont {O.}~\bibnamefont
  {Gritsenko}}, \bibinfo {author} {\bibfnamefont {R.}~\bibnamefont {van
  Leeuwen}}, \bibinfo {author} {\bibfnamefont {E.}~\bibnamefont {van Lenthe}},
  \ and\ \bibinfo {author} {\bibfnamefont {E.~J.}\ \bibnamefont {Baerends}},\
  }\href@noop {} {\bibfield  {journal} {\bibinfo  {journal} {Phys. Rev. A}\
  }\textbf {\bibinfo {volume} {51}},\ \bibinfo {pages} {1944} (\bibinfo {year}
  {1995})}\BibitemShut {NoStop}%
\bibitem [{\citenamefont {Gritsenko}\ \emph {et~al.}(1997)\citenamefont
  {Gritsenko}, \citenamefont {van Leeuwen},\ and\ \citenamefont
  {Baerends}}]{GritsenkoIJQC97}%
  \BibitemOpen
  \bibfield  {author} {\bibinfo {author} {\bibfnamefont {O.~V.}\ \bibnamefont
  {Gritsenko}}, \bibinfo {author} {\bibfnamefont {R.}~\bibnamefont {van
  Leeuwen}}, \ and\ \bibinfo {author} {\bibfnamefont {E.~J.}\ \bibnamefont
  {Baerends}},\ }\href@noop {} {\bibfield  {journal} {\bibinfo  {journal} {Int.
  J. Quantum Chem.}\ }\textbf {\bibinfo {volume} {61}},\ \bibinfo {pages} {231}
  (\bibinfo {year} {1997})}\BibitemShut {NoStop}%
\bibitem [{\citenamefont {Krieger}\ \emph {et~al.}(1992)\citenamefont
  {Krieger}, \citenamefont {Li},\ and\ \citenamefont
  {Iafrate}}]{KriegerPRA92a}%
  \BibitemOpen
  \bibfield  {author} {\bibinfo {author} {\bibfnamefont {J.~B.}\ \bibnamefont
  {Krieger}}, \bibinfo {author} {\bibfnamefont {Y.}~\bibnamefont {Li}}, \ and\
  \bibinfo {author} {\bibfnamefont {G.~J.}\ \bibnamefont {Iafrate}},\
  }\href@noop {} {\bibfield  {journal} {\bibinfo  {journal} {Phys. Rev. A}\
  }\textbf {\bibinfo {volume} {45}},\ \bibinfo {pages} {101} (\bibinfo {year}
  {1992})}\BibitemShut {NoStop}%
\bibitem [{\citenamefont {Sharp}\ and\ \citenamefont
  {Horton}(1953)}]{SharpPR53}%
  \BibitemOpen
  \bibfield  {author} {\bibinfo {author} {\bibfnamefont {R.~T.}\ \bibnamefont
  {Sharp}}\ and\ \bibinfo {author} {\bibfnamefont {G.~K.}\ \bibnamefont
  {Horton}},\ }\href@noop {} {\bibfield  {journal} {\bibinfo  {journal} {Phys.
  Rev.}\ }\textbf {\bibinfo {volume} {90}},\ \bibinfo {pages} {317} (\bibinfo
  {year} {1953})}\BibitemShut {NoStop}%
\bibitem [{\citenamefont {Castelli}\ \emph {et~al.}(2012)\citenamefont
  {Castelli}, \citenamefont {Olsen}, \citenamefont {Datta}, \citenamefont
  {Landis}, \citenamefont {Dahl}, \citenamefont {Thygesen},\ and\ \citenamefont
  {Jacobsen}}]{CastelliEES12}%
  \BibitemOpen
  \bibfield  {author} {\bibinfo {author} {\bibfnamefont {I.~E.}\ \bibnamefont
  {Castelli}}, \bibinfo {author} {\bibfnamefont {T.}~\bibnamefont {Olsen}},
  \bibinfo {author} {\bibfnamefont {S.}~\bibnamefont {Datta}}, \bibinfo
  {author} {\bibfnamefont {D.~D.}\ \bibnamefont {Landis}}, \bibinfo {author}
  {\bibfnamefont {S.}~\bibnamefont {Dahl}}, \bibinfo {author} {\bibfnamefont
  {K.~S.}\ \bibnamefont {Thygesen}}, \ and\ \bibinfo {author} {\bibfnamefont
  {K.~W.}\ \bibnamefont {Jacobsen}},\ }\href@noop {} {\bibfield  {journal}
  {\bibinfo  {journal} {Energy Environ. Sci.}\ }\textbf {\bibinfo {volume}
  {5}},\ \bibinfo {pages} {5814} (\bibinfo {year} {2012})}\BibitemShut
  {NoStop}%
\bibitem [{\citenamefont {Yan}\ \emph {et~al.}(2012)\citenamefont {Yan},
  \citenamefont {Jacobsen},\ and\ \citenamefont {Thygesen}}]{YanPRB12}%
  \BibitemOpen
  \bibfield  {author} {\bibinfo {author} {\bibfnamefont {J.}~\bibnamefont
  {Yan}}, \bibinfo {author} {\bibfnamefont {K.~W.}\ \bibnamefont {Jacobsen}}, \
  and\ \bibinfo {author} {\bibfnamefont {K.~S.}\ \bibnamefont {Thygesen}},\
  }\href@noop {} {\bibfield  {journal} {\bibinfo  {journal} {Phys. Rev. B}\
  }\textbf {\bibinfo {volume} {86}},\ \bibinfo {pages} {045208} (\bibinfo
  {year} {2012})}\BibitemShut {NoStop}%
\bibitem [{\citenamefont {H\"{u}ser}\ \emph {et~al.}(2013)\citenamefont
  {H\"{u}ser}, \citenamefont {Olsen},\ and\ \citenamefont
  {Thygesen}}]{HuserPRB13}%
  \BibitemOpen
  \bibfield  {author} {\bibinfo {author} {\bibfnamefont {F.}~\bibnamefont
  {H\"{u}ser}}, \bibinfo {author} {\bibfnamefont {T.}~\bibnamefont {Olsen}}, \
  and\ \bibinfo {author} {\bibfnamefont {K.~S.}\ \bibnamefont {Thygesen}},\
  }\href@noop {} {\bibfield  {journal} {\bibinfo  {journal} {Phys. Rev. B}\
  }\textbf {\bibinfo {volume} {87}},\ \bibinfo {pages} {235132} (\bibinfo
  {year} {2013})}\BibitemShut {NoStop}%
\bibitem [{\citenamefont {Mir\'{o}}\ \emph {et~al.}(2014)\citenamefont
  {Mir\'{o}}, \citenamefont {Audiffred},\ and\ \citenamefont
  {Heine}}]{MiroCSR14}%
  \BibitemOpen
  \bibfield  {author} {\bibinfo {author} {\bibfnamefont {P.}~\bibnamefont
  {Mir\'{o}}}, \bibinfo {author} {\bibfnamefont {M.}~\bibnamefont {Audiffred}},
  \ and\ \bibinfo {author} {\bibfnamefont {T.}~\bibnamefont {Heine}},\
  }\href@noop {} {\bibfield  {journal} {\bibinfo  {journal} {Chem. Soc. Rev.}\
  }\textbf {\bibinfo {volume} {43}},\ \bibinfo {pages} {6537} (\bibinfo {year}
  {2014})}\BibitemShut {NoStop}%
\bibitem [{\citenamefont {Pilania}\ \emph {et~al.}(2016)\citenamefont
  {Pilania}, \citenamefont {Mannodi-Kanakkithodi}, \citenamefont {Uberuaga},
  \citenamefont {Ramprasad}, \citenamefont {Gubernatis},\ and\ \citenamefont
  {Lookman}}]{PilaniaSR16}%
  \BibitemOpen
  \bibfield  {author} {\bibinfo {author} {\bibfnamefont {G.}~\bibnamefont
  {Pilania}}, \bibinfo {author} {\bibfnamefont {A.}~\bibnamefont
  {Mannodi-Kanakkithodi}}, \bibinfo {author} {\bibfnamefont {B.~P.}\
  \bibnamefont {Uberuaga}}, \bibinfo {author} {\bibfnamefont {R.}~\bibnamefont
  {Ramprasad}}, \bibinfo {author} {\bibfnamefont {J.~E.}\ \bibnamefont
  {Gubernatis}}, \ and\ \bibinfo {author} {\bibfnamefont {T.}~\bibnamefont
  {Lookman}},\ }\href@noop {} {\bibfield  {journal} {\bibinfo  {journal} {Sci.
  Rep.}\ }\textbf {\bibinfo {volume} {6}},\ \bibinfo {pages} {19375} (\bibinfo
  {year} {2016})}\BibitemShut {NoStop}%
\bibitem [{\citenamefont {Kim}\ \emph {et~al.}(2016)\citenamefont {Kim},
  \citenamefont {Pilania},\ and\ \citenamefont {Ramprasad}}]{KimJPCC16}%
  \BibitemOpen
  \bibfield  {author} {\bibinfo {author} {\bibfnamefont {C.}~\bibnamefont
  {Kim}}, \bibinfo {author} {\bibfnamefont {G.}~\bibnamefont {Pilania}}, \ and\
  \bibinfo {author} {\bibfnamefont {R.}~\bibnamefont {Ramprasad}},\ }\href@noop
  {} {\bibfield  {journal} {\bibinfo  {journal} {J. Phys. Chem. C}\ }\textbf
  {\bibinfo {volume} {120}},\ \bibinfo {pages} {14575} (\bibinfo {year}
  {2016})}\BibitemShut {NoStop}%
\bibitem [{\citenamefont {Pandey}\ \emph {et~al.}(2017)\citenamefont {Pandey},
  \citenamefont {Kuhar},\ and\ \citenamefont {Jacobsen}}]{PandeyJPCC17}%
  \BibitemOpen
  \bibfield  {author} {\bibinfo {author} {\bibfnamefont {M.}~\bibnamefont
  {Pandey}}, \bibinfo {author} {\bibfnamefont {K.}~\bibnamefont {Kuhar}}, \
  and\ \bibinfo {author} {\bibfnamefont {K.~W.}\ \bibnamefont {Jacobsen}},\
  }\href@noop {} {\bibfield  {journal} {\bibinfo  {journal} {J. Phys. Chem. C}\
  }\textbf {\bibinfo {volume} {121}},\ \bibinfo {pages} {17780} (\bibinfo
  {year} {2017})}\BibitemShut {NoStop}%
\bibitem [{\citenamefont {Baerends}(2017)}]{BaerendsPCCP17}%
  \BibitemOpen
  \bibfield  {author} {\bibinfo {author} {\bibfnamefont {E.~J.}\ \bibnamefont
  {Baerends}},\ }\href@noop {} {\bibfield  {journal} {\bibinfo  {journal}
  {Phys. Chem. Chem. Phys.}\ }\textbf {\bibinfo {volume} {19}},\ \bibinfo
  {pages} {15639} (\bibinfo {year} {2017})}\BibitemShut {NoStop}%
\bibitem [{\citenamefont {Slater}(1951)}]{SlaterPR51}%
  \BibitemOpen
  \bibfield  {author} {\bibinfo {author} {\bibfnamefont {J.~C.}\ \bibnamefont
  {Slater}},\ }\href@noop {} {\bibfield  {journal} {\bibinfo  {journal} {Phys.
  Rev.}\ }\textbf {\bibinfo {volume} {81}},\ \bibinfo {pages} {385} (\bibinfo
  {year} {1951})}\BibitemShut {NoStop}%
\bibitem [{\citenamefont {Becke}(1988)}]{BeckePRA88}%
  \BibitemOpen
  \bibfield  {author} {\bibinfo {author} {\bibfnamefont {A.~D.}\ \bibnamefont
  {Becke}},\ }\href@noop {} {\bibfield  {journal} {\bibinfo  {journal} {Phys.
  Rev. A}\ }\textbf {\bibinfo {volume} {38}},\ \bibinfo {pages} {3098}
  (\bibinfo {year} {1988})}\BibitemShut {NoStop}%
\bibitem [{\citenamefont {Levy}\ and\ \citenamefont
  {Perdew}(1985)}]{LevyPRA85}%
  \BibitemOpen
  \bibfield  {author} {\bibinfo {author} {\bibfnamefont {M.}~\bibnamefont
  {Levy}}\ and\ \bibinfo {author} {\bibfnamefont {J.~P.}\ \bibnamefont
  {Perdew}},\ }\href@noop {} {\bibfield  {journal} {\bibinfo  {journal} {Phys.
  Rev. A}\ }\textbf {\bibinfo {volume} {32}},\ \bibinfo {pages} {2010}
  (\bibinfo {year} {1985})}\BibitemShut {NoStop}%
\bibitem [{\citenamefont {Kohut}\ \emph {et~al.}(2014)\citenamefont {Kohut},
  \citenamefont {Ryabinkin},\ and\ \citenamefont {Staroverov}}]{KohutJCP14}%
  \BibitemOpen
  \bibfield  {author} {\bibinfo {author} {\bibfnamefont {S.~V.}\ \bibnamefont
  {Kohut}}, \bibinfo {author} {\bibfnamefont {I.~G.}\ \bibnamefont
  {Ryabinkin}}, \ and\ \bibinfo {author} {\bibfnamefont {V.~N.}\ \bibnamefont
  {Staroverov}},\ }\href@noop {} {\bibfield  {journal} {\bibinfo  {journal} {J.
  Chem. Phys.}\ }\textbf {\bibinfo {volume} {140}},\ \bibinfo {pages} {18A535}
  (\bibinfo {year} {2014})}\BibitemShut {NoStop}%
\bibitem [{\citenamefont {Perdew}\ \emph {et~al.}(1992)\citenamefont {Perdew},
  \citenamefont {Chevary}, \citenamefont {Vosko}, \citenamefont {Jackson},
  \citenamefont {Pederson}, \citenamefont {Singh},\ and\ \citenamefont
  {Fiolhais}}]{PerdewPRB92b}%
  \BibitemOpen
  \bibfield  {author} {\bibinfo {author} {\bibfnamefont {J.~P.}\ \bibnamefont
  {Perdew}}, \bibinfo {author} {\bibfnamefont {J.~A.}\ \bibnamefont {Chevary}},
  \bibinfo {author} {\bibfnamefont {S.~H.}\ \bibnamefont {Vosko}}, \bibinfo
  {author} {\bibfnamefont {K.~A.}\ \bibnamefont {Jackson}}, \bibinfo {author}
  {\bibfnamefont {M.~R.}\ \bibnamefont {Pederson}}, \bibinfo {author}
  {\bibfnamefont {D.~J.}\ \bibnamefont {Singh}}, \ and\ \bibinfo {author}
  {\bibfnamefont {C.}~\bibnamefont {Fiolhais}},\ }\href@noop {} {\bibfield
  {journal} {\bibinfo  {journal} {Phys. Rev. B}\ }\textbf {\bibinfo {volume}
  {46}},\ \bibinfo {pages} {6671} (\bibinfo {year} {1992})},\ \bibinfo {note}
  {\textbf{48}, 4978(E) (1993)}\BibitemShut {NoStop}%
\bibitem [{\citenamefont {Perdew}\ \emph {et~al.}(2008)\citenamefont {Perdew},
  \citenamefont {Ruzsinszky}, \citenamefont {Csonka}, \citenamefont {Vydrov},
  \citenamefont {Scuseria}, \citenamefont {Constantin}, \citenamefont {Zhou},\
  and\ \citenamefont {Burke}}]{PerdewPRL08}%
  \BibitemOpen
  \bibfield  {author} {\bibinfo {author} {\bibfnamefont {J.~P.}\ \bibnamefont
  {Perdew}}, \bibinfo {author} {\bibfnamefont {A.}~\bibnamefont {Ruzsinszky}},
  \bibinfo {author} {\bibfnamefont {G.~I.}\ \bibnamefont {Csonka}}, \bibinfo
  {author} {\bibfnamefont {O.~A.}\ \bibnamefont {Vydrov}}, \bibinfo {author}
  {\bibfnamefont {G.~E.}\ \bibnamefont {Scuseria}}, \bibinfo {author}
  {\bibfnamefont {L.~A.}\ \bibnamefont {Constantin}}, \bibinfo {author}
  {\bibfnamefont {X.}~\bibnamefont {Zhou}}, \ and\ \bibinfo {author}
  {\bibfnamefont {K.}~\bibnamefont {Burke}},\ }\href@noop {} {\bibfield
  {journal} {\bibinfo  {journal} {Phys. Rev. Lett.}\ }\textbf {\bibinfo
  {volume} {100}},\ \bibinfo {pages} {136406} (\bibinfo {year} {2008})},\
  \bibinfo {note} {\textbf{102}, 039902(E) (2009); A. E. Mattsson, R. Armiento,
  and T. R. Mattsson, \textit{ibid}. \textbf{101}, 239701 (2008); J. P. Perdew,
  A. Ruzsinszky, G. I. Csonka, O. A. Vydrov, G. E. Scuseria, L. A. Constantin,
  X. Zhou, and K. Burke, \textit{ibid}. \textbf{101}, 239702
  (2008)}\BibitemShut {NoStop}%
\bibitem [{\citenamefont {Armiento}\ \emph {et~al.}(2008)\citenamefont
  {Armiento}, \citenamefont {K\"ummel},\ and\ \citenamefont
  {K\"orzd\"orfer}}]{ArmientoPRB08}%
  \BibitemOpen
  \bibfield  {author} {\bibinfo {author} {\bibfnamefont {R.}~\bibnamefont
  {Armiento}}, \bibinfo {author} {\bibfnamefont {S.}~\bibnamefont {K\"ummel}},
  \ and\ \bibinfo {author} {\bibfnamefont {T.}~\bibnamefont {K\"orzd\"orfer}},\
  }\href@noop {} {\bibfield  {journal} {\bibinfo  {journal} {Phys. Rev. B}\
  }\textbf {\bibinfo {volume} {77}},\ \bibinfo {pages} {165106} (\bibinfo
  {year} {2008})}\BibitemShut {NoStop}%
\bibitem [{\citenamefont {Blaha}\ \emph {et~al.}(2001)\citenamefont {Blaha},
  \citenamefont {Schwarz}, \citenamefont {Madsen}, \citenamefont {Kvasnicka},\
  and\ \citenamefont {Luitz}}]{WIEN2k}%
  \BibitemOpen
  \bibfield  {author} {\bibinfo {author} {\bibfnamefont {P.}~\bibnamefont
  {Blaha}}, \bibinfo {author} {\bibfnamefont {K.}~\bibnamefont {Schwarz}},
  \bibinfo {author} {\bibfnamefont {G.~K.~H.}\ \bibnamefont {Madsen}}, \bibinfo
  {author} {\bibfnamefont {D.}~\bibnamefont {Kvasnicka}}, \ and\ \bibinfo
  {author} {\bibfnamefont {J.}~\bibnamefont {Luitz}},\ }\href@noop {} {\emph
  {\bibinfo {title} {WIEN2K: An Augmented Plane Wave plus Local Orbitals
  Program for Calculating Crystal Properties}}}\ (\bibinfo  {publisher} {Vienna
  University of Technology},\ \bibinfo {address} {Austria},\ \bibinfo {year}
  {2001})\BibitemShut {NoStop}%
\bibitem [{\citenamefont {Andersen}(1975)}]{AndersenPRB75}%
  \BibitemOpen
  \bibfield  {author} {\bibinfo {author} {\bibfnamefont {O.~K.}\ \bibnamefont
  {Andersen}},\ }\href@noop {} {\bibfield  {journal} {\bibinfo  {journal}
  {Phys. Rev. B}\ }\textbf {\bibinfo {volume} {12}},\ \bibinfo {pages} {3060}
  (\bibinfo {year} {1975})}\BibitemShut {NoStop}%
\bibitem [{\citenamefont {Singh}\ and\ \citenamefont
  {Nordstr{\"{o}}m}(2006)}]{Singh}%
  \BibitemOpen
  \bibfield  {author} {\bibinfo {author} {\bibfnamefont {D.~J.}\ \bibnamefont
  {Singh}}\ and\ \bibinfo {author} {\bibfnamefont {L.}~\bibnamefont
  {Nordstr{\"{o}}m}},\ }\href@noop {} {\emph {\bibinfo {title} {Planewaves,
  Pseudopotentials, and the LAPW Method, 2nd ed.}}}\ (\bibinfo  {publisher}
  {Springer},\ \bibinfo {address} {New York},\ \bibinfo {year}
  {2006})\BibitemShut {NoStop}%
\bibitem [{\citenamefont {Perdew}\ and\ \citenamefont
  {Wang}(1992)}]{PerdewPRB92a}%
  \BibitemOpen
  \bibfield  {author} {\bibinfo {author} {\bibfnamefont {J.~P.}\ \bibnamefont
  {Perdew}}\ and\ \bibinfo {author} {\bibfnamefont {Y.}~\bibnamefont {Wang}},\
  }\href@noop {} {\bibfield  {journal} {\bibinfo  {journal} {Phys. Rev. B}\
  }\textbf {\bibinfo {volume} {45}},\ \bibinfo {pages} {13244} (\bibinfo {year}
  {1992})}\BibitemShut {NoStop}%
\bibitem [{\citenamefont {Vl\v{c}ek}\ \emph {et~al.}(2015)\citenamefont
  {Vl\v{c}ek}, \citenamefont {Steinle-Neumann}, \citenamefont {Leppert},
  \citenamefont {Armiento},\ and\ \citenamefont {K\"ummel}}]{VlcekPRB15}%
  \BibitemOpen
  \bibfield  {author} {\bibinfo {author} {\bibfnamefont {V.}~\bibnamefont
  {Vl\v{c}ek}}, \bibinfo {author} {\bibfnamefont {G.}~\bibnamefont
  {Steinle-Neumann}}, \bibinfo {author} {\bibfnamefont {L.}~\bibnamefont
  {Leppert}}, \bibinfo {author} {\bibfnamefont {R.}~\bibnamefont {Armiento}}, \
  and\ \bibinfo {author} {\bibfnamefont {S.}~\bibnamefont {K\"ummel}},\
  }\href@noop {} {\bibfield  {journal} {\bibinfo  {journal} {Phys. Rev. B}\
  }\textbf {\bibinfo {volume} {91}},\ \bibinfo {pages} {035107} (\bibinfo
  {year} {2015})}\BibitemShut {NoStop}%
\bibitem [{\citenamefont {Lindmaa}\ and\ \citenamefont
  {Armiento}(2016)}]{LindmaaPRB16}%
  \BibitemOpen
  \bibfield  {author} {\bibinfo {author} {\bibfnamefont {A.}~\bibnamefont
  {Lindmaa}}\ and\ \bibinfo {author} {\bibfnamefont {R.}~\bibnamefont
  {Armiento}},\ }\href@noop {} {\bibfield  {journal} {\bibinfo  {journal}
  {Phys. Rev. B}\ }\textbf {\bibinfo {volume} {94}},\ \bibinfo {pages} {155143}
  (\bibinfo {year} {2016})}\BibitemShut {NoStop}%
\bibitem [{\citenamefont {Boese}\ and\ \citenamefont
  {Handy}(2001)}]{BoeseJCP01}%
  \BibitemOpen
  \bibfield  {author} {\bibinfo {author} {\bibfnamefont {A.~D.}\ \bibnamefont
  {Boese}}\ and\ \bibinfo {author} {\bibfnamefont {N.~C.}\ \bibnamefont
  {Handy}},\ }\href@noop {} {\bibfield  {journal} {\bibinfo  {journal} {J.
  Chem. Phys.}\ }\textbf {\bibinfo {volume} {114}},\ \bibinfo {pages} {5497}
  (\bibinfo {year} {2001})}\BibitemShut {NoStop}%
\bibitem [{\citenamefont {van Leeuwen}\ and\ \citenamefont
  {Baerends}(1994)}]{vanLeeuwenPRA94}%
  \BibitemOpen
  \bibfield  {author} {\bibinfo {author} {\bibfnamefont {R.}~\bibnamefont {van
  Leeuwen}}\ and\ \bibinfo {author} {\bibfnamefont {E.~J.}\ \bibnamefont
  {Baerends}},\ }\href@noop {} {\bibfield  {journal} {\bibinfo  {journal}
  {Phys. Rev. A}\ }\textbf {\bibinfo {volume} {49}},\ \bibinfo {pages} {2421}
  (\bibinfo {year} {1994})}\BibitemShut {NoStop}%
\bibitem [{\citenamefont {Karolewski}\ \emph {et~al.}(2009)\citenamefont
  {Karolewski}, \citenamefont {Armiento},\ and\ \citenamefont
  {K\"{u}mmel}}]{KarolewskiJCTC09}%
  \BibitemOpen
  \bibfield  {author} {\bibinfo {author} {\bibfnamefont {A.}~\bibnamefont
  {Karolewski}}, \bibinfo {author} {\bibfnamefont {R.}~\bibnamefont
  {Armiento}}, \ and\ \bibinfo {author} {\bibfnamefont {S.}~\bibnamefont
  {K\"{u}mmel}},\ }\href@noop {} {\bibfield  {journal} {\bibinfo  {journal} {J.
  Chem. Theory Comput.}\ }\textbf {\bibinfo {volume} {5}},\ \bibinfo {pages}
  {712} (\bibinfo {year} {2009})}\BibitemShut {NoStop}%
\bibitem [{\citenamefont {Gaiduk}\ and\ \citenamefont
  {Staroverov}(2009)}]{GaidukJCP09}%
  \BibitemOpen
  \bibfield  {author} {\bibinfo {author} {\bibfnamefont {A.~P.}\ \bibnamefont
  {Gaiduk}}\ and\ \bibinfo {author} {\bibfnamefont {V.~N.}\ \bibnamefont
  {Staroverov}},\ }\href@noop {} {\bibfield  {journal} {\bibinfo  {journal} {J.
  Chem. Phys.}\ }\textbf {\bibinfo {volume} {131}},\ \bibinfo {pages} {044107}
  (\bibinfo {year} {2009})}\BibitemShut {NoStop}%
\bibitem [{\citenamefont {Karolewski}\ \emph {et~al.}(2013)\citenamefont
  {Karolewski}, \citenamefont {Armiento},\ and\ \citenamefont
  {K\"ummel}}]{KarolewskiPRA13}%
  \BibitemOpen
  \bibfield  {author} {\bibinfo {author} {\bibfnamefont {A.}~\bibnamefont
  {Karolewski}}, \bibinfo {author} {\bibfnamefont {R.}~\bibnamefont
  {Armiento}}, \ and\ \bibinfo {author} {\bibfnamefont {S.}~\bibnamefont
  {K\"ummel}},\ }\href@noop {} {\bibfield  {journal} {\bibinfo  {journal}
  {Phys. Rev. A}\ }\textbf {\bibinfo {volume} {88}},\ \bibinfo {pages} {052519}
  (\bibinfo {year} {2013})}\BibitemShut {NoStop}%
\bibitem [{\citenamefont {Aschebrock}\ \emph
  {et~al.}(2017{\natexlab{a}})\citenamefont {Aschebrock}, \citenamefont
  {Armiento},\ and\ \citenamefont {K\"ummel}}]{AschebrockPRB17a}%
  \BibitemOpen
  \bibfield  {author} {\bibinfo {author} {\bibfnamefont {T.}~\bibnamefont
  {Aschebrock}}, \bibinfo {author} {\bibfnamefont {R.}~\bibnamefont
  {Armiento}}, \ and\ \bibinfo {author} {\bibfnamefont {S.}~\bibnamefont
  {K\"ummel}},\ }\href@noop {} {\bibfield  {journal} {\bibinfo  {journal}
  {Phys. Rev. B}\ }\textbf {\bibinfo {volume} {95}},\ \bibinfo {pages} {245118}
  (\bibinfo {year} {2017}{\natexlab{a}})}\BibitemShut {NoStop}%
\bibitem [{\citenamefont {Aschebrock}\ \emph
  {et~al.}(2017{\natexlab{b}})\citenamefont {Aschebrock}, \citenamefont
  {Armiento},\ and\ \citenamefont {K\"ummel}}]{AschebrockPRB17b}%
  \BibitemOpen
  \bibfield  {author} {\bibinfo {author} {\bibfnamefont {T.}~\bibnamefont
  {Aschebrock}}, \bibinfo {author} {\bibfnamefont {R.}~\bibnamefont
  {Armiento}}, \ and\ \bibinfo {author} {\bibfnamefont {S.}~\bibnamefont
  {K\"ummel}},\ }\href@noop {} {\bibfield  {journal} {\bibinfo  {journal}
  {Phys. Rev. B}\ }\textbf {\bibinfo {volume} {96}},\ \bibinfo {pages} {075140}
  (\bibinfo {year} {2017}{\natexlab{b}})}\BibitemShut {NoStop}%
\bibitem [{\citenamefont {Tran}\ and\ \citenamefont {Blaha}(2011)}]{TranPRB11}%
  \BibitemOpen
  \bibfield  {author} {\bibinfo {author} {\bibfnamefont {F.}~\bibnamefont
  {Tran}}\ and\ \bibinfo {author} {\bibfnamefont {P.}~\bibnamefont {Blaha}},\
  }\href@noop {} {\bibfield  {journal} {\bibinfo  {journal} {Phys. Rev. B}\
  }\textbf {\bibinfo {volume} {83}},\ \bibinfo {pages} {235118} (\bibinfo
  {year} {2011})}\BibitemShut {NoStop}%
\bibitem [{\citenamefont {Shimazaki}\ and\ \citenamefont
  {Asai}(2008)}]{ShimazakiCPL08}%
  \BibitemOpen
  \bibfield  {author} {\bibinfo {author} {\bibfnamefont {T.}~\bibnamefont
  {Shimazaki}}\ and\ \bibinfo {author} {\bibfnamefont {Y.}~\bibnamefont
  {Asai}},\ }\href@noop {} {\bibfield  {journal} {\bibinfo  {journal} {Chem.
  Phys. Lett.}\ }\textbf {\bibinfo {volume} {466}},\ \bibinfo {pages} {91}
  (\bibinfo {year} {2008})}\BibitemShut {NoStop}%
\bibitem [{\citenamefont {Heyd}\ \emph {et~al.}(2003)\citenamefont {Heyd},
  \citenamefont {Scuseria},\ and\ \citenamefont {Ernzerhof}}]{HeydJCP03}%
  \BibitemOpen
  \bibfield  {author} {\bibinfo {author} {\bibfnamefont {J.}~\bibnamefont
  {Heyd}}, \bibinfo {author} {\bibfnamefont {G.~E.}\ \bibnamefont {Scuseria}},
  \ and\ \bibinfo {author} {\bibfnamefont {M.}~\bibnamefont {Ernzerhof}},\
  }\href@noop {} {\bibfield  {journal} {\bibinfo  {journal} {J. Chem. Phys.}\
  }\textbf {\bibinfo {volume} {118}},\ \bibinfo {pages} {8207} (\bibinfo {year}
  {2003})},\ \bibinfo {note} {\textbf{124}, 219906 (2006)}\BibitemShut
  {NoStop}%
\bibitem [{\citenamefont {Krukau}\ \emph {et~al.}(2006)\citenamefont {Krukau},
  \citenamefont {Vydrov}, \citenamefont {Izmaylov},\ and\ \citenamefont
  {Scuseria}}]{KrukauJCP06}%
  \BibitemOpen
  \bibfield  {author} {\bibinfo {author} {\bibfnamefont {A.~V.}\ \bibnamefont
  {Krukau}}, \bibinfo {author} {\bibfnamefont {O.~A.}\ \bibnamefont {Vydrov}},
  \bibinfo {author} {\bibfnamefont {A.~F.}\ \bibnamefont {Izmaylov}}, \ and\
  \bibinfo {author} {\bibfnamefont {G.~E.}\ \bibnamefont {Scuseria}},\
  }\href@noop {} {\bibfield  {journal} {\bibinfo  {journal} {J. Chem. Phys.}\
  }\textbf {\bibinfo {volume} {125}},\ \bibinfo {pages} {224106} (\bibinfo
  {year} {2006})}\BibitemShut {NoStop}%
\bibitem [{\citenamefont {Becke}(1993)}]{BeckeJCP93}%
  \BibitemOpen
  \bibfield  {author} {\bibinfo {author} {\bibfnamefont {A.~D.}\ \bibnamefont
  {Becke}},\ }\href@noop {} {\bibfield  {journal} {\bibinfo  {journal} {J.
  Chem. Phys.}\ }\textbf {\bibinfo {volume} {98}},\ \bibinfo {pages} {5648}
  (\bibinfo {year} {1993})}\BibitemShut {NoStop}%
\bibitem [{\citenamefont {Perdew}\ \emph {et~al.}(2017)\citenamefont {Perdew},
  \citenamefont {Yang}, \citenamefont {Burke}, \citenamefont {Yang},
  \citenamefont {Gross}, \citenamefont {Scheffler}, \citenamefont {Scuseria},
  \citenamefont {Henderson}, \citenamefont {Zhang}, \citenamefont {Ruzsinszky},
  \citenamefont {Peng}, \citenamefont {Sun}, \citenamefont {Trushin},\ and\
  \citenamefont {G\"{o}rling}}]{PerdewPNAS17}%
  \BibitemOpen
  \bibfield  {author} {\bibinfo {author} {\bibfnamefont {J.~P.}\ \bibnamefont
  {Perdew}}, \bibinfo {author} {\bibfnamefont {W.}~\bibnamefont {Yang}},
  \bibinfo {author} {\bibfnamefont {K.}~\bibnamefont {Burke}}, \bibinfo
  {author} {\bibfnamefont {Z.}~\bibnamefont {Yang}}, \bibinfo {author}
  {\bibfnamefont {E.~K.~U.}\ \bibnamefont {Gross}}, \bibinfo {author}
  {\bibfnamefont {M.}~\bibnamefont {Scheffler}}, \bibinfo {author}
  {\bibfnamefont {G.~E.}\ \bibnamefont {Scuseria}}, \bibinfo {author}
  {\bibfnamefont {T.~M.}\ \bibnamefont {Henderson}}, \bibinfo {author}
  {\bibfnamefont {I.~Y.}\ \bibnamefont {Zhang}}, \bibinfo {author}
  {\bibfnamefont {A.}~\bibnamefont {Ruzsinszky}}, \bibinfo {author}
  {\bibfnamefont {H.}~\bibnamefont {Peng}}, \bibinfo {author} {\bibfnamefont
  {J.}~\bibnamefont {Sun}}, \bibinfo {author} {\bibfnamefont {E.}~\bibnamefont
  {Trushin}}, \ and\ \bibinfo {author} {\bibfnamefont {A.}~\bibnamefont
  {G\"{o}rling}},\ }\href@noop {} {\bibfield  {journal} {\bibinfo  {journal}
  {Proc. Natl. Acad. Sci. U.S.A.}\ }\textbf {\bibinfo {volume} {114}},\
  \bibinfo {pages} {2801} (\bibinfo {year} {2017})}\BibitemShut {NoStop}%
\bibitem [{SM_()}]{SM_GLLB}%
  \BibitemOpen
  \href@noop {} {}\bibinfo {howpublished} {See Supplemental Material at
  http://link.aps.org/supplemental/ for information about the solids in the
  test set and the results for the fundamental band gap and form factors of
  Si.}\BibitemShut {Stop}%
\bibitem [{\citenamefont {Koelling}\ and\ \citenamefont
  {Harmon}(1977)}]{KoellingJPC77}%
  \BibitemOpen
  \bibfield  {author} {\bibinfo {author} {\bibfnamefont {D.~D.}\ \bibnamefont
  {Koelling}}\ and\ \bibinfo {author} {\bibfnamefont {B.~N.}\ \bibnamefont
  {Harmon}},\ }\href@noop {} {\bibfield  {journal} {\bibinfo  {journal} {J.
  Phys. C: Solid State Phys.}\ }\textbf {\bibinfo {volume} {10}},\ \bibinfo
  {pages} {3107} (\bibinfo {year} {1977})}\BibitemShut {NoStop}%
\bibitem [{\citenamefont {MacDonald}\ \emph {et~al.}(1980)\citenamefont
  {MacDonald}, \citenamefont {Pickett},\ and\ \citenamefont
  {Koelling}}]{MacDonaldJPC80}%
  \BibitemOpen
  \bibfield  {author} {\bibinfo {author} {\bibfnamefont {A.~H.}\ \bibnamefont
  {MacDonald}}, \bibinfo {author} {\bibfnamefont {W.~E.}\ \bibnamefont
  {Pickett}}, \ and\ \bibinfo {author} {\bibfnamefont {D.~D.}\ \bibnamefont
  {Koelling}},\ }\href@noop {} {\bibfield  {journal} {\bibinfo  {journal} {J.
  Phys. C: Solid State Phys.}\ }\textbf {\bibinfo {volume} {13}},\ \bibinfo
  {pages} {2675} (\bibinfo {year} {1980})}\BibitemShut {NoStop}%
\bibitem [{\citenamefont {Schipper}\ \emph {et~al.}(2000)\citenamefont
  {Schipper}, \citenamefont {Gritsenko}, \citenamefont {van Gisbergen},\ and\
  \citenamefont {Baerends}}]{SchipperJCP00}%
  \BibitemOpen
  \bibfield  {author} {\bibinfo {author} {\bibfnamefont {P.~R.~T.}\
  \bibnamefont {Schipper}}, \bibinfo {author} {\bibfnamefont {O.~V.}\
  \bibnamefont {Gritsenko}}, \bibinfo {author} {\bibfnamefont {S.~J.~A.}\
  \bibnamefont {van Gisbergen}}, \ and\ \bibinfo {author} {\bibfnamefont
  {E.~J.}\ \bibnamefont {Baerends}},\ }\href@noop {} {\bibfield  {journal}
  {\bibinfo  {journal} {J. Chem. Phys.}\ }\textbf {\bibinfo {volume} {112}},\
  \bibinfo {pages} {1344} (\bibinfo {year} {2000})}\BibitemShut {NoStop}%
\bibitem [{\citenamefont {Terakura}\ \emph {et~al.}(1984)\citenamefont
  {Terakura}, \citenamefont {Oguchi}, \citenamefont {Williams},\ and\
  \citenamefont {K\"{u}bler}}]{TerakuraPRB84}%
  \BibitemOpen
  \bibfield  {author} {\bibinfo {author} {\bibfnamefont {K.}~\bibnamefont
  {Terakura}}, \bibinfo {author} {\bibfnamefont {T.}~\bibnamefont {Oguchi}},
  \bibinfo {author} {\bibfnamefont {A.~R.}\ \bibnamefont {Williams}}, \ and\
  \bibinfo {author} {\bibfnamefont {J.}~\bibnamefont {K\"{u}bler}},\
  }\href@noop {} {\bibfield  {journal} {\bibinfo  {journal} {Phys. Rev. B}\
  }\textbf {\bibinfo {volume} {30}},\ \bibinfo {pages} {4734} (\bibinfo {year}
  {1984})}\BibitemShut {NoStop}%
\bibitem [{\citenamefont {Vurgaftman}\ \emph {et~al.}(2001)\citenamefont
  {Vurgaftman}, \citenamefont {Meyer},\ and\ \citenamefont
  {Ram-Mohan}}]{VurgaftmanJAP01}%
  \BibitemOpen
  \bibfield  {author} {\bibinfo {author} {\bibfnamefont {I.}~\bibnamefont
  {Vurgaftman}}, \bibinfo {author} {\bibfnamefont {J.~R.}\ \bibnamefont
  {Meyer}}, \ and\ \bibinfo {author} {\bibfnamefont {L.~R.}\ \bibnamefont
  {Ram-Mohan}},\ }\href@noop {} {\bibfield  {journal} {\bibinfo  {journal} {J.
  Appl. Phys.}\ }\textbf {\bibinfo {volume} {89}},\ \bibinfo {pages} {5815}
  (\bibinfo {year} {2001})}\BibitemShut {NoStop}%
\bibitem [{\citenamefont {Kim}\ \emph {et~al.}(2010)\citenamefont {Kim},
  \citenamefont {Marsman}, \citenamefont {Kresse}, \citenamefont {Tran},\ and\
  \citenamefont {Blaha}}]{KimPRB10}%
  \BibitemOpen
  \bibfield  {author} {\bibinfo {author} {\bibfnamefont {Y.-S.}\ \bibnamefont
  {Kim}}, \bibinfo {author} {\bibfnamefont {M.}~\bibnamefont {Marsman}},
  \bibinfo {author} {\bibfnamefont {G.}~\bibnamefont {Kresse}}, \bibinfo
  {author} {\bibfnamefont {F.}~\bibnamefont {Tran}}, \ and\ \bibinfo {author}
  {\bibfnamefont {P.}~\bibnamefont {Blaha}},\ }\href@noop {} {\bibfield
  {journal} {\bibinfo  {journal} {Phys. Rev. B}\ }\textbf {\bibinfo {volume}
  {82}},\ \bibinfo {pages} {205212} (\bibinfo {year} {2010})}\BibitemShut
  {NoStop}%
\bibitem [{\citenamefont {See}\ and\ \citenamefont
  {Klebanoff}(1995)}]{SeeSS95}%
  \BibitemOpen
  \bibfield  {author} {\bibinfo {author} {\bibfnamefont {A.~K.}\ \bibnamefont
  {See}}\ and\ \bibinfo {author} {\bibfnamefont {L.~E.}\ \bibnamefont
  {Klebanoff}},\ }\href@noop {} {\bibfield  {journal} {\bibinfo  {journal}
  {Surf. Sci.}\ }\textbf {\bibinfo {volume} {340}},\ \bibinfo {pages} {309}
  (\bibinfo {year} {1995})}\BibitemShut {NoStop}%
\bibitem [{\citenamefont {Paier}\ \emph {et~al.}(2006)\citenamefont {Paier},
  \citenamefont {Marsman}, \citenamefont {Hummer}, \citenamefont {Kresse},
  \citenamefont {Gerber},\ and\ \citenamefont {\'{A}ngy\'{a}n}}]{PaierJCP06}%
  \BibitemOpen
  \bibfield  {author} {\bibinfo {author} {\bibfnamefont {J.}~\bibnamefont
  {Paier}}, \bibinfo {author} {\bibfnamefont {M.}~\bibnamefont {Marsman}},
  \bibinfo {author} {\bibfnamefont {K.}~\bibnamefont {Hummer}}, \bibinfo
  {author} {\bibfnamefont {G.}~\bibnamefont {Kresse}}, \bibinfo {author}
  {\bibfnamefont {I.~C.}\ \bibnamefont {Gerber}}, \ and\ \bibinfo {author}
  {\bibfnamefont {J.~G.}\ \bibnamefont {\'{A}ngy\'{a}n}},\ }\href@noop {}
  {\bibfield  {journal} {\bibinfo  {journal} {J. Chem. Phys.}\ }\textbf
  {\bibinfo {volume} {124}},\ \bibinfo {pages} {154709} (\bibinfo {year}
  {2006})},\ \bibinfo {note} {\textbf{125}, 249901 (2006)}\BibitemShut
  {NoStop}%
\bibitem [{\citenamefont {Jang}\ and\ \citenamefont {Yu}(2012)}]{JangJPSJ12}%
  \BibitemOpen
  \bibfield  {author} {\bibinfo {author} {\bibfnamefont {Y.-R.}\ \bibnamefont
  {Jang}}\ and\ \bibinfo {author} {\bibfnamefont {B.~D.}\ \bibnamefont {Yu}},\
  }\href@noop {} {\bibfield  {journal} {\bibinfo  {journal} {J. Phys. Soc.
  Jpn.}\ }\textbf {\bibinfo {volume} {81}},\ \bibinfo {pages} {114715}
  (\bibinfo {year} {2012})}\BibitemShut {NoStop}%
\bibitem [{\citenamefont {Janthon}\ \emph {et~al.}(2014)\citenamefont
  {Janthon}, \citenamefont {Luo}, \citenamefont {Kozlov}, \citenamefont
  {Vi\~{n}es}, \citenamefont {Limtrakul}, \citenamefont {Truhlar},\ and\
  \citenamefont {Illas}}]{JanthonJCTC14}%
  \BibitemOpen
  \bibfield  {author} {\bibinfo {author} {\bibfnamefont {P.}~\bibnamefont
  {Janthon}}, \bibinfo {author} {\bibfnamefont {S.}~\bibnamefont {Luo}},
  \bibinfo {author} {\bibfnamefont {S.~M.}\ \bibnamefont {Kozlov}}, \bibinfo
  {author} {\bibfnamefont {F.}~\bibnamefont {Vi\~{n}es}}, \bibinfo {author}
  {\bibfnamefont {J.}~\bibnamefont {Limtrakul}}, \bibinfo {author}
  {\bibfnamefont {D.~G.}\ \bibnamefont {Truhlar}}, \ and\ \bibinfo {author}
  {\bibfnamefont {F.}~\bibnamefont {Illas}},\ }\href@noop {} {\bibfield
  {journal} {\bibinfo  {journal} {J. Chem. Theory Comput.}\ }\textbf {\bibinfo
  {volume} {10}},\ \bibinfo {pages} {3832} (\bibinfo {year}
  {2014})}\BibitemShut {NoStop}%
\bibitem [{\citenamefont {Gao}\ \emph {et~al.}(2016)\citenamefont {Gao},
  \citenamefont {Abtew}, \citenamefont {Cai}, \citenamefont {Sun},
  \citenamefont {Zhang},\ and\ \citenamefont {Zhang}}]{GaoSSC16}%
  \BibitemOpen
  \bibfield  {author} {\bibinfo {author} {\bibfnamefont {W.}~\bibnamefont
  {Gao}}, \bibinfo {author} {\bibfnamefont {T.~A.}\ \bibnamefont {Abtew}},
  \bibinfo {author} {\bibfnamefont {T.}~\bibnamefont {Cai}}, \bibinfo {author}
  {\bibfnamefont {Y.-Y.}\ \bibnamefont {Sun}}, \bibinfo {author} {\bibfnamefont
  {S.}~\bibnamefont {Zhang}}, \ and\ \bibinfo {author} {\bibfnamefont
  {P.}~\bibnamefont {Zhang}},\ }\href@noop {} {\bibfield  {journal} {\bibinfo
  {journal} {Solid State Commun.}\ }\textbf {\bibinfo {volume} {234-235}},\
  \bibinfo {pages} {10} (\bibinfo {year} {2016})}\BibitemShut {NoStop}%
\bibitem [{\citenamefont {Svane}\ and\ \citenamefont
  {Gunnarsson}(1990)}]{SvanePRL90}%
  \BibitemOpen
  \bibfield  {author} {\bibinfo {author} {\bibfnamefont {A.}~\bibnamefont
  {Svane}}\ and\ \bibinfo {author} {\bibfnamefont {O.}~\bibnamefont
  {Gunnarsson}},\ }\href@noop {} {\bibfield  {journal} {\bibinfo  {journal}
  {Phys. Rev. Lett.}\ }\textbf {\bibinfo {volume} {65}},\ \bibinfo {pages}
  {1148} (\bibinfo {year} {1990})}\BibitemShut {NoStop}%
\bibitem [{\citenamefont {Tran}\ \emph {et~al.}(2006)\citenamefont {Tran},
  \citenamefont {Blaha}, \citenamefont {Schwarz},\ and\ \citenamefont
  {Nov\'{a}k}}]{TranPRB06}%
  \BibitemOpen
  \bibfield  {author} {\bibinfo {author} {\bibfnamefont {F.}~\bibnamefont
  {Tran}}, \bibinfo {author} {\bibfnamefont {P.}~\bibnamefont {Blaha}},
  \bibinfo {author} {\bibfnamefont {K.}~\bibnamefont {Schwarz}}, \ and\
  \bibinfo {author} {\bibfnamefont {P.}~\bibnamefont {Nov\'{a}k}},\ }\href@noop
  {} {\bibfield  {journal} {\bibinfo  {journal} {Phys. Rev. B}\ }\textbf
  {\bibinfo {volume} {74}},\ \bibinfo {pages} {155108} (\bibinfo {year}
  {2006})}\BibitemShut {NoStop}%
\bibitem [{\citenamefont {Radwanski}\ and\ \citenamefont
  {Ropka}(2008)}]{RadwanskiPB08}%
  \BibitemOpen
  \bibfield  {author} {\bibinfo {author} {\bibfnamefont {R.~J.}\ \bibnamefont
  {Radwanski}}\ and\ \bibinfo {author} {\bibfnamefont {Z.}~\bibnamefont
  {Ropka}},\ }\href@noop {} {\bibfield  {journal} {\bibinfo  {journal} {Physica
  B}\ }\textbf {\bibinfo {volume} {403}},\ \bibinfo {pages} {1453} (\bibinfo
  {year} {2008})}\BibitemShut {NoStop}%
\bibitem [{\citenamefont {Schr\"{o}n}\ and\ \citenamefont
  {Bechstedt}(2013)}]{SchronJPCM13}%
  \BibitemOpen
  \bibfield  {author} {\bibinfo {author} {\bibfnamefont {A.}~\bibnamefont
  {Schr\"{o}n}}\ and\ \bibinfo {author} {\bibfnamefont {F.}~\bibnamefont
  {Bechstedt}},\ }\href@noop {} {\bibfield  {journal} {\bibinfo  {journal} {J.
  Phys.: Condens. Matter}\ }\textbf {\bibinfo {volume} {25}},\ \bibinfo {pages}
  {486002} (\bibinfo {year} {2013})}\BibitemShut {NoStop}%
\bibitem [{\citenamefont {Solovyev}\ \emph {et~al.}(1998)\citenamefont
  {Solovyev}, \citenamefont {Liechtenstein},\ and\ \citenamefont
  {Terakura}}]{SolovyevPRL98}%
  \BibitemOpen
  \bibfield  {author} {\bibinfo {author} {\bibfnamefont {I.~V.}\ \bibnamefont
  {Solovyev}}, \bibinfo {author} {\bibfnamefont {A.~I.}\ \bibnamefont
  {Liechtenstein}}, \ and\ \bibinfo {author} {\bibfnamefont {K.}~\bibnamefont
  {Terakura}},\ }\href@noop {} {\bibfield  {journal} {\bibinfo  {journal}
  {Phys. Rev. Lett.}\ }\textbf {\bibinfo {volume} {80}},\ \bibinfo {pages}
  {5758} (\bibinfo {year} {1998})}\BibitemShut {NoStop}%
\bibitem [{\citenamefont {Shishidou}\ and\ \citenamefont
  {Jo}(1998)}]{ShishidouJPSJ98}%
  \BibitemOpen
  \bibfield  {author} {\bibinfo {author} {\bibfnamefont {T.}~\bibnamefont
  {Shishidou}}\ and\ \bibinfo {author} {\bibfnamefont {T.}~\bibnamefont {Jo}},\
  }\href@noop {} {\bibfield  {journal} {\bibinfo  {journal} {J. Phys. Soc.
  Jpn.}\ }\textbf {\bibinfo {volume} {67}},\ \bibinfo {pages} {2637} (\bibinfo
  {year} {1998})}\BibitemShut {NoStop}%
\bibitem [{\citenamefont {Neubeck}\ \emph {et~al.}(2001)\citenamefont
  {Neubeck}, \citenamefont {Vettier}, \citenamefont {de~Bergevin},
  \citenamefont {Yakhou}, \citenamefont {Mannix}, \citenamefont {Ranno},\ and\
  \citenamefont {Chatterji}}]{NeubeckJPCS01}%
  \BibitemOpen
  \bibfield  {author} {\bibinfo {author} {\bibfnamefont {W.}~\bibnamefont
  {Neubeck}}, \bibinfo {author} {\bibfnamefont {C.}~\bibnamefont {Vettier}},
  \bibinfo {author} {\bibfnamefont {F.}~\bibnamefont {de~Bergevin}}, \bibinfo
  {author} {\bibfnamefont {F.}~\bibnamefont {Yakhou}}, \bibinfo {author}
  {\bibfnamefont {D.}~\bibnamefont {Mannix}}, \bibinfo {author} {\bibfnamefont
  {L.}~\bibnamefont {Ranno}}, \ and\ \bibinfo {author} {\bibfnamefont
  {T.}~\bibnamefont {Chatterji}},\ }\href@noop {} {\bibfield  {journal}
  {\bibinfo  {journal} {J. Phys. Chem. Solids}\ }\textbf {\bibinfo {volume}
  {62}},\ \bibinfo {pages} {2173} (\bibinfo {year} {2001})}\BibitemShut
  {NoStop}%
\bibitem [{\citenamefont {Jauch}\ and\ \citenamefont
  {Reehuis}(2002)}]{JauchPRB02}%
  \BibitemOpen
  \bibfield  {author} {\bibinfo {author} {\bibfnamefont {W.}~\bibnamefont
  {Jauch}}\ and\ \bibinfo {author} {\bibfnamefont {M.}~\bibnamefont
  {Reehuis}},\ }\href@noop {} {\bibfield  {journal} {\bibinfo  {journal} {Phys.
  Rev. B}\ }\textbf {\bibinfo {volume} {65}},\ \bibinfo {pages} {125111}
  (\bibinfo {year} {2002})}\BibitemShut {NoStop}%
\bibitem [{\citenamefont {Ghiringhelli}\ \emph {et~al.}(2002)\citenamefont
  {Ghiringhelli}, \citenamefont {Tjeng}, \citenamefont {Tanaka}, \citenamefont
  {Tjernberg}, \citenamefont {Mizokawa}, \citenamefont {de~Boer},\ and\
  \citenamefont {Brookes}}]{GhiringhelliPRB02}%
  \BibitemOpen
  \bibfield  {author} {\bibinfo {author} {\bibfnamefont {G.}~\bibnamefont
  {Ghiringhelli}}, \bibinfo {author} {\bibfnamefont {L.~H.}\ \bibnamefont
  {Tjeng}}, \bibinfo {author} {\bibfnamefont {A.}~\bibnamefont {Tanaka}},
  \bibinfo {author} {\bibfnamefont {O.}~\bibnamefont {Tjernberg}}, \bibinfo
  {author} {\bibfnamefont {T.}~\bibnamefont {Mizokawa}}, \bibinfo {author}
  {\bibfnamefont {J.~L.}\ \bibnamefont {de~Boer}}, \ and\ \bibinfo {author}
  {\bibfnamefont {N.~B.}\ \bibnamefont {Brookes}},\ }\href@noop {} {\bibfield
  {journal} {\bibinfo  {journal} {Phys. Rev. B}\ }\textbf {\bibinfo {volume}
  {66}},\ \bibinfo {pages} {075101} (\bibinfo {year} {2002})}\BibitemShut
  {NoStop}%
\bibitem [{\citenamefont {Radwanski}\ and\ \citenamefont
  {Ropka}(2004)}]{RadwanskiPB04}%
  \BibitemOpen
  \bibfield  {author} {\bibinfo {author} {\bibfnamefont {R.~J.}\ \bibnamefont
  {Radwanski}}\ and\ \bibinfo {author} {\bibfnamefont {Z.}~\bibnamefont
  {Ropka}},\ }\href@noop {} {\bibfield  {journal} {\bibinfo  {journal} {Physica
  B}\ }\textbf {\bibinfo {volume} {345}},\ \bibinfo {pages} {107} (\bibinfo
  {year} {2004})}\BibitemShut {NoStop}%
\bibitem [{\citenamefont {Boussendel}\ \emph {et~al.}(2010)\citenamefont
  {Boussendel}, \citenamefont {Baadji}, \citenamefont {Haroun}, \citenamefont
  {Dreyss\'e},\ and\ \citenamefont {Alouani}}]{BoussendelPRB10}%
  \BibitemOpen
  \bibfield  {author} {\bibinfo {author} {\bibfnamefont {A.}~\bibnamefont
  {Boussendel}}, \bibinfo {author} {\bibfnamefont {N.}~\bibnamefont {Baadji}},
  \bibinfo {author} {\bibfnamefont {A.}~\bibnamefont {Haroun}}, \bibinfo
  {author} {\bibfnamefont {H.}~\bibnamefont {Dreyss\'e}}, \ and\ \bibinfo
  {author} {\bibfnamefont {M.}~\bibnamefont {Alouani}},\ }\href@noop {}
  {\bibfield  {journal} {\bibinfo  {journal} {Phys. Rev. B}\ }\textbf {\bibinfo
  {volume} {81}},\ \bibinfo {pages} {184432} (\bibinfo {year}
  {2010})}\BibitemShut {NoStop}%
\bibitem [{\citenamefont {Fernandez}\ \emph {et~al.}(1998)\citenamefont
  {Fernandez}, \citenamefont {Vettier}, \citenamefont {de~Bergevin},
  \citenamefont {Giles},\ and\ \citenamefont {Neubeck}}]{FernandezPRB98}%
  \BibitemOpen
  \bibfield  {author} {\bibinfo {author} {\bibfnamefont {V.}~\bibnamefont
  {Fernandez}}, \bibinfo {author} {\bibfnamefont {C.}~\bibnamefont {Vettier}},
  \bibinfo {author} {\bibfnamefont {F.}~\bibnamefont {de~Bergevin}}, \bibinfo
  {author} {\bibfnamefont {C.}~\bibnamefont {Giles}}, \ and\ \bibinfo {author}
  {\bibfnamefont {W.}~\bibnamefont {Neubeck}},\ }\href@noop {} {\bibfield
  {journal} {\bibinfo  {journal} {Phys. Rev. B}\ }\textbf {\bibinfo {volume}
  {57}},\ \bibinfo {pages} {7870} (\bibinfo {year} {1998})}\BibitemShut
  {NoStop}%
\bibitem [{\citenamefont {Cheetham}\ and\ \citenamefont
  {Hope}(1983)}]{CheethamPRB83}%
  \BibitemOpen
  \bibfield  {author} {\bibinfo {author} {\bibfnamefont {A.~K.}\ \bibnamefont
  {Cheetham}}\ and\ \bibinfo {author} {\bibfnamefont {D.~A.~O.}\ \bibnamefont
  {Hope}},\ }\href@noop {} {\bibfield  {journal} {\bibinfo  {journal} {Phys.
  Rev. B}\ }\textbf {\bibinfo {volume} {27}},\ \bibinfo {pages} {6964}
  (\bibinfo {year} {1983})}\BibitemShut {NoStop}%
\bibitem [{\citenamefont {Roth}(1958)}]{RothPR58}%
  \BibitemOpen
  \bibfield  {author} {\bibinfo {author} {\bibfnamefont {W.~L.}\ \bibnamefont
  {Roth}},\ }\href@noop {} {\bibfield  {journal} {\bibinfo  {journal} {Phys.
  Rev.}\ }\textbf {\bibinfo {volume} {110}},\ \bibinfo {pages} {1333} (\bibinfo
  {year} {1958})}\BibitemShut {NoStop}%
\bibitem [{\citenamefont {Battle}\ and\ \citenamefont
  {Cheetham}(1979)}]{BattleJPC79}%
  \BibitemOpen
  \bibfield  {author} {\bibinfo {author} {\bibfnamefont {P.~D.}\ \bibnamefont
  {Battle}}\ and\ \bibinfo {author} {\bibfnamefont {A.~K.}\ \bibnamefont
  {Cheetham}},\ }\href@noop {} {\bibfield  {journal} {\bibinfo  {journal} {J.
  Phys. C: Solid State Phys.}\ }\textbf {\bibinfo {volume} {12}},\ \bibinfo
  {pages} {337} (\bibinfo {year} {1979})}\BibitemShut {NoStop}%
\bibitem [{\citenamefont {Fjellv\r{a}g}\ \emph {et~al.}(1996)\citenamefont
  {Fjellv\r{a}g}, \citenamefont {Gr{\o}nvold}, \citenamefont {St{\o}len},\ and\
  \citenamefont {Hauback}}]{FjellvagJSSC96}%
  \BibitemOpen
  \bibfield  {author} {\bibinfo {author} {\bibfnamefont {H.}~\bibnamefont
  {Fjellv\r{a}g}}, \bibinfo {author} {\bibfnamefont {F.}~\bibnamefont
  {Gr{\o}nvold}}, \bibinfo {author} {\bibfnamefont {S.}~\bibnamefont
  {St{\o}len}}, \ and\ \bibinfo {author} {\bibfnamefont {B.}~\bibnamefont
  {Hauback}},\ }\href@noop {} {\bibfield  {journal} {\bibinfo  {journal} {J.
  Solid State Chem.}\ }\textbf {\bibinfo {volume} {124}},\ \bibinfo {pages}
  {52} (\bibinfo {year} {1996})}\BibitemShut {NoStop}%
\bibitem [{\citenamefont {Khan}\ and\ \citenamefont
  {Erickson}(1970)}]{KhanPRB70}%
  \BibitemOpen
  \bibfield  {author} {\bibinfo {author} {\bibfnamefont {D.~C.}\ \bibnamefont
  {Khan}}\ and\ \bibinfo {author} {\bibfnamefont {R.~A.}\ \bibnamefont
  {Erickson}},\ }\href@noop {} {\bibfield  {journal} {\bibinfo  {journal}
  {Phys. Rev. B}\ }\textbf {\bibinfo {volume} {1}},\ \bibinfo {pages} {2243}
  (\bibinfo {year} {1970})}\BibitemShut {NoStop}%
\bibitem [{\citenamefont {Herrmann-Ronzaud}\ \emph {et~al.}(1978)\citenamefont
  {Herrmann-Ronzaud}, \citenamefont {Burlet},\ and\ \citenamefont
  {Rossat-Mignod}}]{HerrmannRonzaudJPC78}%
  \BibitemOpen
  \bibfield  {author} {\bibinfo {author} {\bibfnamefont {D.}~\bibnamefont
  {Herrmann-Ronzaud}}, \bibinfo {author} {\bibfnamefont {P.}~\bibnamefont
  {Burlet}}, \ and\ \bibinfo {author} {\bibfnamefont {J.}~\bibnamefont
  {Rossat-Mignod}},\ }\href@noop {} {\bibfield  {journal} {\bibinfo  {journal}
  {J. Phys. C: Solid State Phys.}\ }\textbf {\bibinfo {volume} {11}},\ \bibinfo
  {pages} {2123} (\bibinfo {year} {1978})}\BibitemShut {NoStop}%
\bibitem [{\citenamefont {Jauch}\ \emph {et~al.}(2001)\citenamefont {Jauch},
  \citenamefont {Reehuis}, \citenamefont {Bleif}, \citenamefont {Kubanek},\
  and\ \citenamefont {Pattison}}]{JauchPRB01}%
  \BibitemOpen
  \bibfield  {author} {\bibinfo {author} {\bibfnamefont {W.}~\bibnamefont
  {Jauch}}, \bibinfo {author} {\bibfnamefont {M.}~\bibnamefont {Reehuis}},
  \bibinfo {author} {\bibfnamefont {H.~J.}\ \bibnamefont {Bleif}}, \bibinfo
  {author} {\bibfnamefont {F.}~\bibnamefont {Kubanek}}, \ and\ \bibinfo
  {author} {\bibfnamefont {P.}~\bibnamefont {Pattison}},\ }\href@noop {}
  {\bibfield  {journal} {\bibinfo  {journal} {Phys. Rev. B}\ }\textbf {\bibinfo
  {volume} {64}},\ \bibinfo {pages} {052102} (\bibinfo {year}
  {2001})}\BibitemShut {NoStop}%
\bibitem [{\citenamefont {Neubeck}\ \emph {et~al.}(1999)\citenamefont
  {Neubeck}, \citenamefont {Vettier}, \citenamefont {Fernandez}, \citenamefont
  {de~Bergevin},\ and\ \citenamefont {Giles}}]{NeubeckJAP99}%
  \BibitemOpen
  \bibfield  {author} {\bibinfo {author} {\bibfnamefont {W.}~\bibnamefont
  {Neubeck}}, \bibinfo {author} {\bibfnamefont {C.}~\bibnamefont {Vettier}},
  \bibinfo {author} {\bibfnamefont {V.}~\bibnamefont {Fernandez}}, \bibinfo
  {author} {\bibfnamefont {F.}~\bibnamefont {de~Bergevin}}, \ and\ \bibinfo
  {author} {\bibfnamefont {C.}~\bibnamefont {Giles}},\ }\href@noop {}
  {\bibfield  {journal} {\bibinfo  {journal} {J. Appl. Phys.}\ }\textbf
  {\bibinfo {volume} {85}},\ \bibinfo {pages} {4847} (\bibinfo {year}
  {1999})}\BibitemShut {NoStop}%
\bibitem [{\citenamefont {Forsyth}\ \emph {et~al.}(1988)\citenamefont
  {Forsyth}, \citenamefont {Brown},\ and\ \citenamefont
  {Wanklyn}}]{ForsythJPC88}%
  \BibitemOpen
  \bibfield  {author} {\bibinfo {author} {\bibfnamefont {J.~B.}\ \bibnamefont
  {Forsyth}}, \bibinfo {author} {\bibfnamefont {P.~J.}\ \bibnamefont {Brown}},
  \ and\ \bibinfo {author} {\bibfnamefont {B.~M.}\ \bibnamefont {Wanklyn}},\
  }\href@noop {} {\bibfield  {journal} {\bibinfo  {journal} {J. Phys. C: Solid
  State Phys.}\ }\textbf {\bibinfo {volume} {21}},\ \bibinfo {pages} {2917}
  (\bibinfo {year} {1988})}\BibitemShut {NoStop}%
\bibitem [{\citenamefont {Golosova}\ \emph {et~al.}(2017)\citenamefont
  {Golosova}, \citenamefont {Kozlenko}, \citenamefont {Kichanov}, \citenamefont
  {Lukin}, \citenamefont {Liermann}, \citenamefont {Glazyrin},\ and\
  \citenamefont {Savenko}}]{GolosovaJAC17}%
  \BibitemOpen
  \bibfield  {author} {\bibinfo {author} {\bibfnamefont {N.~O.}\ \bibnamefont
  {Golosova}}, \bibinfo {author} {\bibfnamefont {D.~P.}\ \bibnamefont
  {Kozlenko}}, \bibinfo {author} {\bibfnamefont {S.~E.}\ \bibnamefont
  {Kichanov}}, \bibinfo {author} {\bibfnamefont {E.~V.}\ \bibnamefont {Lukin}},
  \bibinfo {author} {\bibfnamefont {H.-P.}\ \bibnamefont {Liermann}}, \bibinfo
  {author} {\bibfnamefont {K.~V.}\ \bibnamefont {Glazyrin}}, \ and\ \bibinfo
  {author} {\bibfnamefont {B.~N.}\ \bibnamefont {Savenko}},\ }\href@noop {}
  {\bibfield  {journal} {\bibinfo  {journal} {J. Alloys Compd.}\ }\textbf
  {\bibinfo {volume} {722}},\ \bibinfo {pages} {593} (\bibinfo {year}
  {2017})}\BibitemShut {NoStop}%
\bibitem [{\citenamefont {Brown}\ \emph {et~al.}(2002)\citenamefont {Brown},
  \citenamefont {Forsyth}, \citenamefont {Leli\`{e}vre-Berna},\ and\
  \citenamefont {Tasset}}]{BrownJPCM02}%
  \BibitemOpen
  \bibfield  {author} {\bibinfo {author} {\bibfnamefont {P.~J.}\ \bibnamefont
  {Brown}}, \bibinfo {author} {\bibfnamefont {J.~B.}\ \bibnamefont {Forsyth}},
  \bibinfo {author} {\bibfnamefont {E.}~\bibnamefont {Leli\`{e}vre-Berna}}, \
  and\ \bibinfo {author} {\bibfnamefont {F.}~\bibnamefont {Tasset}},\
  }\href@noop {} {\bibfield  {journal} {\bibinfo  {journal} {J. Phys.: Condens.
  Matter}\ }\textbf {\bibinfo {volume} {14}},\ \bibinfo {pages} {1957}
  (\bibinfo {year} {2002})}\BibitemShut {NoStop}%
\bibitem [{\citenamefont {Corliss}\ \emph {et~al.}(1965)\citenamefont
  {Corliss}, \citenamefont {Hastings}, \citenamefont {Nathans},\ and\
  \citenamefont {Shirane}}]{CorlissJAP65}%
  \BibitemOpen
  \bibfield  {author} {\bibinfo {author} {\bibfnamefont {L.~M.}\ \bibnamefont
  {Corliss}}, \bibinfo {author} {\bibfnamefont {J.~M.}\ \bibnamefont
  {Hastings}}, \bibinfo {author} {\bibfnamefont {R.}~\bibnamefont {Nathans}}, \
  and\ \bibinfo {author} {\bibfnamefont {G.}~\bibnamefont {Shirane}},\
  }\href@noop {} {\bibfield  {journal} {\bibinfo  {journal} {J. Appl. Phys.}\
  }\textbf {\bibinfo {volume} {36}},\ \bibinfo {pages} {1099} (\bibinfo {year}
  {1965})}\BibitemShut {NoStop}%
\bibitem [{\citenamefont {Baron}\ \emph {et~al.}(2005)\citenamefont {Baron},
  \citenamefont {Gutzmer}, \citenamefont {Rundl\"{o}f},\ and\ \citenamefont
  {Tellgren}}]{BaronSSS05}%
  \BibitemOpen
  \bibfield  {author} {\bibinfo {author} {\bibfnamefont {V.}~\bibnamefont
  {Baron}}, \bibinfo {author} {\bibfnamefont {J.}~\bibnamefont {Gutzmer}},
  \bibinfo {author} {\bibfnamefont {H.}~\bibnamefont {Rundl\"{o}f}}, \ and\
  \bibinfo {author} {\bibfnamefont {R.}~\bibnamefont {Tellgren}},\ }\href@noop
  {} {\bibfield  {journal} {\bibinfo  {journal} {Solid State Sci.}\ }\textbf
  {\bibinfo {volume} {7}},\ \bibinfo {pages} {753} (\bibinfo {year}
  {2005})}\BibitemShut {NoStop}%
\bibitem [{\citenamefont {Hill}\ \emph {et~al.}(2008)\citenamefont {Hill},
  \citenamefont {Jiao}, \citenamefont {Bruce}, \citenamefont {Harrison},
  \citenamefont {Kockelmann},\ and\ \citenamefont {Ritter}}]{HillCM08}%
  \BibitemOpen
  \bibfield  {author} {\bibinfo {author} {\bibfnamefont {A.~H.}\ \bibnamefont
  {Hill}}, \bibinfo {author} {\bibfnamefont {F.}~\bibnamefont {Jiao}}, \bibinfo
  {author} {\bibfnamefont {P.~G.}\ \bibnamefont {Bruce}}, \bibinfo {author}
  {\bibfnamefont {A.}~\bibnamefont {Harrison}}, \bibinfo {author}
  {\bibfnamefont {W.}~\bibnamefont {Kockelmann}}, \ and\ \bibinfo {author}
  {\bibfnamefont {C.}~\bibnamefont {Ritter}},\ }\href@noop {} {\bibfield
  {journal} {\bibinfo  {journal} {Chem. Mater.}\ }\textbf {\bibinfo {volume}
  {20}},\ \bibinfo {pages} {4891} (\bibinfo {year} {2008})}\BibitemShut
  {NoStop}%
\bibitem [{\citenamefont {Dufek}\ \emph {et~al.}(1994)\citenamefont {Dufek},
  \citenamefont {Blaha}, \citenamefont {Sliwko},\ and\ \citenamefont
  {Schwarz}}]{DufekPRB94a}%
  \BibitemOpen
  \bibfield  {author} {\bibinfo {author} {\bibfnamefont {P.}~\bibnamefont
  {Dufek}}, \bibinfo {author} {\bibfnamefont {P.}~\bibnamefont {Blaha}},
  \bibinfo {author} {\bibfnamefont {V.}~\bibnamefont {Sliwko}}, \ and\ \bibinfo
  {author} {\bibfnamefont {K.}~\bibnamefont {Schwarz}},\ }\href@noop {}
  {\bibfield  {journal} {\bibinfo  {journal} {Phys. Rev. B}\ }\textbf {\bibinfo
  {volume} {49}},\ \bibinfo {pages} {10170} (\bibinfo {year}
  {1994})}\BibitemShut {NoStop}%
\bibitem [{\citenamefont {Anisimov}\ \emph {et~al.}(1991)\citenamefont
  {Anisimov}, \citenamefont {Zaanen},\ and\ \citenamefont
  {Andersen}}]{AnisimovPRB91}%
  \BibitemOpen
  \bibfield  {author} {\bibinfo {author} {\bibfnamefont {V.~I.}\ \bibnamefont
  {Anisimov}}, \bibinfo {author} {\bibfnamefont {J.}~\bibnamefont {Zaanen}}, \
  and\ \bibinfo {author} {\bibfnamefont {O.~K.}\ \bibnamefont {Andersen}},\
  }\href@noop {} {\bibfield  {journal} {\bibinfo  {journal} {Phys. Rev. B}\
  }\textbf {\bibinfo {volume} {44}},\ \bibinfo {pages} {943} (\bibinfo {year}
  {1991})}\BibitemShut {NoStop}%
\bibitem [{\citenamefont {Bredow}\ and\ \citenamefont
  {Gerson}(2000)}]{BredowPRB00}%
  \BibitemOpen
  \bibfield  {author} {\bibinfo {author} {\bibfnamefont {T.}~\bibnamefont
  {Bredow}}\ and\ \bibinfo {author} {\bibfnamefont {A.~R.}\ \bibnamefont
  {Gerson}},\ }\href@noop {} {\bibfield  {journal} {\bibinfo  {journal} {Phys.
  Rev. B}\ }\textbf {\bibinfo {volume} {61}},\ \bibinfo {pages} {5194}
  (\bibinfo {year} {2000})}\BibitemShut {NoStop}%
\bibitem [{\citenamefont {de~P.~R.~Moreira}\ \emph {et~al.}(2002)\citenamefont
  {de~P.~R.~Moreira}, \citenamefont {Illas},\ and\ \citenamefont
  {Martin}}]{MoreiraPRB02}%
  \BibitemOpen
  \bibfield  {author} {\bibinfo {author} {\bibfnamefont {I.}~\bibnamefont
  {de~P.~R.~Moreira}}, \bibinfo {author} {\bibfnamefont {F.}~\bibnamefont
  {Illas}}, \ and\ \bibinfo {author} {\bibfnamefont {R.~L.}\ \bibnamefont
  {Martin}},\ }\href@noop {} {\bibfield  {journal} {\bibinfo  {journal} {Phys.
  Rev. B}\ }\textbf {\bibinfo {volume} {65}},\ \bibinfo {pages} {155102}
  (\bibinfo {year} {2002})}\BibitemShut {NoStop}%
\bibitem [{\citenamefont {Franchini}\ \emph {et~al.}(2005)\citenamefont
  {Franchini}, \citenamefont {Bayer}, \citenamefont {Podloucky}, \citenamefont
  {Paier},\ and\ \citenamefont {Kresse}}]{FranchiniPRB05}%
  \BibitemOpen
  \bibfield  {author} {\bibinfo {author} {\bibfnamefont {C.}~\bibnamefont
  {Franchini}}, \bibinfo {author} {\bibfnamefont {V.}~\bibnamefont {Bayer}},
  \bibinfo {author} {\bibfnamefont {R.}~\bibnamefont {Podloucky}}, \bibinfo
  {author} {\bibfnamefont {J.}~\bibnamefont {Paier}}, \ and\ \bibinfo {author}
  {\bibfnamefont {G.}~\bibnamefont {Kresse}},\ }\href@noop {} {\bibfield
  {journal} {\bibinfo  {journal} {Phys. Rev. B}\ }\textbf {\bibinfo {volume}
  {72}},\ \bibinfo {pages} {045132} (\bibinfo {year} {2005})}\BibitemShut
  {NoStop}%
\bibitem [{\citenamefont {Marsman}\ \emph {et~al.}(2008)\citenamefont
  {Marsman}, \citenamefont {Paier}, \citenamefont {Stroppa},\ and\
  \citenamefont {Kresse}}]{MarsmanJPCM08}%
  \BibitemOpen
  \bibfield  {author} {\bibinfo {author} {\bibfnamefont {M.}~\bibnamefont
  {Marsman}}, \bibinfo {author} {\bibfnamefont {J.}~\bibnamefont {Paier}},
  \bibinfo {author} {\bibfnamefont {A.}~\bibnamefont {Stroppa}}, \ and\
  \bibinfo {author} {\bibfnamefont {G.}~\bibnamefont {Kresse}},\ }\href@noop {}
  {\bibfield  {journal} {\bibinfo  {journal} {J. Phys.: Condens. Matter}\
  }\textbf {\bibinfo {volume} {20}},\ \bibinfo {pages} {064201} (\bibinfo
  {year} {2008})}\BibitemShut {NoStop}%
\bibitem [{\citenamefont {Chen}\ \emph {et~al.}(1995)\citenamefont {Chen},
  \citenamefont {Idzerda}, \citenamefont {Lin}, \citenamefont {Smith},
  \citenamefont {Meigs}, \citenamefont {Chaban}, \citenamefont {Ho},
  \citenamefont {Pellegrin},\ and\ \citenamefont {Sette}}]{ChenPRL95}%
  \BibitemOpen
  \bibfield  {author} {\bibinfo {author} {\bibfnamefont {C.~T.}\ \bibnamefont
  {Chen}}, \bibinfo {author} {\bibfnamefont {Y.~U.}\ \bibnamefont {Idzerda}},
  \bibinfo {author} {\bibfnamefont {H.-J.}\ \bibnamefont {Lin}}, \bibinfo
  {author} {\bibfnamefont {N.~V.}\ \bibnamefont {Smith}}, \bibinfo {author}
  {\bibfnamefont {G.}~\bibnamefont {Meigs}}, \bibinfo {author} {\bibfnamefont
  {E.}~\bibnamefont {Chaban}}, \bibinfo {author} {\bibfnamefont {G.~H.}\
  \bibnamefont {Ho}}, \bibinfo {author} {\bibfnamefont {E.}~\bibnamefont
  {Pellegrin}}, \ and\ \bibinfo {author} {\bibfnamefont {F.}~\bibnamefont
  {Sette}},\ }\href@noop {} {\bibfield  {journal} {\bibinfo  {journal} {Phys.
  Rev. Lett.}\ }\textbf {\bibinfo {volume} {75}},\ \bibinfo {pages} {152}
  (\bibinfo {year} {1995})}\BibitemShut {NoStop}%
\bibitem [{\citenamefont {Scherz}(2003)}]{Scherz03}%
  \BibitemOpen
  \bibfield  {author} {\bibinfo {author} {\bibfnamefont {A.}~\bibnamefont
  {Scherz}},\ }\href@noop {} {}\bibinfo {howpublished} {Ph.D. thesis, Free
  University of Berlin} (\bibinfo {year} {2003})\BibitemShut {NoStop}%
\bibitem [{\citenamefont {Reck}\ and\ \citenamefont {Fry}(1969)}]{ReckPR69}%
  \BibitemOpen
  \bibfield  {author} {\bibinfo {author} {\bibfnamefont {R.~A.}\ \bibnamefont
  {Reck}}\ and\ \bibinfo {author} {\bibfnamefont {D.~L.}\ \bibnamefont {Fry}},\
  }\href@noop {} {\bibfield  {journal} {\bibinfo  {journal} {Phys. Rev.}\
  }\textbf {\bibinfo {volume} {184}},\ \bibinfo {pages} {492} (\bibinfo {year}
  {1969})}\BibitemShut {NoStop}%
\bibitem [{\citenamefont {Moon}(1964)}]{MoonPR64}%
  \BibitemOpen
  \bibfield  {author} {\bibinfo {author} {\bibfnamefont {R.~M.}\ \bibnamefont
  {Moon}},\ }\href@noop {} {\bibfield  {journal} {\bibinfo  {journal} {Phys.
  Rev.}\ }\textbf {\bibinfo {volume} {136}},\ \bibinfo {pages} {A195} (\bibinfo
  {year} {1964})}\BibitemShut {NoStop}%
\bibitem [{\citenamefont {Mook}(1966)}]{MookJAP66}%
  \BibitemOpen
  \bibfield  {author} {\bibinfo {author} {\bibfnamefont {H.~A.}\ \bibnamefont
  {Mook}},\ }\href@noop {} {\bibfield  {journal} {\bibinfo  {journal} {J. Appl.
  Phys.}\ }\textbf {\bibinfo {volume} {37}},\ \bibinfo {pages} {1034} (\bibinfo
  {year} {1966})}\BibitemShut {NoStop}%
\bibitem [{\citenamefont {Barbiellini}\ \emph {et~al.}(1990)\citenamefont
  {Barbiellini}, \citenamefont {Moroni},\ and\ \citenamefont
  {Jarlborg}}]{BarbielliniJPCM90}%
  \BibitemOpen
  \bibfield  {author} {\bibinfo {author} {\bibfnamefont {B.}~\bibnamefont
  {Barbiellini}}, \bibinfo {author} {\bibfnamefont {E.~G.}\ \bibnamefont
  {Moroni}}, \ and\ \bibinfo {author} {\bibfnamefont {T.}~\bibnamefont
  {Jarlborg}},\ }\href@noop {} {\bibfield  {journal} {\bibinfo  {journal} {J.
  Phys.: Condens. Matter}\ }\textbf {\bibinfo {volume} {2}},\ \bibinfo {pages}
  {7597} (\bibinfo {year} {1990})}\BibitemShut {NoStop}%
\bibitem [{\citenamefont {Singh}\ \emph {et~al.}(1991)\citenamefont {Singh},
  \citenamefont {Pickett},\ and\ \citenamefont {Krakauer}}]{SinghPRB91b}%
  \BibitemOpen
  \bibfield  {author} {\bibinfo {author} {\bibfnamefont {D.~J.}\ \bibnamefont
  {Singh}}, \bibinfo {author} {\bibfnamefont {W.~E.}\ \bibnamefont {Pickett}},
  \ and\ \bibinfo {author} {\bibfnamefont {H.}~\bibnamefont {Krakauer}},\
  }\href@noop {} {\bibfield  {journal} {\bibinfo  {journal} {Phys. Rev. B}\
  }\textbf {\bibinfo {volume} {43}},\ \bibinfo {pages} {11628} (\bibinfo {year}
  {1991})}\BibitemShut {NoStop}%
\bibitem [{\citenamefont {Leung}\ \emph {et~al.}(1991)\citenamefont {Leung},
  \citenamefont {Chan},\ and\ \citenamefont {Harmon}}]{LeungPRB91}%
  \BibitemOpen
  \bibfield  {author} {\bibinfo {author} {\bibfnamefont {T.~C.}\ \bibnamefont
  {Leung}}, \bibinfo {author} {\bibfnamefont {C.~T.}\ \bibnamefont {Chan}}, \
  and\ \bibinfo {author} {\bibfnamefont {B.~N.}\ \bibnamefont {Harmon}},\
  }\href@noop {} {\bibfield  {journal} {\bibinfo  {journal} {Phys. Rev. B}\
  }\textbf {\bibinfo {volume} {44}},\ \bibinfo {pages} {2923} (\bibinfo {year}
  {1991})}\BibitemShut {NoStop}%
\bibitem [{\citenamefont {Amador}\ \emph {et~al.}(1992)\citenamefont {Amador},
  \citenamefont {Lambrecht},\ and\ \citenamefont {Segall}}]{AmadorPRB92}%
  \BibitemOpen
  \bibfield  {author} {\bibinfo {author} {\bibfnamefont {C.}~\bibnamefont
  {Amador}}, \bibinfo {author} {\bibfnamefont {W.~R.~L.}\ \bibnamefont
  {Lambrecht}}, \ and\ \bibinfo {author} {\bibfnamefont {B.}~\bibnamefont
  {Segall}},\ }\href@noop {} {\bibfield  {journal} {\bibinfo  {journal} {Phys.
  Rev. B}\ }\textbf {\bibinfo {volume} {46}},\ \bibinfo {pages} {1870}
  (\bibinfo {year} {1992})}\BibitemShut {NoStop}%
\bibitem [{\citenamefont {Tran}\ \emph {et~al.}(2012)\citenamefont {Tran},
  \citenamefont {Koller},\ and\ \citenamefont {Blaha}}]{TranPRB12}%
  \BibitemOpen
  \bibfield  {author} {\bibinfo {author} {\bibfnamefont {F.}~\bibnamefont
  {Tran}}, \bibinfo {author} {\bibfnamefont {D.}~\bibnamefont {Koller}}, \ and\
  \bibinfo {author} {\bibfnamefont {P.}~\bibnamefont {Blaha}},\ }\href@noop {}
  {\bibfield  {journal} {\bibinfo  {journal} {Phys. Rev. B}\ }\textbf {\bibinfo
  {volume} {86}},\ \bibinfo {pages} {134406} (\bibinfo {year}
  {2012})}\BibitemShut {NoStop}%
\bibitem [{\citenamefont {Kotani}\ and\ \citenamefont
  {Akai}(1998)}]{KotaniJMMM98}%
  \BibitemOpen
  \bibfield  {author} {\bibinfo {author} {\bibfnamefont {T.}~\bibnamefont
  {Kotani}}\ and\ \bibinfo {author} {\bibfnamefont {H.}~\bibnamefont {Akai}},\
  }\href@noop {} {\bibfield  {journal} {\bibinfo  {journal} {J. Magn. Magn.
  Mater.}\ }\textbf {\bibinfo {volume} {177-181}},\ \bibinfo {pages} {569}
  (\bibinfo {year} {1998})}\BibitemShut {NoStop}%
\bibitem [{\citenamefont {Schnell}\ \emph {et~al.}(2003)\citenamefont
  {Schnell}, \citenamefont {Czycholl},\ and\ \citenamefont
  {Albers}}]{SchnellPRB03}%
  \BibitemOpen
  \bibfield  {author} {\bibinfo {author} {\bibfnamefont {I.}~\bibnamefont
  {Schnell}}, \bibinfo {author} {\bibfnamefont {G.}~\bibnamefont {Czycholl}}, \
  and\ \bibinfo {author} {\bibfnamefont {R.~C.}\ \bibnamefont {Albers}},\
  }\href@noop {} {\bibfield  {journal} {\bibinfo  {journal} {Phys. Rev. B}\
  }\textbf {\bibinfo {volume} {68}},\ \bibinfo {pages} {245102} (\bibinfo
  {year} {2003})}\BibitemShut {NoStop}%
\bibitem [{\citenamefont {Fukazawa}\ and\ \citenamefont
  {Akai}(2015)}]{FukazawaJPCM15}%
  \BibitemOpen
  \bibfield  {author} {\bibinfo {author} {\bibfnamefont {T.}~\bibnamefont
  {Fukazawa}}\ and\ \bibinfo {author} {\bibfnamefont {H.}~\bibnamefont
  {Akai}},\ }\href@noop {} {\bibfield  {journal} {\bibinfo  {journal} {J.
  Phys.: Condens. Matter}\ }\textbf {\bibinfo {volume} {27}},\ \bibinfo {pages}
  {115502} (\bibinfo {year} {2015})}\BibitemShut {NoStop}%
\bibitem [{\citenamefont {Kutepov}(2017)}]{KutepovJPCM17}%
  \BibitemOpen
  \bibfield  {author} {\bibinfo {author} {\bibfnamefont {A.~L.}\ \bibnamefont
  {Kutepov}},\ }\href@noop {} {\bibfield  {journal} {\bibinfo  {journal} {J.
  Phys.: Condens. Matter}\ }\textbf {\bibinfo {volume} {29}},\ \bibinfo {pages}
  {465503} (\bibinfo {year} {2017})}\BibitemShut {NoStop}%
\bibitem [{\citenamefont {van Schilfgaarde}\ \emph {et~al.}(2006)\citenamefont
  {van Schilfgaarde}, \citenamefont {Kotani},\ and\ \citenamefont
  {Faleev}}]{vanSchilfgaardePRL06}%
  \BibitemOpen
  \bibfield  {author} {\bibinfo {author} {\bibfnamefont {M.}~\bibnamefont {van
  Schilfgaarde}}, \bibinfo {author} {\bibfnamefont {T.}~\bibnamefont {Kotani}},
  \ and\ \bibinfo {author} {\bibfnamefont {S.}~\bibnamefont {Faleev}},\
  }\href@noop {} {\bibfield  {journal} {\bibinfo  {journal} {Phys. Rev. Lett.}\
  }\textbf {\bibinfo {volume} {96}},\ \bibinfo {pages} {226402} (\bibinfo
  {year} {2006})}\BibitemShut {NoStop}%
\bibitem [{\citenamefont {Stone}(2016)}]{StoneADNDT16}%
  \BibitemOpen
  \bibfield  {author} {\bibinfo {author} {\bibfnamefont {N.~J.}\ \bibnamefont
  {Stone}},\ }\href@noop {} {\bibfield  {journal} {\bibinfo  {journal} {At.
  Data Nucl. Data Tables}\ }\textbf {\bibinfo {volume} {111-112}},\ \bibinfo
  {pages} {1} (\bibinfo {year} {2016})}\BibitemShut {NoStop}%
\bibitem [{\citenamefont {Ebert}\ \emph {et~al.}(1986)\citenamefont {Ebert},
  \citenamefont {Abart},\ and\ \citenamefont {Voitl\"{a}nder}}]{EbertJPF86}%
  \BibitemOpen
  \bibfield  {author} {\bibinfo {author} {\bibfnamefont {H.}~\bibnamefont
  {Ebert}}, \bibinfo {author} {\bibfnamefont {J.}~\bibnamefont {Abart}}, \ and\
  \bibinfo {author} {\bibfnamefont {J.}~\bibnamefont {Voitl\"{a}nder}},\
  }\href@noop {} {\bibfield  {journal} {\bibinfo  {journal} {J. Phys. F: Met.
  Phys.}\ }\textbf {\bibinfo {volume} {16}},\ \bibinfo {pages} {1287} (\bibinfo
  {year} {1986})}\BibitemShut {NoStop}%
\bibitem [{\citenamefont {Vianden}(1987)}]{ViandenHI87}%
  \BibitemOpen
  \bibfield  {author} {\bibinfo {author} {\bibfnamefont {R.}~\bibnamefont
  {Vianden}},\ }\href@noop {} {\bibfield  {journal} {\bibinfo  {journal}
  {Hyperfine Interact.}\ }\textbf {\bibinfo {volume} {35}},\ \bibinfo {pages}
  {1079} (\bibinfo {year} {1987})}\BibitemShut {NoStop}%
\bibitem [{\citenamefont {Christiansen}\ \emph {et~al.}(1976)\citenamefont
  {Christiansen}, \citenamefont {Heubes}, \citenamefont {Keitel}, \citenamefont
  {Klinger}, \citenamefont {Loeffler}, \citenamefont {Sandner},\ and\
  \citenamefont {Witthuhn}}]{ChristiansenZPB76}%
  \BibitemOpen
  \bibfield  {author} {\bibinfo {author} {\bibfnamefont {J.}~\bibnamefont
  {Christiansen}}, \bibinfo {author} {\bibfnamefont {P.}~\bibnamefont
  {Heubes}}, \bibinfo {author} {\bibfnamefont {R.}~\bibnamefont {Keitel}},
  \bibinfo {author} {\bibfnamefont {W.}~\bibnamefont {Klinger}}, \bibinfo
  {author} {\bibfnamefont {W.}~\bibnamefont {Loeffler}}, \bibinfo {author}
  {\bibfnamefont {W.}~\bibnamefont {Sandner}}, \ and\ \bibinfo {author}
  {\bibfnamefont {W.}~\bibnamefont {Witthuhn}},\ }\href@noop {} {\bibfield
  {journal} {\bibinfo  {journal} {Z. Phys. B}\ }\textbf {\bibinfo {volume}
  {24}},\ \bibinfo {pages} {177} (\bibinfo {year} {1976})}\BibitemShut
  {NoStop}%
\bibitem [{\citenamefont {Graham}\ \emph {et~al.}(1991)\citenamefont {Graham},
  \citenamefont {Riedi},\ and\ \citenamefont {Wanklyn}}]{GrahamJPCM91}%
  \BibitemOpen
  \bibfield  {author} {\bibinfo {author} {\bibfnamefont {R.~G.}\ \bibnamefont
  {Graham}}, \bibinfo {author} {\bibfnamefont {P.~C.}\ \bibnamefont {Riedi}}, \
  and\ \bibinfo {author} {\bibfnamefont {B.~M.}\ \bibnamefont {Wanklyn}},\
  }\href@noop {} {\bibfield  {journal} {\bibinfo  {journal} {J. Phys.: Condens.
  Matter}\ }\textbf {\bibinfo {volume} {3}},\ \bibinfo {pages} {135} (\bibinfo
  {year} {1991})}\BibitemShut {NoStop}%
\bibitem [{\citenamefont {Azevedo}\ \emph {et~al.}(1981)\citenamefont
  {Azevedo}, \citenamefont {Schirber}, \citenamefont {Switendick},
  \citenamefont {Baughman},\ and\ \citenamefont {Morosin}}]{AzevedoJPF81}%
  \BibitemOpen
  \bibfield  {author} {\bibinfo {author} {\bibfnamefont {L.~J.}\ \bibnamefont
  {Azevedo}}, \bibinfo {author} {\bibfnamefont {J.~E.}\ \bibnamefont
  {Schirber}}, \bibinfo {author} {\bibfnamefont {A.~C.}\ \bibnamefont
  {Switendick}}, \bibinfo {author} {\bibfnamefont {R.~J.}\ \bibnamefont
  {Baughman}}, \ and\ \bibinfo {author} {\bibfnamefont {B.}~\bibnamefont
  {Morosin}},\ }\href@noop {} {\bibfield  {journal} {\bibinfo  {journal} {J.
  Phys. F: Met. Phys.}\ }\textbf {\bibinfo {volume} {11}},\ \bibinfo {pages}
  {1521} (\bibinfo {year} {1981})}\BibitemShut {NoStop}%
\bibitem [{\citenamefont {Tran}\ \emph
  {et~al.}(2015{\natexlab{a}})\citenamefont {Tran}, \citenamefont {Blaha},
  \citenamefont {Betzinger},\ and\ \citenamefont {Bl\"{u}gel}}]{TranPRB15}%
  \BibitemOpen
  \bibfield  {author} {\bibinfo {author} {\bibfnamefont {F.}~\bibnamefont
  {Tran}}, \bibinfo {author} {\bibfnamefont {P.}~\bibnamefont {Blaha}},
  \bibinfo {author} {\bibfnamefont {M.}~\bibnamefont {Betzinger}}, \ and\
  \bibinfo {author} {\bibfnamefont {S.}~\bibnamefont {Bl\"{u}gel}},\
  }\href@noop {} {\bibfield  {journal} {\bibinfo  {journal} {Phys. Rev. B}\
  }\textbf {\bibinfo {volume} {91}},\ \bibinfo {pages} {165121} (\bibinfo
  {year} {2015}{\natexlab{a}})}\BibitemShut {NoStop}%
\bibitem [{\citenamefont {Tran}\ \emph {et~al.}(2016)\citenamefont {Tran},
  \citenamefont {Blaha}, \citenamefont {Betzinger},\ and\ \citenamefont
  {Bl\"ugel}}]{TranPRB16}%
  \BibitemOpen
  \bibfield  {author} {\bibinfo {author} {\bibfnamefont {F.}~\bibnamefont
  {Tran}}, \bibinfo {author} {\bibfnamefont {P.}~\bibnamefont {Blaha}},
  \bibinfo {author} {\bibfnamefont {M.}~\bibnamefont {Betzinger}}, \ and\
  \bibinfo {author} {\bibfnamefont {S.}~\bibnamefont {Bl\"ugel}},\ }\href@noop
  {} {\bibfield  {journal} {\bibinfo  {journal} {Phys. Rev. B}\ }\textbf
  {\bibinfo {volume} {94}},\ \bibinfo {pages} {165149} (\bibinfo {year}
  {2016})}\BibitemShut {NoStop}%
\bibitem [{\citenamefont {Rocquefelte}\ \emph {et~al.}(2010)\citenamefont
  {Rocquefelte}, \citenamefont {Whangbo}, \citenamefont {Villesuzanne},
  \citenamefont {Jobic}, \citenamefont {Tran}, \citenamefont {Schwarz},\ and\
  \citenamefont {Blaha}}]{RocquefelteJPCM10}%
  \BibitemOpen
  \bibfield  {author} {\bibinfo {author} {\bibfnamefont {X.}~\bibnamefont
  {Rocquefelte}}, \bibinfo {author} {\bibfnamefont {M.-H.}\ \bibnamefont
  {Whangbo}}, \bibinfo {author} {\bibfnamefont {A.}~\bibnamefont
  {Villesuzanne}}, \bibinfo {author} {\bibfnamefont {S.}~\bibnamefont {Jobic}},
  \bibinfo {author} {\bibfnamefont {F.}~\bibnamefont {Tran}}, \bibinfo {author}
  {\bibfnamefont {K.}~\bibnamefont {Schwarz}}, \ and\ \bibinfo {author}
  {\bibfnamefont {P.}~\bibnamefont {Blaha}},\ }\href@noop {} {\bibfield
  {journal} {\bibinfo  {journal} {J. Phys.: Condens. Matter}\ }\textbf
  {\bibinfo {volume} {22}},\ \bibinfo {pages} {045502} (\bibinfo {year}
  {2010})}\BibitemShut {NoStop}%
\bibitem [{\citenamefont {Haas}\ and\ \citenamefont
  {Correia}(2007)}]{HaasHI07}%
  \BibitemOpen
  \bibfield  {author} {\bibinfo {author} {\bibfnamefont {H.}~\bibnamefont
  {Haas}}\ and\ \bibinfo {author} {\bibfnamefont {J.~G.}\ \bibnamefont
  {Correia}},\ }\href@noop {} {\bibfield  {journal} {\bibinfo  {journal}
  {Hyperfine Interact.}\ }\textbf {\bibinfo {volume} {176}},\ \bibinfo {pages}
  {9} (\bibinfo {year} {2007})}\BibitemShut {NoStop}%
\bibitem [{\citenamefont {Haas}\ \emph {et~al.}(2016)\citenamefont {Haas},
  \citenamefont {Barbosa},\ and\ \citenamefont {Correia}}]{HaasHI16}%
  \BibitemOpen
  \bibfield  {author} {\bibinfo {author} {\bibfnamefont {H.}~\bibnamefont
  {Haas}}, \bibinfo {author} {\bibfnamefont {M.~B.}\ \bibnamefont {Barbosa}}, \
  and\ \bibinfo {author} {\bibfnamefont {J.~G.}\ \bibnamefont {Correia}},\
  }\href@noop {} {\bibfield  {journal} {\bibinfo  {journal} {Hyperfine
  Interact.}\ }\textbf {\bibinfo {volume} {237}},\ \bibinfo {pages} {115}
  (\bibinfo {year} {2016})}\BibitemShut {NoStop}%
\bibitem [{\citenamefont {Blaha}\ \emph {et~al.}(1988)\citenamefont {Blaha},
  \citenamefont {Schwarz},\ and\ \citenamefont {Dederichs}}]{BlahaPRB88a}%
  \BibitemOpen
  \bibfield  {author} {\bibinfo {author} {\bibfnamefont {P.}~\bibnamefont
  {Blaha}}, \bibinfo {author} {\bibfnamefont {K.}~\bibnamefont {Schwarz}}, \
  and\ \bibinfo {author} {\bibfnamefont {P.~H.}\ \bibnamefont {Dederichs}},\
  }\href@noop {} {\bibfield  {journal} {\bibinfo  {journal} {Phys. Rev. B}\
  }\textbf {\bibinfo {volume} {37}},\ \bibinfo {pages} {2792} (\bibinfo {year}
  {1988})}\BibitemShut {NoStop}%
\bibitem [{\citenamefont {Zuo}\ \emph {et~al.}(1997)\citenamefont {Zuo},
  \citenamefont {Blaha},\ and\ \citenamefont {Schwarz}}]{ZuoJPCM97}%
  \BibitemOpen
  \bibfield  {author} {\bibinfo {author} {\bibfnamefont {J.~M.}\ \bibnamefont
  {Zuo}}, \bibinfo {author} {\bibfnamefont {P.}~\bibnamefont {Blaha}}, \ and\
  \bibinfo {author} {\bibfnamefont {K.}~\bibnamefont {Schwarz}},\ }\href@noop
  {} {\bibfield  {journal} {\bibinfo  {journal} {J. Phys.: Condens. Matter}\
  }\textbf {\bibinfo {volume} {9}},\ \bibinfo {pages} {7541} (\bibinfo {year}
  {1997})}\BibitemShut {NoStop}%
\bibitem [{\citenamefont {Saka}\ and\ \citenamefont {Kato}(1986)}]{SakaAC86}%
  \BibitemOpen
  \bibfield  {author} {\bibinfo {author} {\bibfnamefont {T.}~\bibnamefont
  {Saka}}\ and\ \bibinfo {author} {\bibfnamefont {N.}~\bibnamefont {Kato}},\
  }\href@noop {} {\bibfield  {journal} {\bibinfo  {journal} {Acta Cryst.}\
  }\textbf {\bibinfo {volume} {A42}},\ \bibinfo {pages} {469} (\bibinfo {year}
  {1986})}\BibitemShut {NoStop}%
\bibitem [{\citenamefont {Cumming}\ and\ \citenamefont
  {Hart}(1988)}]{CummingAJP88}%
  \BibitemOpen
  \bibfield  {author} {\bibinfo {author} {\bibfnamefont {S.}~\bibnamefont
  {Cumming}}\ and\ \bibinfo {author} {\bibfnamefont {M.}~\bibnamefont {Hart}},\
  }\href@noop {} {\bibfield  {journal} {\bibinfo  {journal} {Aust. J. Phys.}\
  }\textbf {\bibinfo {volume} {41}},\ \bibinfo {pages} {423} (\bibinfo {year}
  {1988})}\BibitemShut {NoStop}%
\bibitem [{\citenamefont {Lu}\ \emph {et~al.}(1993)\citenamefont {Lu},
  \citenamefont {Zunger},\ and\ \citenamefont {Deutsch}}]{LuPRB93}%
  \BibitemOpen
  \bibfield  {author} {\bibinfo {author} {\bibfnamefont {Z.~W.}\ \bibnamefont
  {Lu}}, \bibinfo {author} {\bibfnamefont {A.}~\bibnamefont {Zunger}}, \ and\
  \bibinfo {author} {\bibfnamefont {M.}~\bibnamefont {Deutsch}},\ }\href@noop
  {} {\bibfield  {journal} {\bibinfo  {journal} {Phys. Rev. B}\ }\textbf
  {\bibinfo {volume} {47}},\ \bibinfo {pages} {9385} (\bibinfo {year}
  {1993})}\BibitemShut {NoStop}%
\bibitem [{\citenamefont {St\"adele}\ \emph {et~al.}(1997)\citenamefont
  {St\"adele}, \citenamefont {Majewski}, \citenamefont {Vogl},\ and\
  \citenamefont {G\"orling}}]{StaedelePRL97}%
  \BibitemOpen
  \bibfield  {author} {\bibinfo {author} {\bibfnamefont {M.}~\bibnamefont
  {St\"adele}}, \bibinfo {author} {\bibfnamefont {J.~A.}\ \bibnamefont
  {Majewski}}, \bibinfo {author} {\bibfnamefont {P.}~\bibnamefont {Vogl}}, \
  and\ \bibinfo {author} {\bibfnamefont {A.}~\bibnamefont {G\"orling}},\
  }\href@noop {} {\bibfield  {journal} {\bibinfo  {journal} {Phys. Rev. Lett.}\
  }\textbf {\bibinfo {volume} {79}},\ \bibinfo {pages} {2089} (\bibinfo {year}
  {1997})}\BibitemShut {NoStop}%
\bibitem [{\citenamefont {St\"adele}\ \emph {et~al.}(1999)\citenamefont
  {St\"adele}, \citenamefont {Moukara}, \citenamefont {Majewski}, \citenamefont
  {Vogl},\ and\ \citenamefont {G\"orling}}]{StaedelePRB99}%
  \BibitemOpen
  \bibfield  {author} {\bibinfo {author} {\bibfnamefont {M.}~\bibnamefont
  {St\"adele}}, \bibinfo {author} {\bibfnamefont {M.}~\bibnamefont {Moukara}},
  \bibinfo {author} {\bibfnamefont {J.~A.}\ \bibnamefont {Majewski}}, \bibinfo
  {author} {\bibfnamefont {P.}~\bibnamefont {Vogl}}, \ and\ \bibinfo {author}
  {\bibfnamefont {A.}~\bibnamefont {G\"orling}},\ }\href@noop {} {\bibfield
  {journal} {\bibinfo  {journal} {Phys. Rev. B}\ }\textbf {\bibinfo {volume}
  {59}},\ \bibinfo {pages} {10031} (\bibinfo {year} {1999})}\BibitemShut
  {NoStop}%
\bibitem [{\citenamefont {Aulbur}\ \emph {et~al.}(2000)\citenamefont {Aulbur},
  \citenamefont {St\"adele},\ and\ \citenamefont {G\"orling}}]{AulburPRB00}%
  \BibitemOpen
  \bibfield  {author} {\bibinfo {author} {\bibfnamefont {W.~G.}\ \bibnamefont
  {Aulbur}}, \bibinfo {author} {\bibfnamefont {M.}~\bibnamefont {St\"adele}}, \
  and\ \bibinfo {author} {\bibfnamefont {A.}~\bibnamefont {G\"orling}},\
  }\href@noop {} {\bibfield  {journal} {\bibinfo  {journal} {Phys. Rev. B}\
  }\textbf {\bibinfo {volume} {62}},\ \bibinfo {pages} {7121} (\bibinfo {year}
  {2000})}\BibitemShut {NoStop}%
\bibitem [{\citenamefont {Qteish}\ \emph {et~al.}(2006)\citenamefont {Qteish},
  \citenamefont {Rinke}, \citenamefont {Scheffler},\ and\ \citenamefont
  {Neugebauer}}]{QteishPRB06}%
  \BibitemOpen
  \bibfield  {author} {\bibinfo {author} {\bibfnamefont {A.}~\bibnamefont
  {Qteish}}, \bibinfo {author} {\bibfnamefont {P.}~\bibnamefont {Rinke}},
  \bibinfo {author} {\bibfnamefont {M.}~\bibnamefont {Scheffler}}, \ and\
  \bibinfo {author} {\bibfnamefont {J.}~\bibnamefont {Neugebauer}},\
  }\href@noop {} {\bibfield  {journal} {\bibinfo  {journal} {Phys. Rev. B}\
  }\textbf {\bibinfo {volume} {74}},\ \bibinfo {pages} {245208} (\bibinfo
  {year} {2006})}\BibitemShut {NoStop}%
\bibitem [{\citenamefont {Betzinger}\ \emph {et~al.}(2011)\citenamefont
  {Betzinger}, \citenamefont {Friedrich}, \citenamefont {Bl\"ugel},\ and\
  \citenamefont {G\"orling}}]{BetzingerPRB11}%
  \BibitemOpen
  \bibfield  {author} {\bibinfo {author} {\bibfnamefont {M.}~\bibnamefont
  {Betzinger}}, \bibinfo {author} {\bibfnamefont {C.}~\bibnamefont
  {Friedrich}}, \bibinfo {author} {\bibfnamefont {S.}~\bibnamefont {Bl\"ugel}},
  \ and\ \bibinfo {author} {\bibfnamefont {A.}~\bibnamefont {G\"orling}},\
  }\href@noop {} {\bibfield  {journal} {\bibinfo  {journal} {Phys. Rev. B}\
  }\textbf {\bibinfo {volume} {83}},\ \bibinfo {pages} {045105} (\bibinfo
  {year} {2011})}\BibitemShut {NoStop}%
\bibitem [{\citenamefont {Tran}\ \emph
  {et~al.}(2015{\natexlab{b}})\citenamefont {Tran}, \citenamefont {Blaha},\
  and\ \citenamefont {Schwarz}}]{TranJCTC15}%
  \BibitemOpen
  \bibfield  {author} {\bibinfo {author} {\bibfnamefont {F.}~\bibnamefont
  {Tran}}, \bibinfo {author} {\bibfnamefont {P.}~\bibnamefont {Blaha}}, \ and\
  \bibinfo {author} {\bibfnamefont {K.}~\bibnamefont {Schwarz}},\ }\href@noop
  {} {\bibfield  {journal} {\bibinfo  {journal} {J. Chem. Theory Comput.}\
  }\textbf {\bibinfo {volume} {11}},\ \bibinfo {pages} {4717} (\bibinfo {year}
  {2015}{\natexlab{b}})}\BibitemShut {NoStop}%
\bibitem [{\citenamefont {Sun}\ \emph {et~al.}(2015{\natexlab{a}})\citenamefont
  {Sun}, \citenamefont {Ruzsinszky},\ and\ \citenamefont {Perdew}}]{SunPRL15}%
  \BibitemOpen
  \bibfield  {author} {\bibinfo {author} {\bibfnamefont {J.}~\bibnamefont
  {Sun}}, \bibinfo {author} {\bibfnamefont {A.}~\bibnamefont {Ruzsinszky}}, \
  and\ \bibinfo {author} {\bibfnamefont {J.~P.}\ \bibnamefont {Perdew}},\
  }\href@noop {} {\bibfield  {journal} {\bibinfo  {journal} {Phys. Rev. Lett.}\
  }\textbf {\bibinfo {volume} {115}},\ \bibinfo {pages} {036402} (\bibinfo
  {year} {2015}{\natexlab{a}})}\BibitemShut {NoStop}%
\bibitem [{\citenamefont {Becke}\ and\ \citenamefont
  {Roussel}(1989)}]{BeckePRA89}%
  \BibitemOpen
  \bibfield  {author} {\bibinfo {author} {\bibfnamefont {A.~D.}\ \bibnamefont
  {Becke}}\ and\ \bibinfo {author} {\bibfnamefont {M.~R.}\ \bibnamefont
  {Roussel}},\ }\href@noop {} {\bibfield  {journal} {\bibinfo  {journal} {Phys.
  Rev. A}\ }\textbf {\bibinfo {volume} {39}},\ \bibinfo {pages} {3761}
  (\bibinfo {year} {1989})}\BibitemShut {NoStop}%
\bibitem [{\citenamefont {Perdew}\ \emph {et~al.}(2009)\citenamefont {Perdew},
  \citenamefont {Ruzsinszky}, \citenamefont {Csonka}, \citenamefont
  {Constantin},\ and\ \citenamefont {Sun}}]{PerdewPRL09}%
  \BibitemOpen
  \bibfield  {author} {\bibinfo {author} {\bibfnamefont {J.~P.}\ \bibnamefont
  {Perdew}}, \bibinfo {author} {\bibfnamefont {A.}~\bibnamefont {Ruzsinszky}},
  \bibinfo {author} {\bibfnamefont {G.~I.}\ \bibnamefont {Csonka}}, \bibinfo
  {author} {\bibfnamefont {L.~A.}\ \bibnamefont {Constantin}}, \ and\ \bibinfo
  {author} {\bibfnamefont {J.}~\bibnamefont {Sun}},\ }\href@noop {} {\bibfield
  {journal} {\bibinfo  {journal} {Phys. Rev. Lett.}\ }\textbf {\bibinfo
  {volume} {103}},\ \bibinfo {pages} {026403} (\bibinfo {year} {2009})},\
  \bibinfo {note} {\textbf{106}, 179902 (2011)}\BibitemShut {NoStop}%
\bibitem [{\citenamefont {Sun}\ \emph {et~al.}(2015{\natexlab{b}})\citenamefont
  {Sun}, \citenamefont {Perdew},\ and\ \citenamefont {Ruzsinszky}}]{SunPNAS15}%
  \BibitemOpen
  \bibfield  {author} {\bibinfo {author} {\bibfnamefont {J.}~\bibnamefont
  {Sun}}, \bibinfo {author} {\bibfnamefont {J.~P.}\ \bibnamefont {Perdew}}, \
  and\ \bibinfo {author} {\bibfnamefont {A.}~\bibnamefont {Ruzsinszky}},\
  }\href@noop {} {\bibfield  {journal} {\bibinfo  {journal} {Proc. Natl. Acad.
  Sci. U.S.A.}\ }\textbf {\bibinfo {volume} {112}},\ \bibinfo {pages} {685}
  (\bibinfo {year} {2015}{\natexlab{b}})}\BibitemShut {NoStop}%
\bibitem [{\citenamefont {Tao}\ and\ \citenamefont {Mo}(2016)}]{TaoPRL16}%
  \BibitemOpen
  \bibfield  {author} {\bibinfo {author} {\bibfnamefont {J.}~\bibnamefont
  {Tao}}\ and\ \bibinfo {author} {\bibfnamefont {Y.}~\bibnamefont {Mo}},\
  }\href@noop {} {\bibfield  {journal} {\bibinfo  {journal} {Phys. Rev. Lett.}\
  }\textbf {\bibinfo {volume} {117}},\ \bibinfo {pages} {073001} (\bibinfo
  {year} {2016})}\BibitemShut {NoStop}%
\bibitem [{\citenamefont {Ou-Yang}\ and\ \citenamefont
  {Levy}(1990)}]{Ou-YangPRL90}%
  \BibitemOpen
  \bibfield  {author} {\bibinfo {author} {\bibfnamefont {H.}~\bibnamefont
  {Ou-Yang}}\ and\ \bibinfo {author} {\bibfnamefont {M.}~\bibnamefont {Levy}},\
  }\href@noop {} {\bibfield  {journal} {\bibinfo  {journal} {Phys. Rev. Lett.}\
  }\textbf {\bibinfo {volume} {65}},\ \bibinfo {pages} {1036} (\bibinfo {year}
  {1990})}\BibitemShut {NoStop}%
\bibitem [{\citenamefont {von Weizs\"{a}cker}(1935)}]{vonWeizsackerZP35}%
  \BibitemOpen
  \bibfield  {author} {\bibinfo {author} {\bibfnamefont {C.~F.}\ \bibnamefont
  {von Weizs\"{a}cker}},\ }\href@noop {} {\bibfield  {journal} {\bibinfo
  {journal} {Z. Phys.}\ }\textbf {\bibinfo {volume} {96}},\ \bibinfo {pages}
  {431} (\bibinfo {year} {1935})}\BibitemShut {NoStop}%
\bibitem [{\citenamefont {Thomas}(1927)}]{ThomasPCPS27}%
  \BibitemOpen
  \bibfield  {author} {\bibinfo {author} {\bibfnamefont {L.~H.}\ \bibnamefont
  {Thomas}},\ }\href@noop {} {\bibfield  {journal} {\bibinfo  {journal} {Proc.
  Cambridge Philos. Soc.}\ }\textbf {\bibinfo {volume} {23}},\ \bibinfo {pages}
  {542} (\bibinfo {year} {1927})}\BibitemShut {NoStop}%
\bibitem [{\citenamefont {Fermi}(1927)}]{FermiRANL27}%
  \BibitemOpen
  \bibfield  {author} {\bibinfo {author} {\bibfnamefont {E.}~\bibnamefont
  {Fermi}},\ }\href@noop {} {\bibfield  {journal} {\bibinfo  {journal} {Rend.
  Accad. Naz. Lincei}\ }\textbf {\bibinfo {volume} {6}},\ \bibinfo {pages}
  {602} (\bibinfo {year} {1927})}\BibitemShut {NoStop}%
\bibitem [{\citenamefont {R\"{a}s\"{a}nen}\ \emph {et~al.}(2010)\citenamefont
  {R\"{a}s\"{a}nen}, \citenamefont {Pittalis},\ and\ \citenamefont
  {Proetto}}]{RasanenJCP10}%
  \BibitemOpen
  \bibfield  {author} {\bibinfo {author} {\bibfnamefont {E.}~\bibnamefont
  {R\"{a}s\"{a}nen}}, \bibinfo {author} {\bibfnamefont {S.}~\bibnamefont
  {Pittalis}}, \ and\ \bibinfo {author} {\bibfnamefont {C.~R.}\ \bibnamefont
  {Proetto}},\ }\href@noop {} {\bibfield  {journal} {\bibinfo  {journal} {J.
  Chem. Phys.}\ }\textbf {\bibinfo {volume} {132}},\ \bibinfo {pages} {044112}
  (\bibinfo {year} {2010})}\BibitemShut {NoStop}%
\bibitem [{\citenamefont {Peverati}\ and\ \citenamefont
  {Truhlar}(2012)}]{PeveratiJCP12}%
  \BibitemOpen
  \bibfield  {author} {\bibinfo {author} {\bibfnamefont {R.}~\bibnamefont
  {Peverati}}\ and\ \bibinfo {author} {\bibfnamefont {D.~G.}\ \bibnamefont
  {Truhlar}},\ }\href@noop {} {\bibfield  {journal} {\bibinfo  {journal} {J.
  Chem. Phys.}\ }\textbf {\bibinfo {volume} {136}},\ \bibinfo {pages} {134704}
  (\bibinfo {year} {2012})}\BibitemShut {NoStop}%
\bibitem [{\citenamefont {Yang}\ \emph {et~al.}(2016)\citenamefont {Yang},
  \citenamefont {Peng}, \citenamefont {Sun},\ and\ \citenamefont
  {Perdew}}]{YangPRB16}%
  \BibitemOpen
  \bibfield  {author} {\bibinfo {author} {\bibfnamefont {Z.-h.}\ \bibnamefont
  {Yang}}, \bibinfo {author} {\bibfnamefont {H.}~\bibnamefont {Peng}}, \bibinfo
  {author} {\bibfnamefont {J.}~\bibnamefont {Sun}}, \ and\ \bibinfo {author}
  {\bibfnamefont {J.~P.}\ \bibnamefont {Perdew}},\ }\href@noop {} {\bibfield
  {journal} {\bibinfo  {journal} {Phys. Rev. B}\ }\textbf {\bibinfo {volume}
  {93}},\ \bibinfo {pages} {205205} (\bibinfo {year} {2016})}\BibitemShut
  {NoStop}%
\bibitem [{\citenamefont {Grabowski}\ \emph {et~al.}(2014)\citenamefont
  {Grabowski}, \citenamefont {Fabiano}, \citenamefont {Teale}, \citenamefont
  {\'{S}miga}, \citenamefont {Buksztel},\ and\ \citenamefont
  {Della~Sala}}]{GrabowskiJCP14}%
  \BibitemOpen
  \bibfield  {author} {\bibinfo {author} {\bibfnamefont {I.}~\bibnamefont
  {Grabowski}}, \bibinfo {author} {\bibfnamefont {E.}~\bibnamefont {Fabiano}},
  \bibinfo {author} {\bibfnamefont {A.~M.}\ \bibnamefont {Teale}}, \bibinfo
  {author} {\bibfnamefont {S.}~\bibnamefont {\'{S}miga}}, \bibinfo {author}
  {\bibfnamefont {A.}~\bibnamefont {Buksztel}}, \ and\ \bibinfo {author}
  {\bibfnamefont {F.}~\bibnamefont {Della~Sala}},\ }\href@noop {} {\bibfield
  {journal} {\bibinfo  {journal} {J. Chem. Phys.}\ }\textbf {\bibinfo {volume}
  {141}},\ \bibinfo {pages} {024113} (\bibinfo {year} {2014})}\BibitemShut
  {NoStop}%
\bibitem [{\citenamefont {Ospadov}\ \emph {et~al.}(2017)\citenamefont
  {Ospadov}, \citenamefont {Ryabinkin},\ and\ \citenamefont
  {Staroverov}}]{OspadovJCP17}%
  \BibitemOpen
  \bibfield  {author} {\bibinfo {author} {\bibfnamefont {E.}~\bibnamefont
  {Ospadov}}, \bibinfo {author} {\bibfnamefont {I.~G.}\ \bibnamefont
  {Ryabinkin}}, \ and\ \bibinfo {author} {\bibfnamefont {V.~N.}\ \bibnamefont
  {Staroverov}},\ }\href@noop {} {\bibfield  {journal} {\bibinfo  {journal} {J.
  Chem. Phys.}\ }\textbf {\bibinfo {volume} {146}},\ \bibinfo {pages} {084103}
  (\bibinfo {year} {2017})}\BibitemShut {NoStop}%
\bibitem [{\citenamefont {Bylander}\ and\ \citenamefont
  {Kleinman}(1990)}]{BylanderPRB90}%
  \BibitemOpen
  \bibfield  {author} {\bibinfo {author} {\bibfnamefont {D.~M.}\ \bibnamefont
  {Bylander}}\ and\ \bibinfo {author} {\bibfnamefont {L.}~\bibnamefont
  {Kleinman}},\ }\href@noop {} {\bibfield  {journal} {\bibinfo  {journal}
  {Phys. Rev. B}\ }\textbf {\bibinfo {volume} {41}},\ \bibinfo {pages} {7868}
  (\bibinfo {year} {1990})}\BibitemShut {NoStop}%
\bibitem [{\citenamefont {Marques}\ \emph {et~al.}(2011)\citenamefont
  {Marques}, \citenamefont {Vidal}, \citenamefont {Oliveira}, \citenamefont
  {Reining},\ and\ \citenamefont {Botti}}]{MarquesPRB11}%
  \BibitemOpen
  \bibfield  {author} {\bibinfo {author} {\bibfnamefont {M.~A.~L.}\
  \bibnamefont {Marques}}, \bibinfo {author} {\bibfnamefont {J.}~\bibnamefont
  {Vidal}}, \bibinfo {author} {\bibfnamefont {M.~J.~T.}\ \bibnamefont
  {Oliveira}}, \bibinfo {author} {\bibfnamefont {L.}~\bibnamefont {Reining}}, \
  and\ \bibinfo {author} {\bibfnamefont {S.}~\bibnamefont {Botti}},\
  }\href@noop {} {\bibfield  {journal} {\bibinfo  {journal} {Phys. Rev. B}\
  }\textbf {\bibinfo {volume} {83}},\ \bibinfo {pages} {035119} (\bibinfo
  {year} {2011})}\BibitemShut {NoStop}%
\bibitem [{\citenamefont {Koller}\ \emph {et~al.}(2013)\citenamefont {Koller},
  \citenamefont {Blaha},\ and\ \citenamefont {Tran}}]{KollerJPCM13}%
  \BibitemOpen
  \bibfield  {author} {\bibinfo {author} {\bibfnamefont {D.}~\bibnamefont
  {Koller}}, \bibinfo {author} {\bibfnamefont {P.}~\bibnamefont {Blaha}}, \
  and\ \bibinfo {author} {\bibfnamefont {F.}~\bibnamefont {Tran}},\ }\href@noop
  {} {\bibfield  {journal} {\bibinfo  {journal} {J. Phys.: Condens. Matter}\
  }\textbf {\bibinfo {volume} {25}},\ \bibinfo {pages} {435503} (\bibinfo
  {year} {2013})}\BibitemShut {NoStop}%
\bibitem [{\citenamefont {Skone}\ \emph {et~al.}(2014)\citenamefont {Skone},
  \citenamefont {Govoni},\ and\ \citenamefont {Galli}}]{SkonePRB14}%
  \BibitemOpen
  \bibfield  {author} {\bibinfo {author} {\bibfnamefont {J.~H.}\ \bibnamefont
  {Skone}}, \bibinfo {author} {\bibfnamefont {M.}~\bibnamefont {Govoni}}, \
  and\ \bibinfo {author} {\bibfnamefont {G.}~\bibnamefont {Galli}},\
  }\href@noop {} {\bibfield  {journal} {\bibinfo  {journal} {Phys. Rev. B}\
  }\textbf {\bibinfo {volume} {89}},\ \bibinfo {pages} {195112} (\bibinfo
  {year} {2014})}\BibitemShut {NoStop}%
\bibitem [{\citenamefont {Shimazaki}\ and\ \citenamefont
  {Nakajima}(2015)}]{ShimazakiJCP15}%
  \BibitemOpen
  \bibfield  {author} {\bibinfo {author} {\bibfnamefont {T.}~\bibnamefont
  {Shimazaki}}\ and\ \bibinfo {author} {\bibfnamefont {T.}~\bibnamefont
  {Nakajima}},\ }\href@noop {} {\bibfield  {journal} {\bibinfo  {journal} {J.
  Chem. Phys.}\ }\textbf {\bibinfo {volume} {142}},\ \bibinfo {pages} {074109}
  (\bibinfo {year} {2015})}\BibitemShut {NoStop}%
\end{thebibliography}%

\end{document}